\documentclass[showpacs,amsmath,amssymb,twocolumn,superscriptaddress,notitlepage,preprintnumbers,longbibliography]{revtex4-1}

\usepackage{braket}
\usepackage{physics}
\usepackage{mathtools}
\usepackage{diagbox}
\usepackage[utf8]{inputenc}
\usepackage{algorithm}
\usepackage{color}
\usepackage{amssymb}
\usepackage{amsmath}%
\usepackage{algpseudocode}
\usepackage{algorithm}
\usepackage{orcidlink}
\usepackage{titlesec}
\usepackage{cprotect}

\newtheorem{definition}{Definition}[section]

\newtheorem{corollary}{Corollary}

\newtheorem{innercustomthm}{Theorem}
\newenvironment{customthm}[1]
  {\renewcommand\theinnercustomthm{#1}\innercustomthm}
  {\endinnercustomthm}

  \newtheorem{innercustomprop}{Proposition}
\newenvironment{customprop}[1]
  {\renewcommand\theinnercustomprop{#1}\innercustomprop}
  {\endinnercustomprop}

\renewcommand{\tr}{\text{Tr}}
\newcommand{\qed}{\hfill\blacksquare}
\newcommand{\OO}{\mathcal{O}}
\newcommand{\UU}{\mathcal{U}}

\newcommand{\LL}{\mathcal{L}}

\newcommand{\bos}{\boldsymbol}
\newcommand{\shortxymatrix}[1]{\xymatrix@1@C=.6cm{#1}}
\newcommand{\EZero}{\shortxymatrix{\ar@{-}[r]&}}

\newcommand{\EE}{\mathcal{E}}
\newcommand{\VV}{\mathcal{V}}
\newcommand{\CC}{\mathcal{C}}

\usepackage{hyperref}
\hypersetup{colorlinks=true, linkcolor=blue, citecolor=blue, urlcolor=magenta }

\newtheorem{lemma}{Lemma}
\newtheorem{proposition}{Proposition}
\newtheorem{theorem}{Theorem}

\newcommand{\PKU}{Center on Frontiers of Computing Studies, Peking University, Beijing 100871, China}

\newcommand{\PKUCS}{School of Computer Science, Peking University, Beijing 100871, China}

\begin{document}

\title{Performance guarantees of light-cone variational quantum algorithms for the maximum cut problem}

\author{Xiaoyang Wang\orcidlink{0000-0002-2667-1879} }

\affiliation{RIKEN Center for Interdisciplinary Theoretical and Mathematical Sciences (iTHEMS), Wako 351-0198, Japan}
\affiliation{RIKEN Center for Computational Science (R-CCS), Kobe 650-0047, Japan}
\author{Yuexin Su\orcidlink{0009-0005-6098-3197}}

\affiliation{\PKU}
\affiliation{\PKUCS}

\author{Tongyang Li\orcidlink{0000-0002-0338-413X} }
\thanks{Corresponding author. Email: tongyangli@pku.edu.cn}

\affiliation{\PKU}
\affiliation{\PKUCS}

\date{\today}
\begin{abstract}
Variational quantum algorithms (VQAs) are promising to demonstrate the advantage of quantum computing over classical computing in practical applications, such as the maximum cut (MaxCut) problem. However, current VQAs such as the quantum approximate optimization algorithm (QAOA) have lower performance guarantees compared to the best-known classical algorithm, and suffer from hard optimization processes due to the barren plateau problem. We propose a light-cone VQA by choosing an optimal gate sequence of the standard VQAs, which enables a significant improvement in solution accuracy while avoiding the barren plateau problem. Specifically, we prove that the light-cone VQA with one round achieves an approximation ratio of 0.7926 for the MaxCut problem in the worst case of $3$-regular graphs, which is higher than that of the 3-round QAOA, and can be further improved to 0.8333 by a multi-angle relaxation. We conduct systematic experiments to verify our theory. On the one hand, numerical simulations demonstrate that the light-cone VQA achieves better performance over the classical Goemans-Williamson algorithm and the \verb|CPLEX| solver. On the other hand, we demonstrate on IBM’s quantum devices that the single-round light-cone VQA exceeds the known classical hardness threshold in both 72- and 148-qubit demonstrations, whereas $\text{QAOA}$ fails in the 148-qubit one. This work highlights a promising route towards solving classically hard problems on practical quantum devices.
\end{abstract}

\maketitle


\section{Introduction}
One of the central challenges for current noisy intermediate-scale quantum (NISQ) computers~\cite{Preskill_2018} is identifying how they can demonstrate practical advantages over classical computers~\cite{Daley_22,Bharti22,doi:10.1126/science.ado6285}. Various quantum algorithms have been proposed for this purpose, aiming to leverage the unique capabilities of currently available quantum devices. A prominent example is variational quantum algorithms (VQAs)~\cite{McArdle_20,Cerezo_vqa,TILLY2022,Di_Meglio_2024,Abbas_2024}, which perform optimization of a given problem using the exponentially large Hilbert space possessed by quantum computers. Quantum approximate optimization algorithm (QAOA)~\cite{farhi2014quantum,crooks2018performance, Bravyi_2020,farhi2020quantum, PhysRevA.103.042612,blekos2023review,montanezbarrera2025evaluatingperformancequantumprocess} is a representative VQA to solve the maximum cut (MaxCut) problem.
Given a connected graph $G$, the MaxCut problem asks for a bi-partition of graph nodes such that the number of edges between the two bipartite sets is maximized. The MaxCut problem can be regarded as an optimization problem that is suitable to be solved using VQAs and is hard for classical computers due to its NP-hardness~\cite{Bodlaender_1994,10.1145/502090.502098}, even if the graph $G$ is restricted to 3-regular graphs~\cite{10.1145/800133.804355,HALPERIN2004169}.

The key to VQA’s high solution accuracy relies on its parametrized quantum circuit, i.e., the ansatz, which is constructed by repeated quantum gates for many rounds. For example, the QAOA ansatz encodes one Trotter evolution step of the adiabatic Hamiltonian as one round, which promises to solve the MaxCut problem exactly as the number of rounds goes to infinity~\cite{farhi2014quantum}. However, many practical challenges are posed when the number of rounds is large. First, the barren plateau (BP) problem states that when the number of circuit rounds increases polynomially to the system size, the optimization process is hard to proceed due to exponentially vanishing gradients~\cite{McClean_18, Cerezo_21,ragone_2024,fontana2023adjoint}. Second, deep noisy quantum circuits can be simulated more efficiently as the noise rate increases using classical algorithms~\cite{PhysRevLett.133.120603,10.1145/3564246.3585234}, and the possible quantum advantage is hard to approach if the circuits are classically simulatable. Finally, for moderate and shallow depth quantum circuits, their theoretical performance guarantees still have notable distance from that of the best-known classical algorithms~\cite{farhi2014quantum, PhysRevA.103.042612, Bharti22, GoemansWilliamson1995}.

\begin{figure}
    \centering
    \includegraphics[width=0.48\textwidth]{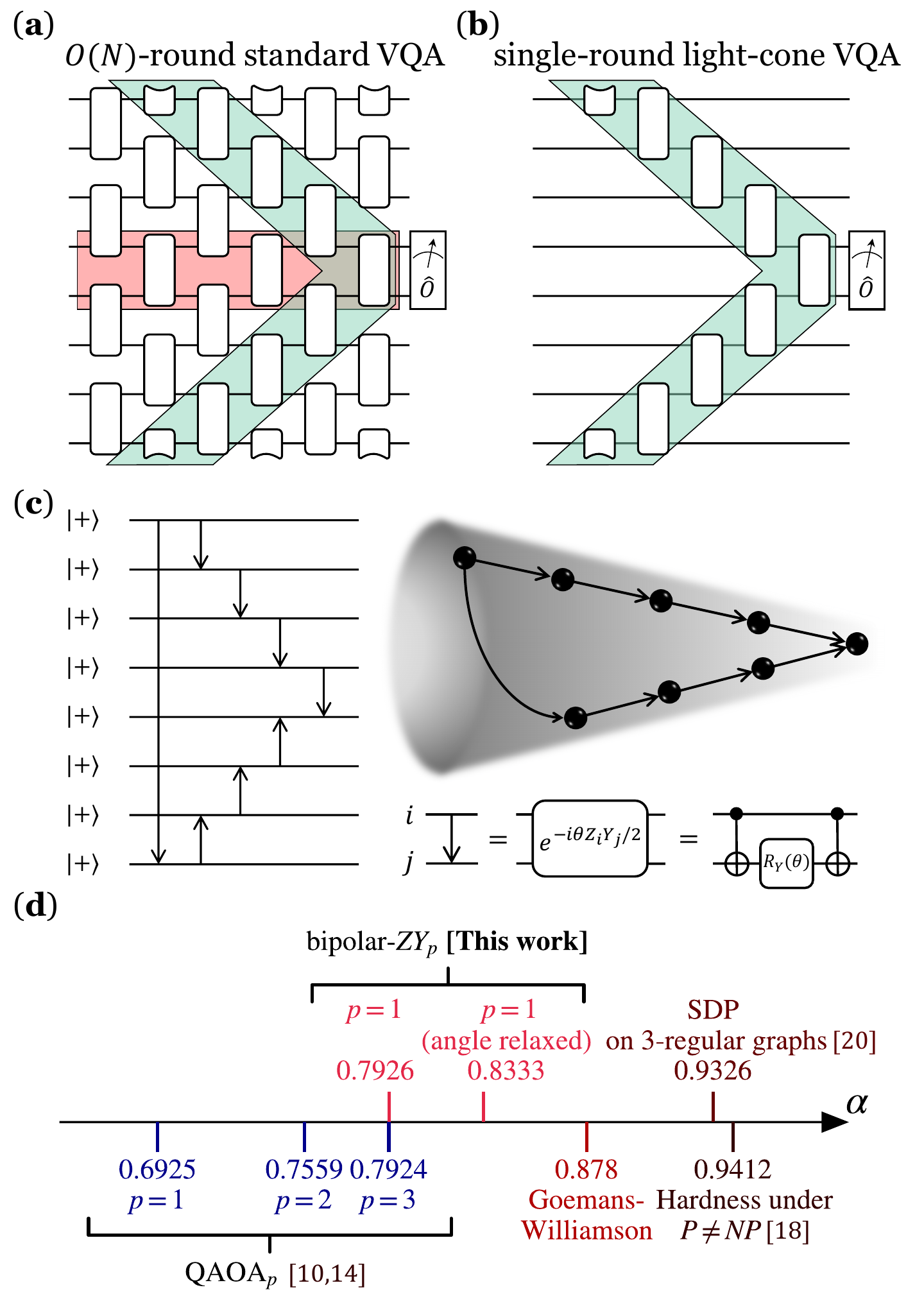}
    \caption{Main ideas and results of this work. (\textbf{a}) and (\textbf{b}) illustrate the standard VQA with $\OO(N)$ rounds and the light-cone VQA with one round, respectively. The light-cone VQA has two-qubit gates expanded by the backward light-cone of the local observable $\hat{O}$. (\textbf{c}) The single-round light-cone VQA for the MaxCut problem. The light-cone VQA can be represented by a directed acyclic graph. The directed edge denotes a $ZY$ gate $e^{-i\theta ZY/2}$, and can be decomposed into a single-qubit gate $R_Y(\theta)\equiv e^{-i\theta Y/2}$ and two CNOT gates. (\textbf{d}) Performance guarantees of the bipolar-$ZY_p$ ansatz, a specific light-cone ansatz with $p$-round, and its comparison with QAOA and classical algorithms. $0.9412$ is the hardness threshold~\cite{10.1145/502090.502098}.}
    \label{fig:intro-figure}
\end{figure}

In this work, we address the above challenges by proposing a VQA with a light-cone ansatz. A schematic illustration of our idea is shown in Fig.~\ref{fig:intro-figure}(\textbf{a},\textbf{b}). For standard VQAs, each round has quantum gates arranged in a brick-wall pattern~\cite{farhi2014quantum,Cerezo_21,Uvarov_2021,Pesah_21}. Fig.~\ref{fig:intro-figure}(\textbf{a}) illustrates the brick-wall pattern consisting of two-qubit gates. This ansatz in general suffers from the BP problem when the number of rounds $p$ increases linearly with the system size~\cite{Cerezo_21}. For a local observable $\hat{O}$, the BP problem emerges due to a linearly increasing circuit local depth~\cite{zhang_absence_2024}, i.e., the number of two-qubit gates in the red region of Fig.~\ref{fig:intro-figure}(\textbf{a}). On the other hand, consider the two-qubit gates expanding the backward light-cone of $\hat{O}$ in the green region of Fig.~\ref{fig:intro-figure}(\textbf{a}). This extensive backward light-cone makes the quantum circuits in general hard to simulate by existing classical methods~\cite{doi:10.1137/050644756,slattery2021quantumcircuitstwodimensionalisometric,PhysRevLett.124.037201,PhysRevLett.128.010607}. Thus, we propose a light-cone ansatz whose each round has two-qubit gates expanding the backward light-cones, as shown in Fig.~\ref{fig:intro-figure}(\textbf{b}). This ansatz is free from the BP problem due to a finite local depth~\cite{zhang_absence_2024}, and is in general hard to simulate classically due to the extensive backward light-cone.

We apply the light-cone VQAs to the MaxCut problem. Given a connected graph $G=(\VV,\EE)$ consisting of nodes $i\in\VV$ and edges $(i,j)\in\EE$, the MaxCut problem is equivalent to finding the minimum eigenvalue of the MaxCut Hamiltonian
\begin{align}
H_{\text{MC}}\equiv\sum_{(i,j)\in\EE}(Z_iZ_j-1)/2.
\end{align}
Different from the two-qubit gates $e^{-i\gamma Z_iZ_j/2}$ used in QAOA, the light-cone ansatz uses the $ZY$ gate $e^{-i\theta Z_iY_j/2}$ as the basic gate. We choose $ZY$ gates instead of $ZZ$ because, on the one hand, the resulting circuits cannot be compressed into a brick-wall pattern as in QAOA~\cite{Bravyi_2020,thanh2025hamiltonianreorderingshallowertrotterization} due to the non-commutativity of $ZY$ gates on adjacent edges. On the other hand, the  $ZY$ gate is closely related to the imaginary-time and the counter-diabatic evolution of the $Z_iZ_j$ term in $H_{\text{MC}}$~\cite{PhysRevA.111.032612,Wurtz2022counterdiabaticity,PhysRevA.105.042415,guan2023singlelayerdigitizedcounterdiabaticquantumoptimization}. By denoting $ZY$ gates as directed edges as in the left panel of Fig.~\ref{fig:intro-figure}(\textbf{c}), the light-cone ansatz naturally corresponds to a directed acyclic graph (DAG)~\cite{AlgorithmDesign2005} illustrated in the right panel of Fig.~\ref{fig:intro-figure}(\textbf{c}).

We propose a new analysis method to estimate the performance guarantees of the light-cone VQA. The performance guarantee of a MaxCut algorithm is the worst-case approximation ratio $\alpha\in[0,1]$ obtained by the algorithm among all graph instances, where the exact MaxCut solution has $\alpha=1$. Although the analysis method has been conducted for the $\text{QAOA}_p$ circuits~\cite{farhi2014quantum,PhysRevA.103.042612} where $p$ denotes the number of circuit rounds, the method is only applicable to the standard ans\"atze with the brick-wall pattern. The proposed method in this paper is applied to analyzing ans\"atze with an arbitrary gate sequence by taking local truncations, where the truncation errors are estimated and well controlled. Specifically, among all gate sequences, we rigorously prove that the obtained light-cone sequence, additionally combined with the bipolar orientation~\cite{Tarjan1982,10.1007/11618058_32}, achieves the optimal performance. This leads to a specific light-cone ansatz which we named the \textit{bipolar-$ZY$} ansatz. The performance guarantees of the bipolar-$ZY$ ansatz on $3$-regular graphs are presented in Fig.~\ref{fig:intro-figure}(\textbf{d}). In particular, we prove that the bipolar-$ZY_1$ ansatz is guaranteed to achieve an approximation ratio of 0.7926, better than that of $\text{QAOA}_3$ shown in Ref.~\cite{PhysRevA.103.042612}. By an additional multi-angle relaxation, the performance guarantee can be further improved to 0.8333.

Finally, both numerical simulations and hardware implementations are conducted to verify the performance of the light-cone VQA. For random 3-regular graphs, we numerically show that the angle-relaxed bipolar-$ZY$ ansatz achieves larger approximation ratios than the classical Goemans-Williamson algorithm~\cite{GoemansWilliamson1995}, and the scaling advantage of the multi-angle bipolar-$ZY_1$ ansatz over the classical \verb|CPLEX| solver~\cite{cplex} is observed. On IBM’s quantum hardware, the bipolar-$ZY_1$ ansatz finds the exact MaxCut solution with $\alpha=1$ in a $72$-qubit demonstration, and $\alpha=0.965$ in a $148$-qubit demonstration. Both surpass the hardness threshold $\alpha=0.941$~\cite{10.1145/502090.502098}, whereas $\text{QAOA}_p$ with $p=1,2,3$ fails in the $148$-qubit case.

\section{Light-cone variational ansatz}\label{sec:MaxCut and the non-local ZY ansatz}

For an arbitrary connected graph $G=(\VV,\EE)$ with $N$ nodes, we define its \textit{$ZY_p$ ansatz} with $p$ rounds of gates, which has one $ZY$ gate $e^{-i\theta Z_iY_j/2}$ for every undirected edge $(i,j)\in\EE$ in each round. The variational ansatz state reads
\begin{align}
    \ket{\phi_p(\bos{\theta})} = \UU_p(\bos{\theta})\ket{+}^{\otimes N},
    \label{eq:ZY_p_ansatz}
\end{align}
where $\UU_p(\bos{\theta})= \prod_{l=0}^{p-1} \prod_{(i,j)\in\EE}e^{-i\theta_l Z_iY_j/2}$ and $\ket{+}$ is the eigenstate of Pauli-$X$ with an eigenvalue $1$. For our convenience, we use $i$ to represent both the index of a graph node and the imaginary unit when there is no ambiguity. Each round of $ZY_p$ ansatz is determined by a $ZY$ orientation and a gate sequence defined as follows:
\begin{definition}[$ZY$ orientation]
    For every undirected edge $(i,j)\in\EE$, $ZY$ orientation assigns a direction $i\rightarrow j$ (resp.~$i\leftarrow j$) corresponding to the $ZY$ gate $e^{-i\theta_l Z_iY_j/2}$ (resp.~$e^{-i\theta_l Y_iZ_j/2}$) in one round of the $ZY_p$ ansatz.
\end{definition}
\begin{definition}[gate sequence]
    The gate sequence assigns an order of the product $\prod_{(i,j)\in\EE}$ in one round of the $ZY_p$ ansatz.
\end{definition}

\begin{figure}
    \centering
    \includegraphics[width=0.42\textwidth]{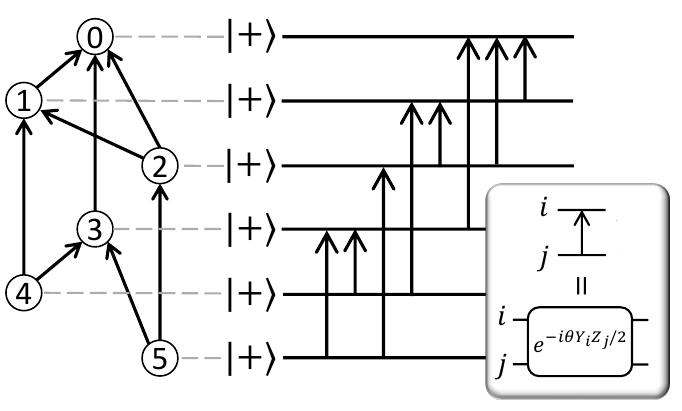}
    \caption{The lightcone-$ZY_1$ ansatz for a graph $G$ using the breadth-first search (BFS) traversal. BFS gives a directed acyclic graph (DAG, left panel) of $G$. The topological order of the DAG and the edge direction give the gate sequence and the $ZY$ orientation of the lightcone-$ZY$ ansatz (right panel), respectively.}
    \label{fig:dag_construction}
\end{figure}

Our idea of constructing the light-cone ansatz provides a primitive $ZY$ orientation and a gate sequence using the breadth-first search (BFS) traversal, which resembles the light propagation in the graph. Starting from an arbitrary node $s\in\VV$, BFS traverses the neighbors of $s$ if the neighbor has never been visited, and then all nodes of $G$ are traversed recursively. The order of traversing nodes gives the order of the product $\prod_{(i,j)\in\EE}$, i.e., the gate sequence, and we choose the $ZY$ orientation $i\leftarrow j$ if $i$ is the parent of $j$ in the BFS traversal. Figure~\ref{fig:dag_construction} shows an example of the $ZY$ oriented graph by the BFS traversal (left panel) and the corresponding \textit{lightcone-$ZY_p$ ansatz} with $p=1$ (right panel), which can be naturally generalized for arbitrary $p$~\cite{supp}.

The lightcone-$ZY_1$ ansatz and the corresponding directed graph have the following properties. First, assuming that an observable $\hat{O}$ is non-trivial at the selected root node $s$, its backward light-cone covers the whole system. For example, the MaxCut Hamiltonian of the graph with $s=0$ in Fig.~\ref{fig:dag_construction} has local observables $Z_0Z_1, Z_0Z_2, Z_0Z_3$ whose backward light-cones cover all the qubits. Second, the directed graph generated by the BFS traversal is a directed acyclic graph (DAG). The directed graph induced by a BFS traversal is acyclic because each directed edge connects a node to a newly discovered node at a strictly greater BFS level, so any directed path is strictly increasing in BFS depth and therefore cannot contain a directed cycle.

We briefly introduce some useful terminologies of the directed graph and DAG. For a directed edge $(i\to j)$, we say edge $(i,j)$ \textit{leaves} node $i$ and \textit{enters} node $j$. The in-degree and out-degree of a node $i$ are the numbers of edges entering and leaving $i$, respectively, denoted by $\deg^-(i)$ and $\deg^+(i)$. A $D$-regular graph, denoted by $G_D$ in this work, has $\deg^-(i)+\deg^+(i)=D, \forall i\in\VV$. A \textit{source} of a directed graph is the node having all adjacent edges leaving it, and a \textit{sink} has all edges entering it. The number of sources and sinks in the directed graph are denoted by $N_+$ and $N_-$, respectively. A DAG must have at least one source and one sink, i.e., $N_+\geq1$ and $N_-\geq1$. In particular, because BFS has a unique root node and we choose the $ZY$ orientation $i\leftarrow j$ if $i$ is the parent of $j$ in the BFS traversal, the directed graph of the lightcone-$ZY$ ansatz has $N_-=1$. A \textit{topological order} of the DAG is an ordering of its nodes as $v_0,\ldots,v_{N-1}$ so that for every edge $v_i\leftarrow v_j$, we have $i<j$. The lightcone-$ZY$ ansatz gate sequence is exactly given by the topological order of the corresponding DAG, as illustrated in Fig.~\ref{fig:dag_construction}. 

The lightcone-$ZY_p$ ansatz for $D$-regular graphs with a constant $p$ is free from the barren plateau (BP) problem. Barren plateaus can be diagnosed by calculating the variance of the energy expectation over the variational parameters
\begin{align}
    \mathrm{Var}(\langle H_{\mathrm{MC}}\rangle) \equiv \mathrm{Var}_{\bos{\theta}}(\bra{\phi_p(\bos{\theta})}H_{\mathrm{MC}}\ket{\phi_p(\bos{\theta})}),
\end{align}
where $\bos{\theta}$ is a random vector with each element picked independently and uniformly from $[0, 2\pi)$. If $\mathrm{Var}(\langle H_{\mathrm{MC}}\rangle)$  does not vanish exponentially with the number of nodes, then the energy landscape
of the ansatz is free from BPs~\cite{Arrasmith_2022}. Since the lightcone-$ZY_p$ ansatz belongs to the $p$-round imaginary Hamiltonian variational ansatz ($i$HVA) defined in Ref.~\cite{PhysRevA.111.032612}, the lightcone-$ZY_p$ ansatz has a lower bounded energy variance, as a corollary of Theorem~2 in Ref.~\cite{PhysRevA.111.032612}.
\begin{corollary}\label{corollary:BP-free}
For the lightcone-$ZY_p$ ansatz solving MaxCut on $D$-regular graphs with $N$ nodes, the variance of the energy expectation is lower bounded by
\begin{align}
    \mathrm{Var}(\langle H_{\mathrm{MC}} \rangle)\geq \frac{DN}{ 2^{D(2\lceil p/2 \rceil+1)-1}}.
    \label{eq:variance-bound}
\end{align}
\end{corollary}
This corollary guarantees that the lightcone-$ZY_p$ ansatz for $D$-regular graphs with a constant $p$ is free from BP. In the following content, we focus on solving the MaxCut problem of $D$-regular graphs. 

Beyond the MaxCut problem, the light-cone VQA is BP-free for general models in quantum many-body systems on finite-dimensional lattices. Because a finite-dimensional lattice has a constant graph degree, leading to a finite local depth of the corresponding light-cone ansatz, and thus the ansatz is BP-free~\cite{zhang_absence_2024}. In Supplemental Material~\cite{supp}, we show that the light-cone ansatz of the $\mathbb{Z}_2$ gauge model on a 2-$d$ lattice is BP-free and can exactly prepare the ground state of the $\mathbb{Z}_2$ gauge model.



\subsection{Performance guarantees by local analysis}\label{sec:k-local-analysis}
\begin{figure}
    \centering
    \includegraphics[width=0.40\textwidth]{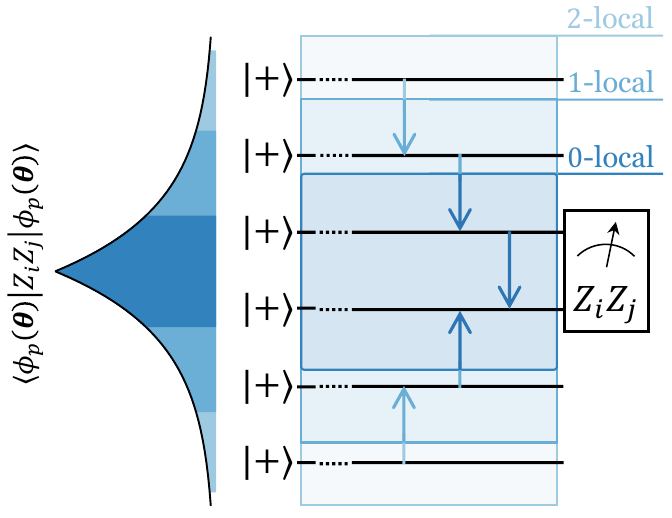}
    \caption{Schematic illustration of $k$-local analysis. $k$-local analysis calculates the expectation of $Z_iZ_i$ by truncating the quantum gates applied on qubits that are all more than $k$ steps away from the central edge $(i,j)$. The truncation error of $k$-local analysis can be well controlled since the distribution of $Z_iZ_j$ expectation decays exponentially with $k$, as shown schematically in the left panel.}
    \label{fig:k_local_analysis}
\end{figure}

For a given connected graph $G$, the performance of the lightcone-$ZY$ ansatz $\ket{\phi_p(\bos{\theta})}$ can be evaluated by the approximation ratio
\begin{align}
    \alpha(G) \equiv\frac{\max_{\bos{\theta}}(-\bra{\phi_p(\bos{\theta})} H_{\text{MC}}\ket{\phi_p(\bos{\theta})})}{C_{\max}},
    \label{eq:approx-ratio-definition}
\end{align}
where the numerator is the expected cut number of the ansatz, and $C_{\max}$ is the maximum cut number of $G$. The performance guarantee of a certain class of graphs is defined as the worst-case approximation ratio $\alpha\equiv \min_G \alpha(G)$ among all instances in this class. Thus, to calculate the performance guarantee $\alpha$, we should be able to evaluate the expectation of the MaxCut Hamiltonian
\begin{equation}
    \begin{aligned}
    &\bra{\phi_p(\bos{\theta})} H_{\text{MC}}\ket{\phi_p(\bos{\theta})} \\
    = &\frac{1}{2}\sum_{(i,j)\in\EE}(\bra{+}^{\otimes N} \UU_p^{\dagger}(\bos{\theta}) Z_iZ_j \UU_p(\bos{\theta})\ket{+}^{\otimes N}-1).\nonumber
\end{aligned}
\end{equation}
However, for the lightcone-$ZY_p$ ansatz, calculating all the $Z_iZ_j$ expectations exactly by considering the backward light-cone of $Z_iZ_j$ is challenging in general. Instead, we can evaluate $\bra{\phi_p(\bos{\theta})} H_{\text{MC}}\ket{\phi_p(\bos{\theta})}$ approximately. This is achieved using the $k$-local analysis defined as follows:

\begin{definition}[$k$-local analysis]
For each observable $Z_iZ_j$ in $H_{\text{MC}}$, $k$-local analysis calculates the expectation $\bra{\phi_p(\bos{\theta})}Z_iZ_j\ket{\phi_p(\bos{\theta})}$ by truncating the quantum gates applied on qubits that are all more than $k$ steps away from the center edge $(i,j)$.
\end{definition}
Figure~\ref{fig:k_local_analysis} illustrates the truncation of the lightcone-$ZY_p$ ansatz. In $k$-local analysis, each truncated quantum circuit has a constant number of qubits, and hence the $Z_iZ_j$ expectation can be calculated approximately by considering only $ZY$ gates inside the $k$-local circuits. For an arbitrary observable $O$, we denote the approximate expectation by $k$-local analysis as  $\bra{\phi_p(\bos{\theta})}O \ket{\phi_p(\bos{\theta})}_k$, and define the approximate performance guarantee
\begin{align}
    \alpha_k \equiv \min_G\frac{\max_{\bos{\theta}}(-\bra{\phi_p(\bos{\theta})} H_{\text{MC}}\ket{\phi_p(\bos{\theta})}_k)}{C_{\max}}.
\end{align}
Note that $k$-local analysis retrieves the exact expectation and the performance guarantee as $k\rightarrow \infty$. 

For a finite $k$, the error of approximating $\bra{\phi_p(\bos{\theta})}Z_iZ_j\ket{\phi_p(\bos{\theta})}$ by $\bra{\phi_p(\bos{\theta})}Z_iZ_j\ket{\phi_p(\bos{\theta})}_k$ can be well controlled. Specifically, for the $ZY_p$ ansatz, we prove the following proposition~\cite{supp}:
\begin{proposition}\label{prop:truncation-error}
    For each observable $Z_iZ_j$ in $H_{\text{MC}}$ and the $ZY_p$ ansatz state $\ket{\phi_p(\bos{\theta})}$, the truncation error of the $k$-local analysis $|\bra{\phi_p(\bos{\theta})}Z_iZ_j\ket{\phi_p(\bos{\theta})}_k- \bra{\phi_p(\bos{\theta})}Z_iZ_j\ket{\phi_p(\bos{\theta})}|$ is bounded by $\OO(\sin^{2k+1}\|\bos{\theta}\|_{\infty})$.
\end{proposition}
Hence, the truncation error of $k$-local analysis decreases exponentially with $k$ as long as $\|\bos{\theta}\|_{\infty}$ is far from $\pm\pi/2$. This decay behavior is illustrated in the left panel of Fig.~\ref{fig:k_local_analysis}. In this proposition, the error bound focuses on its dependence on $k$. Its dependence on the graph degree $D$ and the circuit rounds $p$ is ignored. For example, the truncation error can be larger for larger $p$ due to more complicated circuit structures, which will be observed explicitly in the $k$-local analysis for $3$-regular graphs in Table~\ref{tab:performance-guarantees-by-k-local-analysis} of the next subsection.

\begin{figure}
    \centering
    \includegraphics[width=0.48\textwidth]{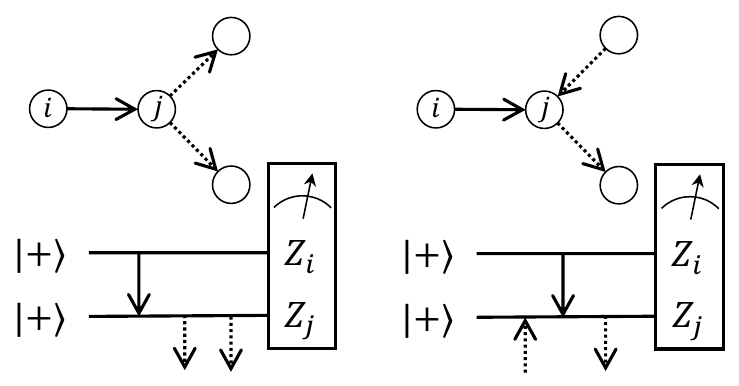}
    \caption{Two kinds of directed edges $(i\to j)$ in $3$-regular graphs with $j$'s in-degree $\deg^-(j)=1$ (left panel) and $\deg^-(j)=2$ (right panel) and their lightcone-$ZY_1$ circuits. In both circuits, $ZY$ gates entering $j$ are arranged to the left of gates leaving $j$. Directed edges with $\deg^-(j)=1$ and $\deg^-(j)=2$ contribute to the terms of $\sin\theta$ and $\cos\theta\sin\theta$ in Eq.~\eqref{eq:alpha_0_lower-bound}, respectively.}
    \label{fig:two-kind-edges}
\end{figure}

According to Proposition~\ref{prop:truncation-error}, for each $Z_iZ_j$ in $H_{\text{MC}}$, the leading contribution of its expectation is the $0$-local $\bra{\phi_p(\bos{\theta})}Z_iZ_j\ket{\phi_p(\bos{\theta})}_0$. Due to the dominance and simplicity of the $0$-local result, it can be used to find the $ZY_p$ ansatz with the optimal $ZY$ orientation and gate sequence by maximizing the approximate performance guarantee $\alpha_0$. We find that to maximize $\alpha_0$, for each qubit $j$ of the $ZY_1$ ansatz, its entering $ZY$ gates $(i\to j)$ should be arranged to the left of its leaving $ZY$ gates $(j\to k)$~\cite{supp}, as illustrated in Fig.~\ref{fig:two-kind-edges}. This requirement cannot be fulfilled simultaneously by all qubits if the $ZY$ orientation contains directed cycles. In other words, the optimal $ZY$ orientation with the maximized $\alpha_0$ should be acyclic, and the optimal gate sequence arranges $ZY$ gates following the topological order of the resulting directed acyclic graph (DAG), as we did in constructing the lightcone-$ZY_1$ ansatz. Additionally, we show that the optimal $ZY$ orientation should have DAG with the minimal number of the sink node $N_-=1$. One can check that the lightcone-$ZY_1$ ansatz in Fig.~\ref{fig:dag_construction} satisfies these requirements.

For $3$-regular graphs, the $ZY_1$ ansatz equipped with these optimal properties has a specific lower bound of $\alpha_0$ depending only on the number of sources $N_+$ of the $3$-regular DAG, whereas the more detailed structure of the DAG is irrelevant. Formally, we prove the following theorem~\cite{supp}.

\begin{theorem}\label{theorem:ZY_1-ansatz-optimal}
    For a 3-regular graph with the number of nodes $N\to \infty$, among all $ZY_1$ ans\"atze with different $ZY$ orientations and gate sequences, $\alpha_0$ has the maximal lower bound
\begin{equation}
\begin{aligned}
    \alpha_0\geq \max_{\theta} \frac{1}{2} [1&+(1-k_{N_+})\sin\theta+k_{N_+}\cos\theta\sin\theta],
    \label{eq:alpha_0_lower-bound}
\end{aligned}        
\end{equation}
where $k_{N_+}\equiv 2/3+4 N_+/(3N)$, and this maximal lower bound is achieved by the lightcone-$ZY_1$ ansatz.
\end{theorem}

\begin{figure*}
    \centering
    \includegraphics[width=0.97\textwidth]{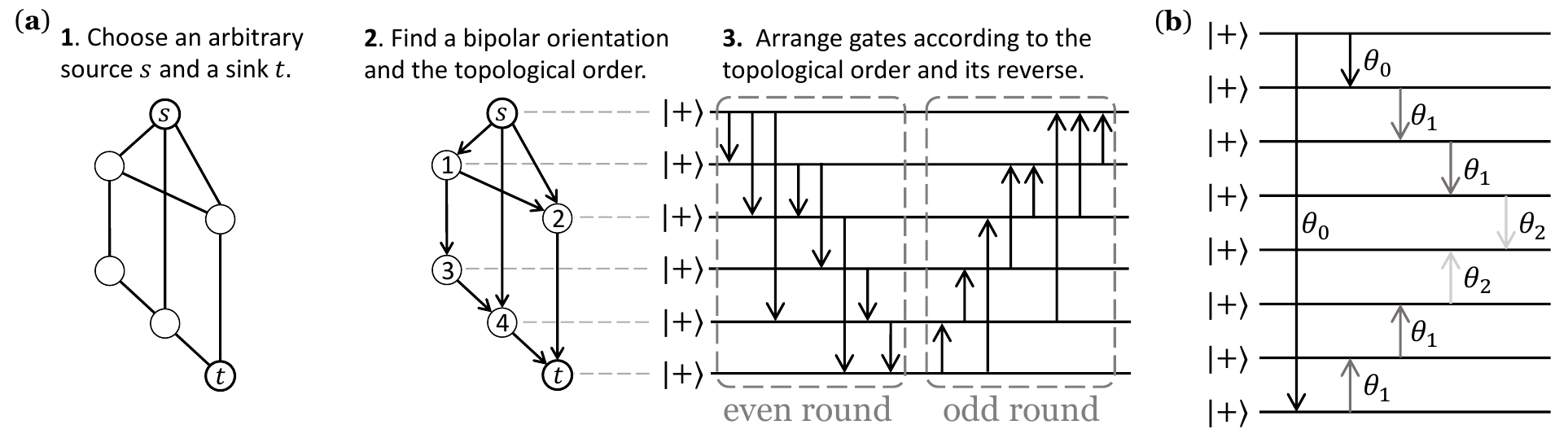}
    \caption{Bipolar-$ZY$ and the angle-relaxed ansatz. (\textbf{a}) Construction procedures of the bipolar-$ZY_p$ ansatz for a biconnected graph described in Algorithm~\ref{alg:dag-ansatz}. The graph is biconnected because it remains connected if any one node were to be removed. $s$ and $t$ are chosen as the source and sink nodes in the bipolar orientation. (\textbf{b}) Illustration of the angle-relaxed bipolar-$ZY_1$ ansatz for a $2$-regular graph of length $8$ with $3$ free angles labeled for each $ZY$ gate.} 
    \label{fig:bipolar_ZY_ansatz}
\end{figure*}

We have three remarks. First, the lower bound in Eq.~\eqref{eq:alpha_0_lower-bound} gets larger as $N_+$ is smaller. If we construct a specific DAG with an $N_+$ constant to $N$, this leads to $\lim_{N\to\infty }k_{N_+}=2/3$ and the $0$-local performance guarantee
\begin{align}
    \alpha_0 \geq \max_{\theta}\frac{1}{2} (1&+\frac{1}{3}\sin\theta+\frac{2}{3}\cos\theta\sin\theta)=0.7934.
    \label{eq:0-local-lower-bounds-ZY1}
\end{align}
We will fabricate this specific DAG later. Second, the $\sin\theta$ and $\cos\theta\sin\theta$ in Eq.~\eqref{eq:alpha_0_lower-bound} come from two kinds of directed edges $(i\to j)$ in $3$-regular graphs. $\sin\theta$ corresponds to $\deg^-(j)=1$ and $\cos\theta\sin\theta$ corresponds to $\deg^-(j)=2$, as shown in Fig.~\ref{fig:two-kind-edges}. We can use this observation to further improve the lower bound as will be detailed later. Third, if we do not take the topological order in the ansatz gate sequence, or $N_-$ of the DAG increases linearly with $N$, then $\cos^{2}\theta\sin\theta$ will appear in Eq.~\eqref{eq:alpha_0_lower-bound}~\cite{supp}, leading to a degradation in $\alpha_0$ of the ansatz. For example, a lower bound consisting only of a $\cos^2\theta\sin\theta$ term gives
\begin{align}
    \alpha_0\geq \max_{\theta} \frac{1}{2} (1&+\cos^2\theta\sin\theta) = 0.6925,
\end{align}
which equals to the performance guarantee of $\text{QAOA}_1$ for $3$-regular graphs~\cite{farhi2014quantum}. Thus, the lightcone-$ZY_1$ ansatz performs better than $\text{QAOA}_1$ for $3$-regular graphs in the sense of the $0$-local analysis.

\subsection{Bipolar-$ZY$ ansatz}\label{sec:Bipolar-$ZY$ ansatz}

According to Theorem~\ref{theorem:ZY_1-ansatz-optimal}, the lower bound of $\alpha_0$ is increased as the number of the source node $N_+$ decreases. Thus, we look for a $ZY$ orientation of the undirected graph $G$ which leads to a DAG with $N_+=N_-=1$. This is known as the \emph{bipolar orientation} of $G$. 
\begin{definition}[Bipolar orientation]
Bipolar orientation (or st-orientation) of an undirected graph $G$ is an assignment of a direction to each edge such that the directed graph under this orientation is a DAG with a single source $s$ and a single sink $t$.
\end{definition}
We call the lightcone-$ZY$ ansatz whose $ZY$ orientation is bipolar as the \textit{bipolar-$ZY$ ansatz}.  

The bipolar orientation of an arbitrary connected graph can be obtained efficiently with a linear time complexity to the system size~\cite{Tarjan1982,PAPAMANTHOU2008224}, with an equally efficient preprocessing of decomposing the connected graph $G$ into biconnected components~\cite{Grotschel_1984}. Thus, to solve the MaxCut problem of $G$, the first step is to decompose $G$ into biconnected components. Second, for each biconnected component $G^{(2)}\subseteq G$, we construct the bipolar-$ZY_p$ ansatz following the steps in Algorithm~\ref{alg:dag-ansatz}. In this algorithm, we arbitrarily choose two nodes of $G^{(2)}$ to be the source and sink node, since different choices do not affect the local structure of the ansatz, and thus they have the same approximate performance guarantees. Additionally, as a multi-round ansatz, we alternatively reverse the topological order of the bipolar DAG for the even and the odd round of the ansatz, such that both the $ZY$ orientation and the gate sequence are alternatively reversed in each round as we did for the lightcone-$ZY_p$ ansatz. A schematic illustration of the algorithm is shown in Fig.~\ref{fig:bipolar_ZY_ansatz}(\textbf{a}). Finally, the MaxCut solution of $G$ is retrieved by combining the MaxCut solutions of all $G^{(2)}$ components, and we prove the retrieved approximation ratio $\alpha$ of $G$ is no smaller than each component $G^{(2)}$~\cite{supp}. Therefore, in the following discussion, we assume by default that the studied graph is biconnected.

\begin{algorithm}[H]
\caption{Construct the bipolar-$ZY_p$ ansatz for the biconnected component $G^{(2)}\subseteq G$}\label{alg:dag-ansatz}
\begin{algorithmic}
 \State Find a bipolar orientation of $G^{(2)}$ with an arbitrary source $s$ and a sink $t$.
 \State Compute the topological ordering $T_0$ of nodes according to the bipolar orientation of $G^{(2)}$. 
 \State Denote $T_1$ as the reverse of the ordering $T_0$.
 \State Initialize the quantum state as $\ket{+}^{\otimes N}$.
 \State Initial $l\gets 0$.
\While{$l < p$}
\For{node $i$ in $T_{(l~\text{mod}~2)}$}
        \State For every directed edge $(i\rightarrow j)$ of the DAG, arrange a $ZY$ gate $e^{-i\theta Z_iY_j/2}$ to the quantum circuit.
\EndFor
\State$l\gets l+1$
\EndWhile
\end{algorithmic}
\end{algorithm}

\begin{table}[]
\begin{tabular}{c|l|l|l}
\hline
 $\alpha_k$             & $p=1$    & $p=2$    & $p=3$    \\
\hline\hline
0-local $(\dagger)$       & 0.7934 & 0.8025 & 0.8402 \\
1-local $(\ddagger)$     & 0.7934 & 0.7988 & 0.8084 \\
\hline
0-local error $|(\dagger)-(\ddagger)|$ & 0      & 0.0037 & 0.0318\\
\hline
\end{tabular}
\caption{Performance guarantees $\alpha_k$ of bipolar-$ZY_p$ ansatz on 3-regular graphs by $k$-local analysis.}
\label{tab:performance-guarantees-by-k-local-analysis}
\end{table}


For bipolar-$ZY_p$ ansatz with $p=1,2,3$, we perform $0$- and $1$-local analysis to provide its performance guarantees on $3$-regular graphs with results listed in Table~\ref{tab:performance-guarantees-by-k-local-analysis}. Let us make a few remarks. First, in the $1$-local row of Table~\ref{tab:performance-guarantees-by-k-local-analysis}, $\alpha_1$ increases marginally as $p$ grows. This indicates that increasing the ansatz rounds is not a proper way to improve the solution accuracy. Second, the truncation error of the $0$-local analysis can be roughly estimated by the difference between $\alpha_0$ and $\alpha_1$, as shown in the last row of Table~\ref{tab:performance-guarantees-by-k-local-analysis}. Larger $p$ has a larger systematic error by the truncation of the $k$-local analysis. Thus, the local analysis may not be suitable to evaluate the performance of the light-cone VQAs with multiple rounds.

For bipolar-$ZY_1$ ansatz on $3$-regular graphs, we provide a rigorous performance guarantee $\alpha \equiv \min_{G_3}(\alpha(G_3))$. $\alpha$ can be derived because the bipolar-$ZY_1$ ansatz has only one $ZY$ gate applied on each edge of the problem graph, such that the expected cut number in Eq.~\eqref{eq:approx-ratio-definition} can be evaluated exactly. For the bipolar-$ZY_1$ ansatz state $\ket{\phi_1(\theta)}$, we find that an odd (even)-length cycle in the graph $G_3$ leads to a negative (positive) contribution to the cut number of the ansatz. Thus, we construct worst-case candidates of $3$-regular graphs with the maximum possible numbers of odd-length cycles, such that the ansatz cut number is minimized. Evaluating the approximation ratios on these candidates and selecting the worst one, we provide the following theorem.
\begin{theorem}\label{theorem:exact-ZY1-bound}
    For infinite $3$-regular graphs, the bipolar-$ZY_1$ ansatz has the performance guarantee lower bounded by 
    \begin{align}
        \alpha \geq 0.7926.
        \label{eq:exact-lower-bounds-ZY1}
    \end{align}
\end{theorem}
This lower bound is larger than the performance guarantees of $\text{QAOA}_3$ with $\alpha\geq 0.7924$ for $3$-regular graphs~\cite{PhysRevA.103.042612}. Thus, the solution accuracy of bipolar-$ZY_1$ is higher than that of $\text{QAOA}_3$ on MaxCut problems of $3$-regular graphs with a large number of nodes. The proof and numerical verification of this theorem is presented in the Supplemental Material~\cite{supp}.


\begin{figure}
    \centering
    \includegraphics[width=0.49\textwidth]{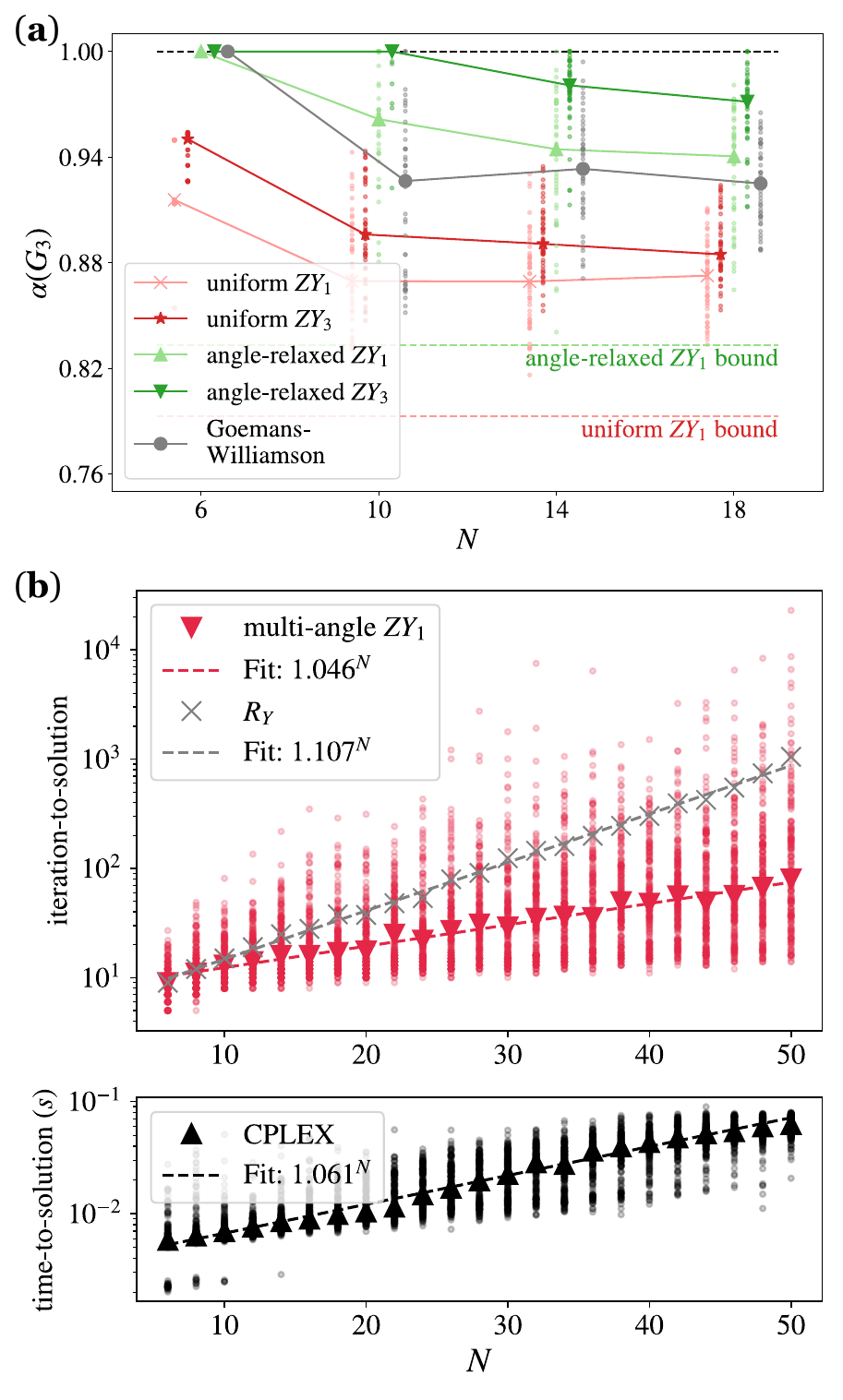}
    \cprotect\caption{(\textbf{a}) The median approximation ratios $\alpha(G_3)$ derived by the bipolar-$ZY_p$ ansatz and the classical Goemans-Williamson algorithm versus the number of nodes of 3-regular graphs. $50$ random $3$-regular graphs are generated for each $N$ with the ratios plotted by dots. The performance bounds of the uniform and angle-relaxed $ZY_1$ ansatz are given by the red and green dashed lines, respectively. (\textbf{b}) The median iteration-to-solution as a function of graph nodes $N$ using the multi-angle bipolar-$ZY_1$ and $R_Y$ ansatz (upper panel), and the time-to-solution (in units of seconds) using the classical solver \verb|CPLEX| (lower panel). $200$ random $3$-regular graphs are generated for each $N$ with the results plotted by dots. These median run times are fitted using exponential functions with respect to $N$. The plots are in the log-$y$ scale.}
    \label{fig:time_to_solution}
\end{figure}

\subsection{Multi-angle bipolar-$ZY$ ansatz}

The bipolar-$ZY$ ansatz can be further improved by relaxing its variational parameters to free angles. Here we consider two schemes. In the first scheme, we assign each $ZY$ gate in the ansatz with one free angle, such that the number of angles in each ansatz round equals to the number of graph edges and grows linearly with the system size. It can be shown that this multi-angle bipolar-$ZY_1$ ansatz can achieve the MaxCut solutions of any graph, i.e., $\alpha=1$~\cite{supp}. However, the linearly growing number of free angles requires high optimization consumption, which is similar to the multi-angle QAOA framework~\cite{herrman2022multi, Gaidai2024,li2024quantum}. 

The second scheme partially relaxes the variational parameters to a constant number of free angles in each round: We label each directed edge $(i\to j)$ in the bipolar-$ZY$ ansatz by its degree pair $(\deg^+(i),\deg^-(j))$; Then the single parameter in each round is relaxed into different parameters for directed edges with different degree pairs. For example, a $D$-regular graph has $\deg^+(i)\in[1,D]$ and $\deg^-(j)\in[1,D]$. Thus, each round of the angle-relaxed bipolar-$ZY$ ansatz has at most $D^2$ free parameters. The angle-relaxed bipolar-$ZY_1$ ansatz for a 2-regular graph with $3$ free parameters $\theta_i, i=0,1,2$ is illustrated in Fig.~\ref{fig:bipolar_ZY_ansatz}(\textbf{b}).

The angle-relaxed bipolar-$ZY_1$ ansatz has higher solution accuracy than the uniform-angle one. For $2$-regular graphs of any lengths, one can exactly calculate the expected cut number of the angle-relaxed ansatz and show that the ansatz achieves the exact MaxCut solution by taking $\theta_0=\theta_1=\pi/2$ and $\theta_2=\pi/4$, while the uniform-angle ansatz can only obtain approximate solutions and $\text{QAOA}_p$ requires $p\sim\OO(N)$~\cite{supp,mbeng2019quantum}. For $3$-regular graphs, since its directed edges with $\deg^-(j)=1$ and $\deg^-(j)=2$ corresponds to different parameters in the angle-relaxed ansatz, Theorem~\ref{theorem:ZY_1-ansatz-optimal} has the following corollary:
\begin{corollary} \label{corollary:angle-relax-result} 
    For the angle-relaxed bipolar-$ZY_1$ ansatz on an infinite $3$-regular graph, the $0$-local performance guarantee is lower bounded by 
        \begin{align}
        \alpha_0&\geq \max_{\theta,\theta'} \frac{1}{2} [1+\frac{1}{3}\sin\theta+\frac{2}{3}\cos\theta'\sin\theta']=0.8333,
    \end{align}
    where $\theta$ and $\theta'$ are variational parameters of directed edges $(i\rightarrow j)$ with $\deg^-(j)=1$ and $\deg^-(j)=2$, respectively.
\end{corollary}
This lower bound is significantly larger than that of the uniform-angle ansatz given by Theorem~\ref{theorem:exact-ZY1-bound}.

Similar to the uniform-angle bipolar-$ZY_p$ ansatz, the multi-angle ansatz and the angle-relaxed ansatz with a constant $p$ are free from the BP problem for $D$-regular graphs, since both ans\"atze belong to the $p$-round $i$HVA and have the variance lower bound in Corollary~\ref{corollary:BP-free}. Although the multi-angle bipolar-$ZY_1$ ansatz can derive any MaxCut solutions and is free from the BP problem, its optimization consumption would be high due to its linearly growing number of parameters and thus numerous local minima. This optimization issue will be studied numerically in the next section.

\section{Numerical and hardware results}\label{sec:Numerical and hardware results}

In this section, we compare the MaxCut performance of the bipolar-$ZY$ ansatz with that of QAOA and the classical solver by numerical simulations on $3$-regular graphs, and by hardware demonstrations on non-planar graphs. The hardware we use is the superconducting chip IBM Heron R2~\cite{Qiskit}, including \verb|ibm_fez| and \verb|ibm_marrakesh|. Details on the circuit transpilation, error suppression, post-processing, and chip calibration are given in Supplemental Material~\cite{supp}. Additionally, the classical simulability of the bipolar-$ZY$ ansatz is discussed to support the utility of the light-cone VQA on practical quantum devices.

\subsection{Numerical simulation}

\begin{figure*}
\centering
\includegraphics[width=1\textwidth]{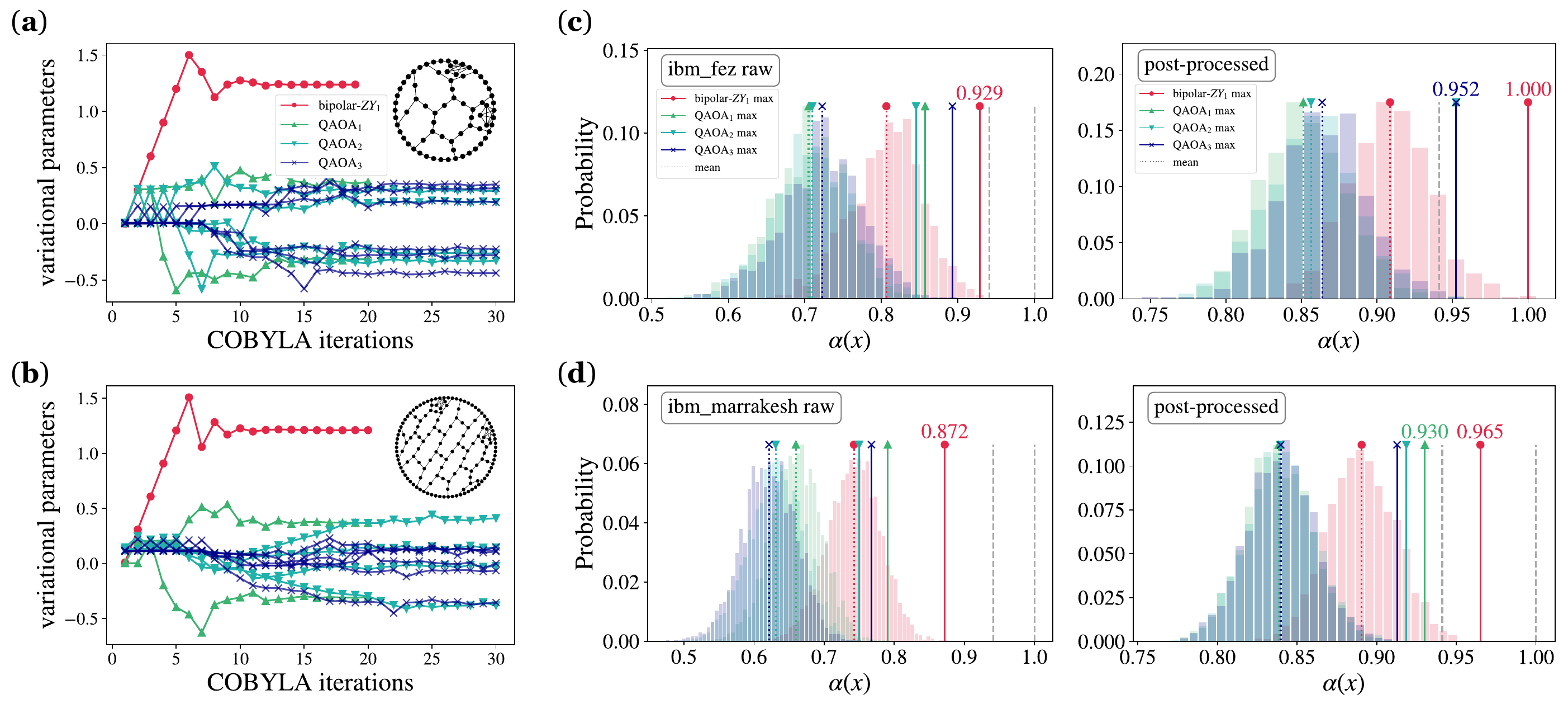}
\cprotect\caption{Hardware demonstration to solve the MaxCut instances of non-planar graphs with $72$ and $148$ nodes, using \verb|ibm_fez| and $\verb|ibm_marrakesh|$, respectively.  (\textbf{a},\textbf{b}) The evolution of variational parameters in the bipolar-$ZY_1$ ansatz and $\text{QAOA}_p$ with $p=1,2,3$. The number of iterations is chosen to ensure their convergence. The subplots of (\textbf{a}) and (\textbf{b}) show the non-planar graphs with $72$ and $148$ nodes, respectively. (\textbf{c},\textbf{d}) The optimized probability distributions of the raw approximation ratios $\alpha(x)$ and the classical post-processed ones. $1024$ and $4096$ bit strings $x$ are sampled from the optimized ans\"atze in (\textbf{c}) and (\textbf{d}), respectively. The maximum and mean of approximation ratio distributions are marked on the plot using solid and dashed vertical lines, respectively. The gray dashed lines denote the hardness threshold $\alpha=0.941$ and the exact $\alpha=1$.}
\label{fig:hardware-results}
\end{figure*}

We first compare the approximation ratio derived by the bipolar-$ZY$ ansatz and the classical Goemans-Williamson (GW) algorithm, which is a polynomial-time approximation algorithm with classically optimal approximation ratio under the unique games conjecture~\cite{Khot2007} (unless P=NP). Figure~\ref{fig:time_to_solution}(\textbf{a}) plots the median approximation ratios among 50 random 3-regular graphs for each graph nodes $N$ using the bipolar-$ZY_p$ ansatz and the GW algorithm, with $p=1,3$ for both uniform and angle-relaxed ansatz. The approximation ratios derived by the uniform and angle-relaxed $ZY_1$ ansatz satisfy the lower bounds given in Theorem~\ref{theorem:exact-ZY1-bound} and Corollary~\ref{corollary:angle-relax-result}, respectively. Compared with the classical GW algorithm, the angle-relaxed $ZY_p$ ansatz with $p=1,3$ derives larger median approximation ratios, showcasing the advantage of the light-cone VQA compared with the classical algorithm.

For the multi-angle bipolar-$ZY$ ansatz, we numerically test its optimization consumption to achieve $\alpha=1$ and compare it with the classical solver \verb|CPLEX|, which can also achieve $\alpha=1$. Since the time consumption of both methods should be exponential to the graph nodes, time-to-solution (TTS) of the bipolar-$ZY_1$ is dominated by the total number of iterations, where the time of each iteration is polynomial to $N$ and thus is not dominated. Thus, we count the bipolar-$ZY_1$ ansatz’s TTS as the summation of iterations over repeated random initial guess $\theta_j\in[-\pi,\pi]$, and subsequently optimize each initial guess using the optimizer \verb|SLSQP|~\cite{2020SciPy-NMeth}, until the exact MaxCut is obtained. For the classical solver \verb|CPLEX|, its TTS is the time duration of executing the algorithm on a 12-core central processing unit (CPU). Figure~\ref{fig:time_to_solution}(\textbf{b}) plots the median TTS versus $N$ among 200 random 3-regular graphs for each $N$. The numerical simulation for large 3-regular graphs is achieved using the Pauli propagation method without truncation~\cite{Angrisani2025,lerch2024efficientquantumenhancedclassicalsimulation,rudolph2025paulipropagationcomputationalframework}. The median time consumptions using the bipolar-$ZY_1$ and \verb|CPLEX| can be fitted using exponential functions $1.046^N$ and $1.061^N$, respectively~—~ indicating a scaling advantage of the light-cone VQA over \verb|CPLEX| on the average behavior of solving the MaxCut problem.

To confirm that the scaling advantage comes from the light-cone structure, Figure~\ref{fig:time_to_solution}(\textbf{b}) also presents the median iteration-to-solution using a naive $R_Y$ ansatz $\ket{\phi_{R_Y}(\bos{\theta})} = \prod_{j\in [1,N]}e^{-i\theta_jY_j/2}\ket{+}^{\otimes N}$. The $R_Y$ ansatz can derive the exact MaxCut solution for any graphs but suffers from the hardness of parameter optimization~\cite{Bittel2021}. Its median iteration-to-solution scales as $1.107^N$, which is worse than both the bipolar-$ZY_1$ and \verb|CPLEX|. These results show that the multi-angle bipolar-$ZY_1$ ansatz mitigates the hardness of parameter optimization and presents the scaling advantage due to its light-cone structure.

\subsection{Hardware implementation}
We implement the uniform-angle bipolar-$ZY_1$ ansatz and $\text{QAOA}_p$ to solve the MaxCut problem on two instances of non-planar graphs with $72$ and $148$ nodes, with each node maps to a qubit of the superconducting chips.  The two instances are shown in Fig.~\ref{fig:hardware-results}(\textbf{a}) and \ref{fig:hardware-results}(\textbf{b}), respectively. We circumvent planar graphs because there exists a polynomial-time algorithm for the MaxCut problem on planar graphs~\cite{Hadlock_75}. On the other hand, Kuratowski's Theorem states that a finite graph is planar if and only if it does not contain a subgraph that is a subdivision of $K_{5}$ or of $K_{3,3}$, and the two graphs we consider here are non-planar due to the existence of 5-node complete graphs. 

To reduce the circuit depth of the bipolar-$ZY_1$ ansatz, we adopt a BFS-type bipolar orientation algorithm proposed by Ref.~\cite{PAPAMANTHOU2008224}, as well as randomly sample 20 pairs of source and sink nodes and choose one pair with the minimum circuit depth. The circuit depth of $\text{QAOA}_p$ is also minimized by the edge-coloring method~\cite{Bravyi_2020}. More details about the circuit depth and compilation procedures are given in Supplemental Material~\cite{supp}.

 The variational parameters of the bipolar-$ZY_1$ and $\text{QAOA}_p$ with $p=1,2,3$ are trained in a hybrid way using the quantum chips and the classical optimizer \verb|COBYLA|. Optimization trajectories of these parameters are presented in Fig.~\ref{fig:hardware-results}(\textbf{a}) and \ref{fig:hardware-results}(\textbf{b}) for the $72$-qubit and $148$-qubit demonstrations, respectively. The bipolar-$ZY_1$ ansatz only has one parameter to be trained and converged within $20$ \verb|COBYLA| iterations, whereas $\text{QAOA}_p$ has $2p$ parameters converged after 30 iterations.
 
 
In the $72$-qubit demonstration, Figure~\ref{fig:hardware-results}(\textbf{c}) shows the raw probability distributions of the approximation ratios $\alpha(x)\equiv C(x)/C_{\max}$, where $C(x)$ is the cut number of bit-string $x$ sampled from the optimized ansatz and $C_{\max}$ is obtained from the classical solver \verb|CPLEX|. The right panel of Fig.~\ref{fig:hardware-results}(\textbf{c}) shows the improved results using the classical post-processing by local bit-flip~\cite{sachdeva2024quantumoptimizationusing127qubit}. The maximum and mean ratios of the sampled $\alpha(x)$ are denoted by solid and dashed lines, respectively. We see that the bipolar-$ZY_1$ achieves the exact $\alpha(x)=1.000$ after the post-processing. Meanwhile, $\text{QAOA}_p$ only finds the approximate solution up to $\alpha(x)=0.952$ using $p=3$.

In the $148$-qubit demonstration, as shown in figure~\ref{fig:hardware-results}(\textbf{d}), the largest approximation ratio derived by the bipolar-$ZY_1$ is $\alpha(x)=0.965$, which surpasses the hardness threshold $\alpha=0.941$. For $\text{QAOA}_p$, increasing the ansatz rounds $p$ degrades the performance due to more severe decoherence noise as $p$ increases, such that $\text{QAOA}_1$ has the best performance of $\alpha(x)=0.930$, but still smaller than the hardness threshold. This degrading of $\text{QAOA}_p$ for larger $p$ showcasing that $\text{QAOA}_p$ is hard to derive more accurate solutions than the bipolar-$ZY_1$ by increasing $p$ and the execution time. Our hardware results demonstrate the applicability of the light-cone VQA to solve large-scale MaxCut problems using practical quantum devices.
\subsection{Classical simulability}
We numerically demonstrate the difficulties of classically simulating the bipolar-$ZY_1$ ansatz using the matrix product state (MPS) and Pauli propagation method. The time complexity of simulating quantum circuits using MPS increases exponentially with respect to the entanglement entropy of the ansatz state~\cite{ORUS2014117}. Figure~\ref{fig:entanglement_entropy}(\textbf{a}) plots the median entropy of the multi-angle bipolar-$ZY_1$ ansatz states for 50 random 3-regular graphs with $N$ nodes, where each data point is the entropy peak during the classical optimization. The median entropy grows linearly with respect to $N$ up to $N=18$, indicating an exponentially increased bond dimension and time complexity of simulating the multi-angle bipolar-$ZY_1$ ansatz using MPS.

\begin{figure}
    \centering
    \includegraphics[width=0.49\textwidth]{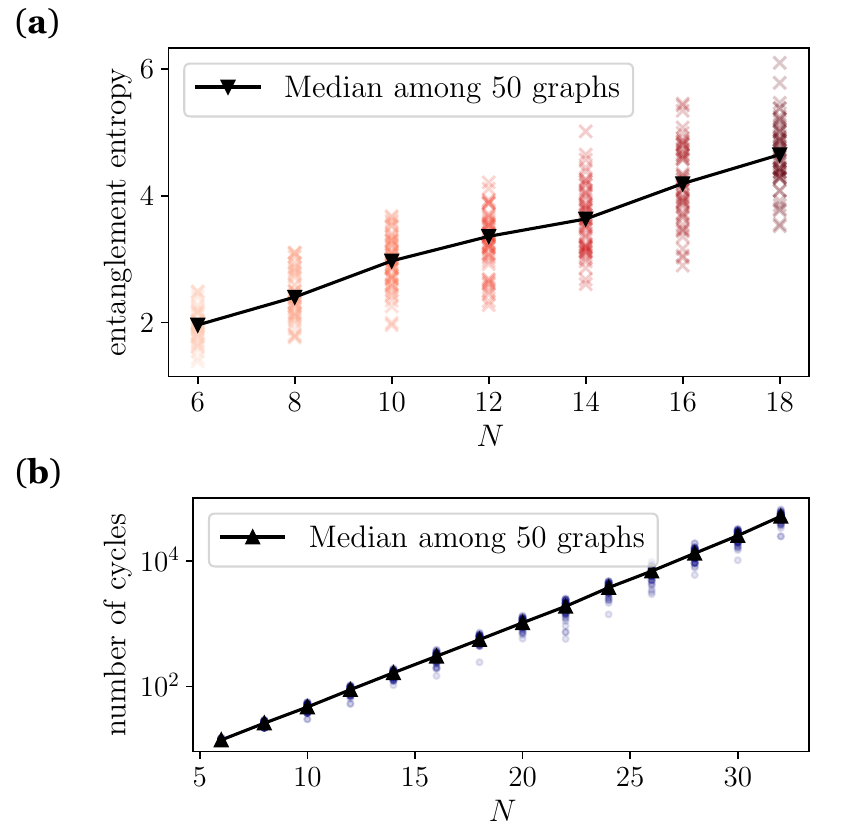}
    \cprotect\caption{(\textbf{a})The entanglement entropy of the multi-angle bipolar-$ZY_1$ ansatz states as a function of graph nodes $N$. The median values among 50 random 3-regular graphs are marked by lower-triangles. (\textbf{b}) The number of cycles of 50 random 3-regular graphs versus graph nodes $N$, with the median values marked by upper-triangles. The plot is in the log-$y$ scale.}
    \label{fig:entanglement_entropy}
\end{figure}

For the Pauli propagation method, we consider the exact simulation without truncation~\cite{Angrisani2025,lerch2024efficientquantumenhancedclassicalsimulation,rudolph2025paulipropagationcomputationalframework}. The time complexity of the Pauli propagation method simulating the bipolar-$ZY_1$ ansatz is at least proportional to the number of cycles in the corresponding graph~\cite{supp}. For 3-regular graphs, it has been shown that the number of cycles can grow exponentially with the number of nodes~\cite{arman2017maximumnumbercyclesgraph}. In Fig.~\ref{fig:entanglement_entropy}(\textbf{b}), we numerically estimate the number of cycles in random 3-regular graphs. Up to $N=32$, the median number of cycles among 50 random graphs grows exponentially with $N$, indicating that the Pauli propagation method without truncation is hard to evaluate the Hamiltonian expectation for very large graphs. Additionally, the Pauli propagation methods with Pauli weight truncation and coefficient magnitude truncation fail to adequately control the truncation errors for the bipolar-$ZY_1$ ansatz, as further discussed in Supplemental Material~\cite{supp}.

\section{Discussion}
The utility of the variational quantum algorithms is challenged by the barren plateau (BP) problem, lacking theoretical performance guarantees, and their classical simulability. In this work, we address these challenges by proposing the light-cone VQA. The light-cone VQA has an extensive backward light-cone and is hence intractable to be simulated classically. Meanwhile, the variational ansatz has a constant number of rounds such that the BP problem is circumvented. The performance of the lightcone-$ZY$ ansatz on the MaxCut problem is analyzed by the local analysis method. Using this method, we provide Theorem~\ref{theorem:ZY_1-ansatz-optimal}, which states that the DAG orientation of the lightcone-$ZY$ ansatz is the optimal $ZY$ orientation and the DAG's topological order provides the optimal gate sequence. This Theorem also guides further enhancement of the lightcone-$ZY$ ansatz into the bipolar-$ZY$ ansatz with a bipolar DAG.

The light-cone VQA has better performance compared with QAOA and classical algorithms. Theorem~\ref{theorem:exact-ZY1-bound} presents the performance guarantee of the bipolar-$ZY_1$ ansatz better than that of $\text{QAOA}_3$ for infinite $3$-regular graphs, and the guarantee is further improved by the angle relaxation as stated in Corollary~\ref{corollary:angle-relax-result}. The advantage of the light-cone VQA over the classical algorithms and its performance in the existence of realistic quantum noise are demonstrated through the numerical simulation and the hardware implementation.

The performance guarantees of the single-round bipolar-$ZY$ ansatz we derived in this work still have a distance from those of the best-known classical solver. Angle-relaxed bipolar-$ZY_p$ ansatz with a few constant rounds $p>1$ is a promising candidate to have larger performance guarantees and theoretically show the possible quantum advantage over classical solver, which remains to be further investigated. In principle, the performance of the multi-round bipolar-$ZY_p$ ansatz can be estimated using the local analysis method developed in this work, but in practice, the local analysis error is larger as the ansatz rounds $p$ grows. It is intrinsically difficult to control the analysis error due to the combinatorial explosion in the number of possible local subgraphs. Therefore, more advanced analysis methods remain to be developed to estimate the performance of the light-cone VQAs with multiple rounds.



The light-cone VQA is primarily designed for sparse graphs with bounded degree, while dense graphs may still suffer from barren plateaus due to the exponentially vanishing gradient variance. This limitation could potentially be alleviated by incorporating problem-specific initialization strategies~\cite{PhysRevResearch.2.043246,Park2024hamiltonian,chai2024structureinspired}. Moreover, the potentially linear circuit-depth scaling of the light-cone structure introduces sensitivity to decoherence on NISQ devices. However, the impact of decoherence can be partially suppressed using techniques such as dynamic decoupling~\cite{Ezzel2023}, benefiting from the long idle periods naturally present in the light-cone circuits. These considerations reveal the current limitations of the light-cone VQA and suggest promising directions for extending its applicability to larger and more challenging quantum optimization problems.


\section*{Acknowledgements}
We thank Lukas Broers and Han Xu for insightful discussions and valuable comments. XW was supported by the RIKEN TRIP initiative (RIKEN Quantum) and the UTokyo Quantum Initiative. YS and TL were supported by the National Natural Science Foundation of China (Grant Numbers 92365117 and 62372006), and The Fundamental Research Funds for the Central Universities, Peking University.

\section*{Author Contribution}
X.W. and T.L. conceived the original study. X.W. and Y.S. carried out the numerical calculations to support the study. All authors discussed the results of the manuscript, and all authors contributed to the writing of the manuscript.

\section*{Competing Interest}
We declare no competing interest.

\section*{Data Availability}
Our codes of constructing the light-cone ansatz, hardware results, and calibration details are available at~\cite{Wang2025_lightcone}.

\nocite{shao2024simulatingnoisyvariationalquantum, gunderson2010handbook,GORING2000295,Hopcroft1973,cormen2001introduction,min_max_theorem,Wormald_1999,PhysRevB.107.L041109,Cochran_2025,PhysRevX.11.041039,riveradean2021avoiding,Barkoutsos2020improving,sachdeva2024quantumoptimizationusing127qubit}

\bibliography{main.bib}
\bibliographystyle{IEEEtran}

\onecolumngrid
\appendix

\newpage

\titleformat{\section}{\normalfont\large\bfseries}{Supplementary Note \thesection:}{1em}{}

\section{Construction of the lightcone-$ZY$ ansatz}\label{app:lightcone-ZY ansatz}

\begin{figure}[b]
    \centering
    \includegraphics[width=0.9\textwidth]{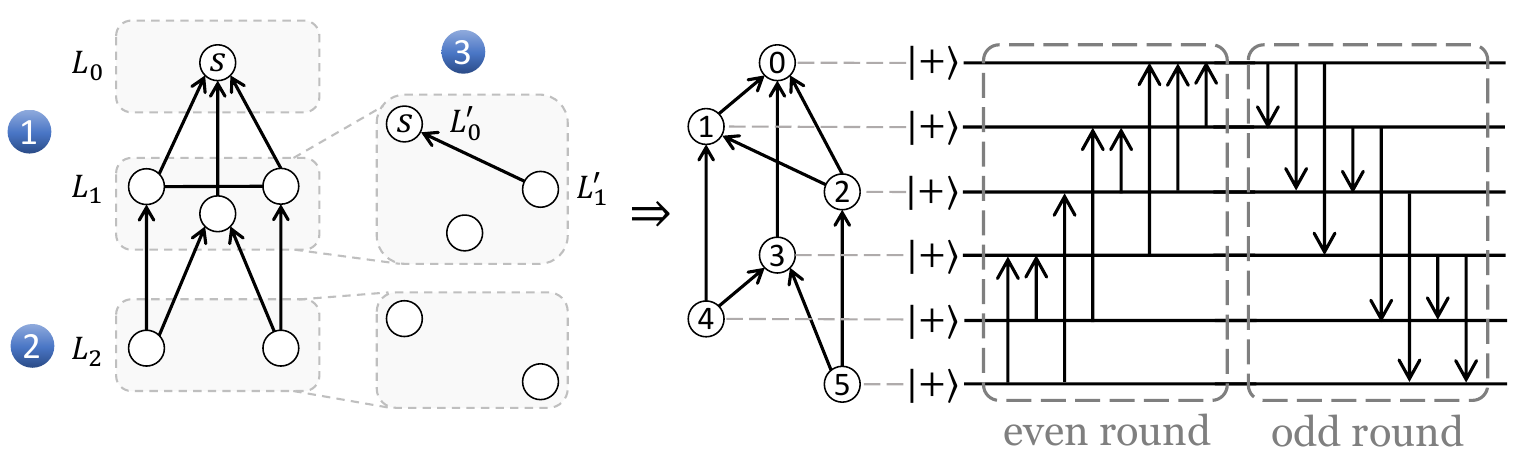}
    \caption{Construction procedures of the lightcone-$ZY_p$ ansatz by a recursive breadth-first search (BFS) process. Nodes labeled by $s$ are selected as the root nodes during the recursive BFS. The BFS traversal gives a directed acyclic graph (DAG) in the right panel. The edge orientation and the topological order of the DAG give a $ZY$ orientation and a gate sequence of the even round of the lightcone-$ZY_p$ ansatz. The odd round is derived by reversing the $ZY$ orientation and gate sequence of the even round.}
    \label{fig:dag_construction_app}
\end{figure}

In this note, we describe the procedures of constructing the lightcone-$ZY_p$ ansatz with unitary gates $\UU_p(\bos{\theta})= \prod_{l=0}^{p-1} \prod_{(i,j)\in\EE}e^{-i\theta_l Z_iY_j/2}$. The first round ($l=0$) of the ansatz is constructed by a recursive breadth-first search (BFS) process described as follows, and we repeat the same structure for all the even rounds (even $l$ in $\UU_p(\bos{\theta})$).

For the connected graph $G=(\VV,\EE)$, we define the layers $L_0, L_1\ldots$ constructed by the BFS algorithm. First, pick an arbitrary root $s$ and we define $L_0=\{s\}$. Layer $L_1$ consists of all neighbors of $s$. Second, assuming that we have defined layers $L_1,\ldots, L_j$, the layer $L_{j+1}$ consists of all nodes that do not belong to an earlier layer and that have an edge to a node in layer $L_j$. Finally, each layer $L_j$ consists of one or more connected graphs, and we repeat the above two steps recursively for each of these connected graphs until every node is in distinguished layers. The left panel of Fig.~\ref{fig:dag_construction_app} illustrates these three steps. Because every node is in distinguished layers, the orientation of each edge $(i,j)\in\EE$ can be uniquely assigned as $(i\leftarrow j)$ if $i$'s layer subscript is smaller than $j$'s, i.e., $i$ is the parent of $j$ in the BFS traversal. The directed edges determine the $ZY$ orientation of the even rounds, and the resulting oriented graph is a directed acyclic graph (DAG) as shown in the right panel of Fig.~\ref{fig:dag_construction_app}. 

The gate sequence of the $ZY$ ansatz is determined by the topological order of the DAG. Assume that the topological order reads $T=(v_{0},v_{1},\dots, v_{N-1})$. Following the reverse time direction of the quantum circuit (from the right-end to the left-end of the circuit in Fig.~\ref{fig:dag_construction_app}), starting from the node $v_{0}$, we arrange $ZY$ gates corresponding to all edges entering $v_{0}$. Then arrange $ZY$ gates corresponding to all edges entering $v_{1}$, and so on. Note that $ZY$ gates entering the same qubit commute with each other, i.e.,
\begin{align}
    [e^{-i\theta Z_j Y_i/2},e^{-i\theta Z_k Y_i/2}]=0,
\end{align}
for arbitrary nodes $v_i\neq v_j, v_i\neq v_k$. Thus, arranging $ZY$ gates entering a particular node in different orders gives the same $ZY$ ansatz. These procedures construct the even round of the lightcone-$ZY_p$ ansatz. 

The odd round of the lightcone-$ZY_p$ ansatz is constructed by reversing the $ZY$ orientation and the gate sequence in the even round simultaneously. We reverse the $ZY$ orientation because both $e^{-i\theta_l Z_i Y_j/2}$ and $e^{-i\theta_l Y_i Z_j/2}$ contribute to the imaginary-time evolution of $Z_iZ_j$~\cite{PhysRevA.111.032612}, i.e.,
\begin{align}
    e^{-\tau Z_iZ_j}\ket{++}&\propto e^{-i\theta_1(\tau)Z_iY_j/2}\ket{++}=e^{-i\theta_2(\tau)Y_iZ_j/2}\ket{++}.
\end{align}
where $\theta_1(\tau),\theta_2(\tau)$ are functions of the imaginary time $\tau$. Thus, there is no reason to keep one and discard another. The gate sequence is reversed to keep the $Z$-right-$Y$-left structure defined in Supplementary Note~\ref{app:theorem-2-proof}. The resulting odd round circuit by reversing the $ZY$ orientation and the gate sequence of the even round is illustrated in the right panel of Fig.~\ref{fig:dag_construction_app}.

\section{Generalization to quadratic unconstrained binary optimization (QUBO) problems}\label{app:lightcone-ZY ansatz}

The construction of the lightcone-VQA can be generalized to other QUBO problems by considering the symmetry of the target Hamiltonian. In the main text, the lightcone-VQA is constructed for the MaxCut Hamiltonian $H_{\text{MC}}=\sum_{(i,j)\in\EE}(Z_iZ_j-1)/2$. $H_{\text{MC}}$ has the bit-flip symmetry due to the commutation relation $[H_{\text{MC}},\prod_{j\in{\VV}}X_j]=0$, and the time-reversal symmetry since $H_{\text{MC}}$ is a real matrix. Therefore, we use the $ZY$ gate as the basic block for the light-cone construction to mimic the imaginary-time evolution $e^{-\tau H_{\text{MC}}}$~\cite{PhysRevA.111.032612}. For general QUBO problems, up to a global constant, the Hamiltonian reads
    \begin{align}
        H_{\text{QUBO}} = \sum_{(i,j)\in\EE}w_{ij}Z_iZ_j+\sum_{i\in\VV}h_i Z_i,
    \end{align}
    with real coefficients $w_{ij}$ and $h_i$. The single-qubit $Z_i$ term breaks the bit-flip symmetry because $[Z_i,\prod_{j\in{\VV}}X_j]\neq0$, while the element-wise realness of the Hamiltonian is preserved. Thus, the original basic unitary block $e^{-i\theta\sigma/2}$ should be extended as
    \begin{align}
        \text{MaxCut:}~\sigma\in\{ZY, YZ\}\rightarrow \text{QUBO:}~\sigma\in\{ZY, YZ, XY, YX, IY, YI\}.
    \end{align}
    These 6 Pauli strings are all of the purely imaginary two-qubit  Paulis, such that the corresponding $e^{-i\theta\sigma/2}$ is purely real, which mimic the imaginary-time evolution $e^{-\tau H_{\text{QUBO}}}$.
    
    Therefore, one round of the ansatz should be composed of three unitaries
    \begin{align}
        U^{(l)} = \prod_{(i,j)\in\EE}e^{-i\theta_l Z_iY_j/2}e^{-iw_l X_iY_j/2}e^{-i\phi_l I_iY_j/2}
    \end{align}
    for even $l$, and
    \begin{align}
        U^{(l)} = \prod_{(i,j)\in\EE}e^{-i\theta_l Y_iZ_j/2}e^{-iw_l Y_iZ_j/2}e^{-i\phi_l Y_iI_j/2}
    \end{align}
    for odd $l$, where the order of the product $\prod_{(i,j)\in\EE}$ follows the same light-cone gate sequence. Then, the resulting ansatz state 
    \begin{align}
        \ket{\phi_{p}(\boldsymbol{\theta})}=\prod_{l=0}^{p-1}U^{(l)}\ket{+}^{\otimes N}
    \end{align}
    is a natural generalization of Eq.~(2) in the main text. However, the performance analysis and the optimality of the gate sequence for the QUBO problems are not straightforwardly generalized, which remains to be investigated in further work.
    
\section{Truncation error of the $k$-local analysis}\label{app:truncation-error-of-k-local}

In this note, we estimate the error of performing $k$-local analysis of the $ZY$ ansatz with an arbitrary $ZY$ orientation and gate sequence. In the following content, we first explain the definition of $k$-local analysis given in the main text by introducing the Pauli path~\cite{shao2024simulatingnoisyvariationalquantum} and $k$-local subgraph~\cite{PhysRevA.103.042612} and use them to prove the Proposition~1 of the main text.

We use the Pauli path to evaluate the expected cut number of the $ZY_p$ ansatz. For a given connected graph $G=(\VV,\EE)$ with $N$ nodes, the expected cut number of the $ZY_p$ ansatz reads
\begin{align}
    \LL(\bos{\theta},p)\equiv -\bra{\phi_p(\bos{\theta})}  H_{\text{MC}}\ket{\phi_p(\bos{\theta})}=\frac{M}{2}-\frac{1}{2}\sum_{(i,j)\in \EE} \bra{\phi_p(\bos{\theta})}  Z_iZ_j\ket{\phi_p(\bos{\theta})}.
    \label{eq:cost-summation}
\end{align}
We estimate the truncation error of the expected cut number by estimating each $Z_iZ_j$ expectation. The latter expectation can be rewritten as 
\begin{align}
    \bra{\phi_p(\bos{\theta})}  Z_iZ_j\ket{\phi_p(\bos{\theta})}=\bra{+}^{\otimes N}\UU_p^{\dagger}  Z_iZ_j \UU_p(\bos{\theta})\ket{+}^{\otimes N} = \tr(\UU_p^{\dagger}(\bos{\theta}) Z_iZ_j \UU_p(\bos{\theta})\rho_X  ),
    \label{eq:ZZ-trace}
\end{align}
where $\rho_X=\ket{+}\bra{+}^{\otimes N}$ is the density matrix of the $N$-qubit input state. This initial state can be written as a summation of $d \equiv 2^{N}$ Pauli-$X$ strings $\rho_X=\sum_{v\subseteq \VV}\sigma_X|_v/d$, where $\sigma_X|_v$ is a Pauli-$X$ string supported by a subset of nodes $v\subseteq \VV$. For example, if $\VV$ contains two nodes, then $\sigma_X|_v$ can be $\{II,IX,XI,XX\}$. Thus, the trace in Eq.~\eqref{eq:ZZ-trace} is decomposed into 
\begin{align}
    \tr(\UU_p^{\dagger}(\bos{\theta}) Z_iZ_j \UU_p(\bos{\theta})\rho_X  )= \sum_{v\subseteq \VV}\tr(\UU_p^{\dagger}(\bos{\theta}) Z_iZ_j \UU_p(\bos{\theta})\sigma_X|_v  )/d.
    \label{eq:decomposed-ZZ}
\end{align}
Each trace in the summation can be regarded as the back-propagation of the observable $Z_iZ_j$ to $\sigma_X|_v$ along Pauli paths.
\begin{definition}[Pauli path of the $ZY$ ansatz]
A Pauli path is a sequence $s=(s_0,\ldots,s_L)\in \bos{P}_N^{L+1}$, where $\bos{P}_N=\{I/\sqrt{2},X/\sqrt{2},Y/\sqrt{2},Z/\sqrt{2}\}^{\otimes N}$ is the set of all normalized $N$-qubit Pauli strings. $L$ is the total number of $ZY$ gates in $\UU_p(\bos{\theta})$.
\end{definition}
Since the Pauli paths consist of a complete basis, the trace in Eq.~\eqref{eq:decomposed-ZZ} can be expanded by
\begin{align}
    \tr(\UU_p^{\dagger}(\bos{\theta}) Z_iZ_j \UU_p(\bos{\theta})\sigma_X|_v ) =\sum_{s\in \bos{P}_N^{L+1}} \tr(Z_iZ_j s_L)\left[\prod_{m=1}^L \tr(U_m^{\dagger}s_m U_m s_{m-1})\right] \tr(s_0 \sigma_X|_v),
    \label{eq:ZZ-to-Pauli-paths}
\end{align}
where $U_m$ denotes an arbitrary $ZY$ gate. The proof of this formula can be found in Ref.~\cite{shao2024simulatingnoisyvariationalquantum}. Gathering Eq.~(\ref{eq:ZZ-trace}-\ref{eq:ZZ-to-Pauli-paths}), the $Z_iZ_j$ expectation is evaluated by
\begin{align}
    \bra{\phi_p(\bos{\theta})}  Z_iZ_j\ket{\phi_p(\bos{\theta})} = \frac{1}{d}\sum_{v\subseteq \VV}\sum_{s\in \bos{P}_N^{L+1}} \tr(Z_iZ_j s_L)\left[\prod_{m=1}^L \tr(U_m^{\dagger}s_m U_m s_{m-1})\right] \tr(s_0 \sigma_X|_v).
    \label{eq:ZZ-summary}
\end{align}


In the definition of $k$-local analysis given in the main text, the $Z_iZ_j$ expectation is calculated by truncating the quantum circuit at most $k$ steps away from the center edge $(i,j)$. This truncated quantum circuit has the same nodes set and $ZY$ gates connectivity as the \textit{$k$-local subgraph} of the edge $(i,j)\in \EE$.
\begin{definition}[$k$-local subgraph]
$k$-local subgraph $G^{k}_{(i,j)}$ of an edge $(i,j)\in \EE$ is a subgraph of the problem graph $G$, which includes center edge $(i,j)$, as well as edges and vertices within $k$ steps of $i$ and $j$.
\end{definition}
and we define $0$-local subgraph includes only the edge $(i,j)$. In other words, in $k$-local analysis, we discard the contribution of a Pauli path $s$ to the $Z_iZ_j$ expectation in Eq.~\eqref{eq:ZZ-summary}, if $s$ is transformed by a $ZY$ gate $U_m$ not belong to the $k$-local subgraph (a transformation by $U_m$ means $s_{m}\to s_{m-1}\neq s_{m}$ in Eq.~\eqref{eq:ZZ-summary}). A special case is that, if a $ZY$ gate $e^{-i\theta Z_iY_j/2}$ has $i$ and $j$ included in the $k$-local subgraph but the edge $(i,j)$ lies outside of it, we regard this gate as two gates $e^{-i\theta Z_iY_k/2}$ and $e^{-i\theta Z_{k'}Y_j/2}$ with $k,k'$ be arbitrary isolated nodes, but the gate $e^{-i\theta Z_iY_j/2}$ can not be used to transform the Pauli path. The left panel of Fig.~\ref{fig:Pauli_path_truncated}(\textbf{a}) illustrates an example of the $1$-local subgraph with the center edge $(2,3)$ of the graph $G$. Examples of Pauli paths contributing to the $1$-local expectation and the truncated ones are illustrated in the right panel of Fig.~\ref{fig:Pauli_path_truncated}(\textbf{a}). The truncated Pauli path corresponds to the special case that we described. 

For ease of description, in this section, we assume that variational parameters in all rounds are equal, i.e., 
\begin{align}
    \theta_l=\theta, \quad \forall l\in\{1,\ldots,p\}.
\end{align}
Conclusions in this note can be easily generalized to the non-equal case by taking $\theta \equiv  \|\bos{\theta}\|_{\infty}$. Then, Proposition~\ref{prop:truncation-error} in the main text is recalled as follows:

\begin{figure}
    \centering
    \includegraphics[width=0.90\textwidth]{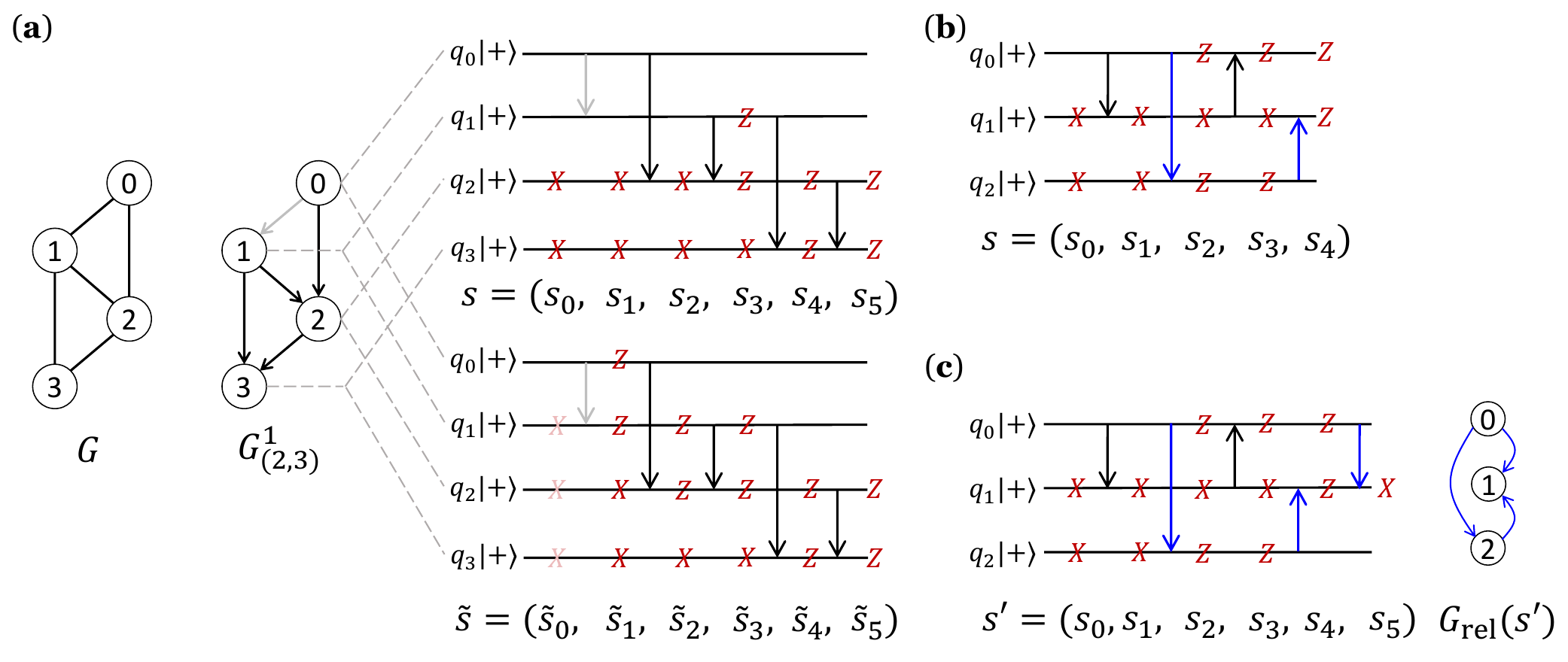}
    \caption{(\textbf{a}) Illustration of Pauli paths in the $ZY$ ansatz. To calculate the expectation of observable $Z_2Z_3$ in the $ZY$ ansatz using $1$-local analysis, a Pauli path is truncated if the path is outside of the $1$-local graph $G_{(2,3)}^1$. For example, the path $s=(s_0,\ldots,s_5)$ in the upper panel should be incorporated in the $1$-local analysis, whereas the path $\tilde{s}=(\tilde{s}_0,\ldots,\tilde{s}_5)$ in the the lower panel is discarded because the edge $(0,1)$ is outside of $G_{(2,3)}^1$. (\textbf{b}) An example of Pauli path $s=(s_0,s_1,\ldots, s_4)$ contributed to the expectation of $Z_0 Z_1$. Blue-directed edges denote relevant $ZY$ gates of the path $s$. (\textbf{c}) The $ZY$ circuit with an extended Pauli path $s'=(s_0,s_1,\ldots, s_4, s_5=X_1/\sqrt{2^3})$ induced from the Pauli path $s$ in (\textbf{b}), and the directed graph $G_{\text{rel}}(s')$ consisted of the relevant gates of $s'$. This graph is 2-edge-connected.}
    \label{fig:Pauli_path_truncated}
\end{figure}

\begin{customprop}{1}\label{prop:truncation-error}
For each observable $Z_iZ_j$ in $H_{\text{MC}}$ and the $ZY_p$ ansatz state $\ket{\phi_p(\theta)}$, the truncation error of the $k$-local analysis $|\bra{\phi_p(\theta)}Z_iZ_j\ket{\phi_p(\theta)}_k- \bra{\phi_p(\theta)}Z_iZ_j\ket{\phi_p(\theta)}|$ is bounded by $\OO(\sin^{2k+1}\theta)$.
\end{customprop}


The truncation error of the proposition is estimated by evaluating the most dominant Pauli path contributions to $Z_iZ_j$ expectation that are truncated in the $k$-local analysis. We denote one Pauli path contribution
\begin{align}
    c_s(\theta)\equiv\tr(Z_iZ_j s_L)\left[\prod_{m=1}^L \tr(U_m^{\dagger} s_m U_m s_{m-1})\right] \tr(s_0 \sigma_X|v).
    \label{eq:Pauli-path-coefficient}
\end{align}
According to the orthogonality of Pauli strings in the traces, $c_s(\theta)$ is non-zero only if $s_0=\sigma_X|v/\sqrt{d}, s_L=Z_iZ_j/\sqrt{d}$, and all $\tr(s_m U_m s_{m-1}U_m^{\dagger})$ are non-zero. Assuming that $U_m=e^{-i\theta Z_k Y_l/2}$, $\tr(U_m^{\dagger} s_m U_m s_{m-1})$ is non-zero for three cases:
\begin{align}
    \tr(U_m^{\dagger} s_m U_m s_{m-1})=\left\{ \begin{array}{ll}
 1 & \textrm{$[s_{m},Z_kY_l]=0$ and $ s_{m-1}=s_{m}$;}\\
  \cos\theta & \textrm{$\{s_{m},Z_kY_l\}=0$ and $ s_{m-1}=s_{m}$;}\\
  -\eta \sin \theta & \textrm{$\{s_{m},Z_kY_l\}=0$ and $s_{m-1}=-i\eta Z_k Y_l s_{m}$.}
  \end{array} \right.
  \label{eq:non-zero-cases}
\end{align}
where $\eta=\pm 1$ is the sign depending on the resulting product $Z_k Y_l s_{m} = \eta i\sigma, \sigma\in \bos{P}_N$. During the back-propagation from $s_L$ to $s_0$, every change of Pauli path from $s_m$ to $s_{m-1}=-i\eta Z_k Y_l s_{m}$ is accompanied by a multiplicative factor $\sin\theta$ in $c_s(\theta)$. We call a $ZY$ gate transforming the Pauli path from $s_m$ to $s_{m-1}\neq s_m$ the \textit{relevant gate} of the path $s$, and we have
\begin{align}
    c_s(\theta)\propto (\sin\theta)^{\text{number of relevant gates of $s$}}.
    \label{eq:cv-magnitude-initial}
\end{align}
Figure~\ref{fig:Pauli_path_truncated}(\textbf{b}) shows an example of non-zero Pauli paths starting from $s_L = Z_0Z_1/\sqrt{2^3}$ to $s_0 = X_1X_2/\sqrt{2^3}$, where the relevant gates are colored in blue. Thus, the magnitude of $c_s(\theta)$ is exponentially suppressed by the number of the relevant gates in $s$, as long as $\theta\neq \pm\pi/2$.  

For the convenience of the following description, we extend the original $ZY$ circuit by adding a relevant gate $e^{-i\theta Z_i Y_j/2}$ at the right end of the circuit, as shown in Fig.~\ref{fig:Pauli_path_truncated}(\textbf{c}).The extended Pauli path $s'$ induced from $s$ has an additional element $s'=(s_0,\ldots,s_L,s_{L+1}=X_j/\sqrt{d})$, and all the relevant gates of $s'$ in this extended circuit compose a directed graph $G_{\text{rel}}(s')$, as illustrated in the right panel of Fig.~\ref{fig:Pauli_path_truncated}(\textbf{c}). Then, the number of relevant gates in $s'$ equals to the number of edges in $G_{\text{rel}}(s')$, and we have
\begin{align}
    c_s(\theta)\propto \sin^{(M_{\text{rel}}-1)}\theta,
    \label{eq:cv-magnitude}
\end{align}
where $M_{\text{rel}}$ is the number of edges in $G_{\text{rel}}(s')$, and the minus one is due to the extended circuit. We prove that $G_{\text{rel}}(s')$ has the following properties.
\begin{lemma}
    Assume that $G'_{\text{rel}}(s')$ is an undirected underlying graph obtained by replacing all directed edges of $G_{\text{rel}}(s')$ with undirected edges. Then, $G'_{\text{rel}}(s')$ is 2-edge-connected.    
\end{lemma}
We say that an undirected graph is 2-edge-connected if it remains connected when any one of its edges is removed.

\noindent \textit{proof.} We first note that $G'_{\text{rel}}(s')$ is connected because all the relevant gates of path $s$ should be connected with the endpoint of the Pauli path $s_{L+1}=X_j/\sqrt{d}$. The lemma is proved by investigating the transformation from the single-qubit $X_j$ to the multi-qubit Pauli-$X$ string $\sigma_X|_v$. For example, the first step of the transformation $X_j\rightarrow Z_iZ_j$ is led by the conjugation
\begin{align}
    e^{i\theta Z_i Y_j/2} X_j e^{-i\theta Z_i Y_j/2} =\cos\theta X_j+\sin\theta Z_i Z_j,
    \label{eq:first-step}
\end{align}
where the new Pauli string $\sigma = Z_iZ_j$ comes from the multiplication
\begin{align}
    Z_iY_j\times X_j=-iZ_iZ_j.
    \label{eq:Pauli-string-multiplication}
\end{align}
This multiplication leads to a Pauli path transformation, and thus the $ZY$ gate $e^{-i\theta Z_i Y_j/2}$ is relevant. All the relevant gates lead to the whole transformation $X_j\rightarrow \sigma_X|_v$, and the result of every single transformation $s_m\to s_{m-1}$ can be read directly by the multiplication of $ZY$ to $s_m$ as in Eq.~\eqref{eq:Pauli-string-multiplication}. Applying the relevant gates in different orders leads to different orders of the multiplications. However, since Pauli strings are commute or anti-commute with each other, different orders of the multiplication give different $s_{m-1}$ up to a $\pm$ sign, which is irrelevant to our analysis of $G'_{\text{rel}}(s')$'s connectivity. Hence, the order of the $ZY$ operators multiplied on $X_j$ will be neglected in the subsequent discussion.

In the $ZY$ ansatz, the only multiplier to $X_j$ is the Pauli $Z$ and $Y$ operators on arbitrary qubits (or nodes). To make up the Pauli string transformation $X_j\rightarrow \sigma_X|_v$, the Pauli letter on each node of the string has the following four cases, which are led by different numbers of $Z$ and $Y$ multipliers
\begin{align}
    \left\{ \begin{array}{lll}
 \text{node $j$}:& X_j\rightarrow X_j & \text{even $Y_j$ and even $Z_j$}; \\
& X_j\rightarrow I_j & \text{odd $Y_j$ and odd $Z_j$}; \\
 \text{node $j'\neq j$}:& I_{j'}\rightarrow X_{j'} & \text{odd $Y_{j'}$ and odd $Z_{j'}$}; \\
& I_{j'}\rightarrow I_{j'} & \text{even $Y_{j'}$ and even $Z_{j'}$} .\\
  \end{array} \right.
\end{align}
For example, in the first case, the initial $X_j$ remains to be $X_j$ in $\sigma_X|_v$ if and only if even number of $Y_j$s and $Z_j$s are multiplied to $X_j$, because
\begin{align}
    (Y_j)^{2k}(Z_j)^{2k'} X_j =X_j, \quad k,k'\in \mathbb{N}.
\end{align}
Then, since the $Y$ and $Z$ operators correspond to the head and tail of a directed edge in $G_{\text{rel}}(s')$, respectively, the numbers of $Y$ and $Z$ operators can be interpreted as the in- and out-degree of the directed graph $G_{\text{rel}}(s')$. Thus we have
\begin{align}
    \left\{ \begin{array}{lll}
 \text{node $j$}:& X_j\rightarrow X_j & \text{even $\deg^-(j)$ and even $\deg^+(j)$}; \\
& X_j\rightarrow I_j & \text{odd $\deg^-(j)$ and odd $\deg^+(j)$}; \\
 \text{node $j'\neq j$}:& I_{j'}\rightarrow X_{j'} & \text{odd $\deg^-(j')$ and odd $\deg^+(j')$}; \\
& I_{j'}\rightarrow I_{j'} & \text{even $\deg^-(j')$ and even $\deg^+(j')$} .\\
  \end{array} \right.
\end{align}
For all the four cases, $\forall i\in G_{\text{rel}}(s')$, the summation $\deg^-(i)+\deg^+(i)$ is even . Thus, all nodes in the undirected graph $G'_{\text{rel}}(s')$ induced by $G_{\text{rel}}(s')$ have
\begin{align}
    \text{$\deg(i)=\deg^-(i)+\deg^+(i)$ is even,  $\quad \forall i \in G'_{\text{rel}}(s')$. }
\end{align}

The connected graph $G'_{\text{rel}}(s')$ is 2-edge-connected if all its nodes have even degrees. Because if removing one edge disconnects $G'_{\text{rel}}(s')$ into two connected graphs, each connected part will have one and only one node with odd degrees. This is in contradiction with the handshaking lemma~\cite{gunderson2010handbook}, which states that the summation of all node degrees should be even. Thus, we prove that the graph $G'_{\text{rel}}(s')$ is 2-edge-connected. $\qed$

A corollary of 2-edge-connectivity is that between
every pair of distinct nodes of the undirected graph, there are at least two edge-disjoint paths between them, as a consequence of Menger’s theorem~\cite{GORING2000295}. Because the center edge $(i,j)$ belongs to $G'_{\text{rel}}(s')$, we conclude that every node in $G'_{\text{rel}}(s')$ has two edge-disjoint paths to $i$ and two edge-disjoint paths to $j$. We can use this fact to prove Proposition~\ref{prop:truncation-error}.

\vspace{0.2cm}

\begin{figure}
    \centering
    \includegraphics[width=0.25\textwidth]{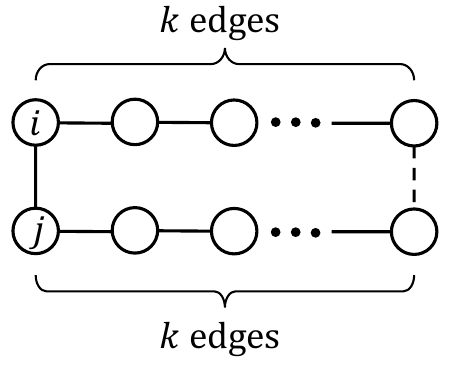}
    \caption{The relevant gate graph $G'_{\text{rel}}(s_{\text{tr}}')$ with the minimum number of edges that has the dashed edge outside of the $k$-local subgraph $G^k_{(i,j)}$ of the center edge $(i,j)$, and every node has two edge-disjoint paths to both $i$ and $j$.}
    \label{fig:minimum_edges}
\end{figure}

\noindent\textbf{Proof of Proposition~\ref{prop:truncation-error} ---} In $k$-local analysis, the expectation of $Z_iZ_j$ of the $ZY_p$ ansatz is a linear combination of $c_s(\theta)$, which has the relevant gate graph $G'_{\text{rel}}(s')$ whose edges and nodes all belong to $G^k_{(i,j)}$. In other words, a truncated Pauli path $s_{\text{tr}}$ has contribution $c_{s_{\text{tr}}}(\theta)$ and the relevant gate graph $G'_{\text{rel}}(s_{\text{tr}}')$ with at least one node or one edge not belong to the $k$-local subgraph $G^k_{(i,j)}$. The truncation error is the magnitude of $c_{s_{\text{tr}}}(\theta)$. Therefore, to estimate the magnitude of the truncation error, we need to find the $G'_{\text{rel}}(s_{\text{tr}}')$ with the minimum number of edges and satisfying the following two conditions.
\begin{itemize}
    \item[1.] $G'_{\text{rel}}(s_{\text{tr}}')$ has at least one node or one edge not belonging to $G^k_{(i,j)}$.
    \item[2.] Every node in $G'_{\text{rel}}(s_{\text{tr}}')$ has two edge-disjoint paths to $i$ and two edge-disjoint paths to $j$.
\end{itemize}
We find such undirected graph $G'_{\text{rel}}(s_{\text{tr}}')$ with the minimum number of edges shown in Fig.~\ref{fig:minimum_edges}. This graph is a cycle with length $2k+2$ and has the dashed edge outside of the $k$-local subgraph of the center edge $(i,j)$. According to Eq.~\eqref{eq:cv-magnitude}, we have $c_{s_{\text{tr}}}(\theta)\propto \sin^{2k+1}\theta$. All $c_{s}(\theta)$ truncated by $k$-local analysis have the relevant gate graph with edges at least $2k+2$, and the corresponding $c_s(\theta)$ is proportional to the higher orders of $\sin\theta$. Therefore, the truncation error of $k$-local analysis is bounded by
\begin{align}
    |\bra{\phi_p(\theta)}Z_iZ_j\ket{\phi_p(\theta)}_k- \bra{\phi_p(\theta)}Z_iZ_j\ket{\phi_p(\theta)}|=\OO(c_{s_{\text{tr}}}(\theta))=\OO(\sin^{2k+1}\theta).
\end{align}
These complete the proof of Proposition~\ref{prop:truncation-error}.

\section{Proof of Theorem~1}\label{app:theorem-2-proof}

\begin{figure}
    \centering
    \includegraphics[width=0.9\textwidth]{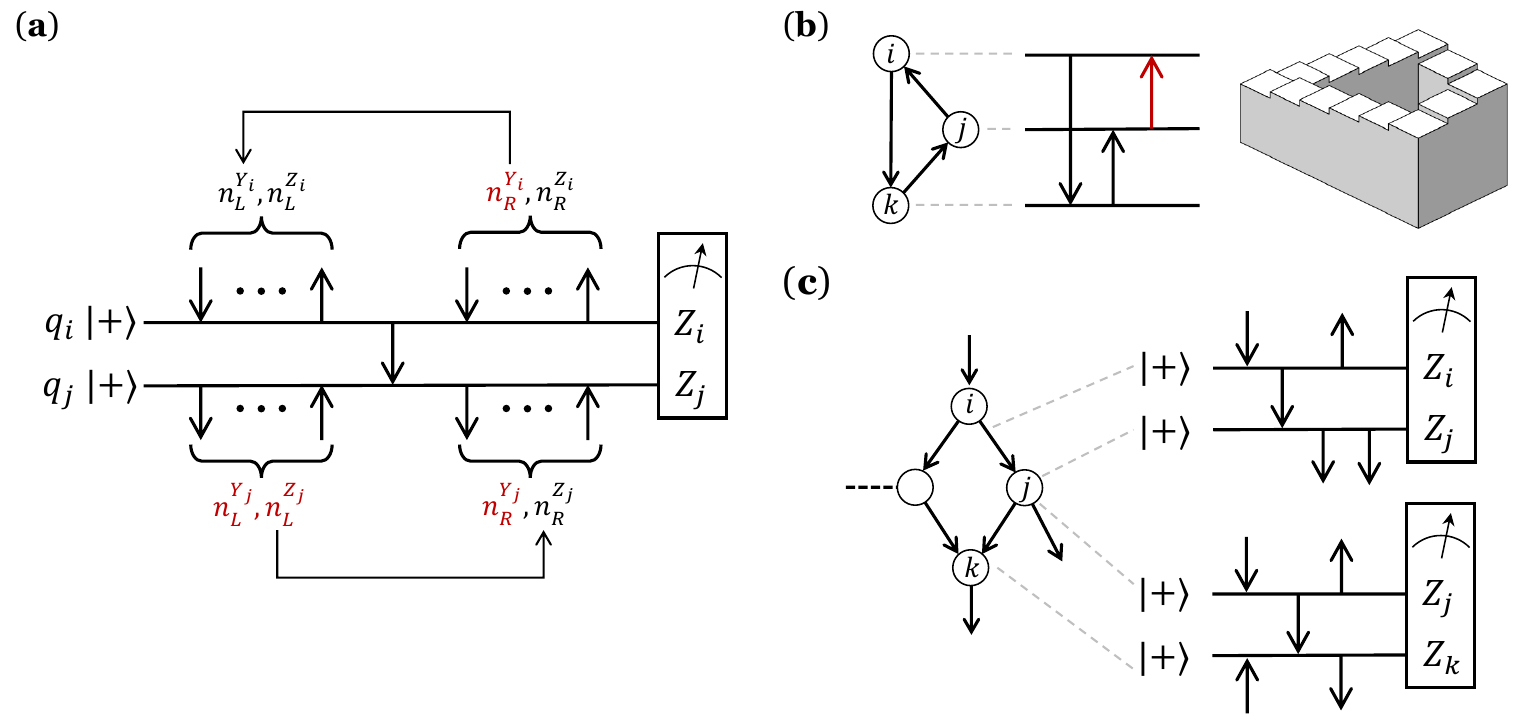}
    \caption{$0$-local analysis of the $ZY_1$ ansatz on $3$-regular graphs. (\textbf{a}) A general $0$-local structure of the $ZY_1$ ansatz. $ZY$ gates adjacent to center qubit $i,j$ are arranged at the left or right side of $e^{-i\theta Z_iY_j/2}$. $n_L^O$ and $n_R^O$ denote the numbers of these $ZY$ gates at the left ($L$) and right ($R$) side, respectively, with the single-qubit Pauli operator $O= Z_{i(j)}, Y_{i(j)}$ attached to qubit $i$ or $j$. The red-highlighted gate numbers contribute to $k_{ij}$. Particularly, $n_R^{Y_i}$ and $n_L^{Z_j}$ can be made $0$ by the left and right gate movement as denoted by the arrows and has the consequent $Z$-right-$Y$-left structure. (\textbf{b}) A directed cycle and the $Z$-right-$Y$-left structure are not compatible. The $Y$ end of the red-highlighted $ZY$ gate $e^{-i\theta Z_j Y_i/2}$ violates the $Z$-right-$Y$-left structure. Intuitively, a directed cycle and the $Z$-right-$Y$-left structure lead to an impossible Penrose stair. (\textbf{c}) Two kinds of directed edges in $3$-regular graphs and their corresponding $0$-local circuits of the lightcone-$ZY$ ansatz. The two kinds of edges have the in-degree $\deg^-(j)=1$ (upper panel) and $\deg^-(k)=2$ (lower panel), respectively.}
    \label{fig:0-local-3-regular}
\end{figure}

Theorem~1 in the main text claims the optimality of a lower bound and gives its explicit expression. We first show that, according to the $0$-local analysis, the optimal $ZY$ orientation should correspond to a DAG with the minimal number of the sink node $N_- = 1$, and the optimal gate sequence should arrange $ZY$ gates following the topological order of the DAG. These facts will be used in the proof of Theorem~1.

$0$-local analysis calculates the $Z_iZ_j$ expectation by considering $ZY$ gates directly adjacent to the qubits $i$ and $j$. Without loss of generality, we assume $(i,j)$ takes $ZY$ orientation $(i\rightarrow j)$ and thus the $ZY$ gate is $e^{-i\theta Z_iY_j/2}$, as shown in Fig.~\ref{fig:0-local-3-regular}(\textbf{a}). Additionally, we use non-negative integer $n_L^{O}, n_R^{O}, O=Z_{i(j)}, Y_{i(j)}$ to represent the number of other $ZY$ gates adjacent to qubit $i$ and $j$. With a direct calculation in the Heisenberg picture, the expectation of $Z_iZ_j$ in the $0$-local circuit depends only on these gate numbers
\begin{align}
    \bra{\phi_1(\theta)}Z_iZ_j\ket{\phi_1(\theta)}_0 = -\cos^{k_{ij}}\theta\sin\theta,
\end{align}
where 
\begin{align}
    k_{ij} \equiv n^{Y_i}_R+n^{Y_j}_R+n^{Y_j}_L+n^{Z_j}_L.
    \label{eq:kij-exp}
\end{align}
These gate numbers are red-highlighted in Fig.~\ref{fig:0-local-3-regular}(\textbf{a}).
Taking this expectation to calculate the approximation ratio $\alpha_0$ of the $ZY_1$ ansatz, since $C_{\max}\leq M$, $\alpha_0$ is lower bounded by 
\begin{align}
    \alpha_0 \geq \max_{\theta} \frac{1}{2}\left(1+\frac{1}{M}\sum_{(i,j)\in\EE} \cos^{k_{ij}}\theta\sin\theta\right).
    \label{eq:alpha0-lower-bound}
\end{align}
When we maximize $\theta$, it can be checked that $\theta\in (0,\pi/2)$. In this region, $\cos^{k_{ij}}\theta\sin\theta$ is monotonically decreasing as $k_{ij}$ increases. Thus, to have a maximal lower bound of $\alpha_0$, $k_{ij}$ should be as small as possible. We aim to maximize the lower bound of $\alpha_0$ by minimizing $k_{ij}$. 

$k_{ij}$ can be minimized according to its expression in Eq.~\eqref{eq:kij-exp}. For a given $ZY$ orientation of the $ZY_1$ ansatz, the total number of $Z$ and $Y$ attached to each qubit is fixed for different gate sequences of the $ZY_1$ ansatz. Specifically, $n_L^{Y_j}+n_R^{Y_j}$ is fixed and can be derived by the in-degree of the head node $j$
\begin{align}
    n_L^{Y_j}+n_R^{Y_j}= \deg^-(j)-1,
\end{align}
where $\deg^{-}(j)\geq 1$ because we have presumed the orientation $(i\rightarrow j)$. However, the other two terms $n_R^{Y_i}$ and $n_L^{Z_j}$ in Eq.~\eqref{eq:kij-exp} can be made zero by moving the $Z$ ends of all $ZY$ gates to be at the far right of the quantum circuit and the $Y$ ends to be at the far left, as illustrated by the two arrows in Fig.~\ref{fig:0-local-3-regular}(\textbf{a}). This \textit{$Z$-right-$Y$-left structure} is equivalent to
\begin{align}
    n_R^{Y_i}=n_L^{Z_j}=0,
\end{align}
such that $k_{ij}$ is minimized. Thus, the optimal $ZY$ orientation and gate sequence of the $ZY$ ansatz should obey the $Z$-right-$Y$-left structure on each qubit.

The $Z$-right-$Y$-left structure leads to the DAG $ZY$ orientation, and the gate sequence should follow the topological order of the DAG.  If a directed graph has a directed cycle, every node in the directed cycle has at least one $Z$ and one $Y$, as illustrated in Fig.~\ref{fig:0-local-3-regular}(\textbf{b}). Then, starting from an arbitrary edge of the cycle, e.g., $(i\to k)$ in Fig.~\ref{fig:0-local-3-regular}(\textbf{b}), the $Z$-right-$Y$-left structure requires that the next $ZY$ gate $(k\to j)$ should be at the right-hand-side of $(i\to k)$, and the third $ZY$ gate $(j\to i)$ should be at the right-hand-side of $(k\to j)$. But the $Y$ end at $i$ violates the $Z$-right-$Y$-left structure. Intuitively, a directed cycle and the $Z$-right-$Y$-left structure lead to an impossible Penrose stair, as shown in the right panel of Fig.~\ref{fig:0-local-3-regular}(\textbf{b}). Thus, the $Z$-right-$Y$-left structure requires the $ZY$ orientation to be acyclic, i.e., the DAG $ZY$ orientation, and the gate sequence should follow the topological order of the DAG. 

In conclusion, the DAG and its topological order give an optimal $ZY$ orientation and gate sequence in the $0$-local sense. With this gate sequence, $k_{ij}$ can be derived directly by observing the DAG
\begin{align}
    k_{ij} = n_L^{Y_j}+n_R^{Y_j} = \deg^{-}(j)-1.
    \label{eq:kij-exp-minimized}
\end{align}
Additionally, each round of the lightcone-$ZY_p$ ansatz obeys the $Z$-right-$Y$-left structure, as described in Supplementary Note~\ref{app:lightcone-ZY ansatz}.

The explicit lower bound in Theorem~1 targets on $3$-regular graphs. Here we briefly explain how it is derived. For $3$-regular graphs, $\deg^{-}(j)$ takes $1,2$ and $3$. $\deg^{-}(j)=3$ corresponds to the maximum $k_{ij}$, which is not preferred to maximize the lower bound of $\alpha_0$. Note that $\deg^{-}(j)=3$ only if $j$ is the sink of the DAG, whereas the number of sinks of the lightcone-$ZY_1$ ansatz is the minimum $N_-=1$. This feature further supports the optimality of the lightcone-$ZY_1$ ansatz. The other two cases with $\deg^{-}(j)=1$ and $\deg^{-}(j)=2$ are illustrated in Fig.~\ref{fig:0-local-3-regular}(\textbf{c}). In a $3$-regular graph with an infinite number of nodes $N\rightarrow\infty$, the numbers of these two cases should satisfy a specific ratio. Taking this ratio to Eq.~\eqref{eq:alpha0-lower-bound} leads to the lower bound in  Theorem~1, which are recalled as follows.

\begin{customthm}{1}
    For a 3-regular graph with the number of nodes $N\to \infty$, among all $ZY_1$ ans\"atze with different $ZY$ orientations and gate sequences, $\alpha_0$ has the maximal lower bound
\begin{equation}
\begin{aligned}
    \alpha_0\geq \max_{\theta} \frac{1}{2} [1&+(1-k_{N_+})\sin\theta+k_{N_+}\cos\theta\sin\theta],
    \label{eq:alpha_0_lower-bound-app}
\end{aligned}        
\end{equation}
where $k_{N_+}\equiv 2/3+4 N_+/(3N)$, and this maximal lower bound is achieved by the lightcone-$ZY_1$ ansatz.
\end{customthm}

\noindent \textit{proof.} We start the proof from the lower bound of $\alpha_0$ given in Eq.~\eqref{eq:alpha0-lower-bound}. 
\begin{align}
    \alpha_0 \geq \max_{\theta} \frac{1}{2} (1+\frac{1}{M}\sum_{(i,j)\in\EE} \cos^{k_{ij}}\theta\sin\theta).
    \label{eq:alpha0-lower-bound-app}
\end{align}
We have shown that this lower bound is maximized by arranging $ZY$ gates following the topological order of a DAG of the undirected graph, and $k_{ij}$ equals the head in-degree of $(i\to j)$
\begin{align}
    k_{ij} = \deg^-(j)-1\equiv I.
\end{align}
For $3$-regular graph, $I\in\{0,1,2\}$. We denote the number of edges that have head in-degree $I$ with $m_I$, and the ratio of this kind of edges as $r_I\equiv m_I/M$. These ratios satisfy the normalization condition
\begin{align}
    \sum_{I=0}^2 r_I=1.
    \label{eq:normalization-condition}
\end{align}
Using these notations, Eq.~\eqref{eq:alpha0-lower-bound-app} can be rewritten as 
\begin{align}
    \alpha_0 \geq \max_{\theta} \frac{1}{2} (1+\sum_{I=0}^2 r_I\cos^{I}\theta\sin\theta).
    \label{eq:alpha_0-ratiolized}
\end{align}
This lower bound is optimized if $r_2$ is minimized, because $\cos^{2}\theta\sin\theta < \cos\theta\sin\theta<\sin\theta$ for $\theta\in (0,\pi/2)$. Note that $I=\deg^-(j)-1=2$ only if $j$ is the sink node of the $3$-regular graph, and every sink node has three entering edges with head in-degree $I=2$. Thus, 
\begin{align}
    r_2 = \frac{3N_-}{M} = \frac{2N_-}{N},
    \label{eq:r2}
\end{align}
where we have used $M=3N/2$ for a $3$-regular graph. We see that $r_2$ is minimized by choosing $N_-=1$, which is a feature of the DAG of the lightcone-$ZY$ ansatz. 

In the following subsection, Corollary~\ref{corollary:averaged-head-in-degree=3=regular} provides another linear relationship of $r_I$ for arbitrary directed $3$-regular graphs
\begin{align}
    \sum_{I=0}^{2} r_I I = \frac{1}{M} \sum_{(i,j)\in\EE} (\deg^-(j)-1)\equiv I_h= \frac{2}{3}+\frac{4(N_++N_-)}{3N}.
    \label{eq:average-constrain}
\end{align}
Combining Eq.~(\ref{eq:normalization-condition}, \ref{eq:r2}) and Eq.~\eqref{eq:average-constrain}, we solve that
\begin{align}
    r_1 = k_{N_+}-\frac{8N_-}{3N},~r_0 = 1-k_{N_+}+\frac{2N_-}{3N}.
\end{align}
where $k_{N_+}\equiv 2/3+4N_+/(3N)$. Taking these ratios to Eq.~\eqref{eq:alpha_0-ratiolized} and $N_-=1, N\to \infty$ give the lower bound of the theorem. 

In the above derivation, $r_I, I=0,1,2$ is uniquely determined by $N_+$ and $N_-$ of the DAG, whereas more internal details of the DAG do not matter. In other words, given a fixed $N_+$, the lightcone-$ZY_1$ ansatz on infinite 3-regular graphs cannot be further optimized by fine-tuning the internal structure of the DAG. Therefore, this lower bound of $\alpha_0$ is optimal. Additionally, it can be achieved by the lightcone-$ZY_1$ ansatz. $\qed$

\subsection{Averaged heads in-degree of directed $D$-regular graphs}\label{app:Geometry constraints}

In this section, we prove some geometric properties of directed $D$-regular graphs. For a directed edge $(i\to j)$, $(\deg^- (j)-1)$ plays an important role in calculating $\alpha_0$'s lower bound. We call this quantity the \textit{head in-degree} of $(i\to j)$. For a directed graph $G=(\VV,\EE)$, we define the \textit{averaged heads in-degree} over all directed edges, which reads
\begin{align}
    I_h \equiv \frac{1}{M} \sum_{(i,j)\in\EE} (\deg^-(j)-1).
    \label{eq:It-Ih-definition}
\end{align}
where $M$ is the number of edges of the directed graph. We can derive the following lemma according to this definition.
\begin{lemma}
    Let $G$ be a directed $D$-regular graph with $N$ nodes. Then
    \begin{align}
        I_h\leq D-3+\frac{2}{D}+(2-\frac{2}{D})\frac{N_++N_-}{N}
    \end{align}
    and 
    \begin{align}
    I_h\geq  \left\{ \begin{array}{ll}
 \frac{D}{2}-1+\frac{D}{2}\frac{N_++N_-}{N} & \textrm{for even D.}\\
  \frac{D}{2}-1+\frac{1}{2D}+(\frac{D}{2}-\frac{1}{2D})\frac{N_++N_-}{N}& \textrm{for odd D.}
  \end{array} \right.
\end{align}
where $N_+$ and $N_-$ is the numbers of the source and sink of $G$, respectively.
\label{lemma:averaged heads in-degree}
\end{lemma}
\textit{proof}. For a directed graph $G$, the summation of the head in-degrees over directed edges in Eq.~\eqref{eq:It-Ih-definition} is equal to a summation over nodes
\begin{equation}
\begin{aligned}
    I_h &= \frac{1}{M} \sum_{i\in \VV} \deg^-(i)(\deg^-(i)-1)\\
    &= \frac{1}{M} \sum_{i\in \VV} \deg^-(i)(D-\deg^+(i)-1)\\
    &= D-1-S(G'),
    \label{eq:averaged-in-degree-of-heads}
\end{aligned}
\end{equation}
where we used $\deg^+(i)+\deg^-(i)=D$ for all nodes $i$ of the $D$-regular graph in the second line. The relation $\sum_{i\in \VV} \deg^-(i)=M$ is used and we define $ S(G')\equiv \sum_{i\in \VV} \deg^-(i)\deg^+(i)$ in the third line. If a node $i$ is a source (sink) of $G'$, then $\deg^-(i)=0$ ($\deg^+(i)=0$). Thus we have
\begin{align}
    S(G') &= \frac{1}{M} \sum_{i\in \VV_{/ss}} \deg^+(i)\deg^-(i)= \frac{1}{M} \sum_{i\in \VV_{/ss}} (D-\deg^-(i) )\deg^-(i),
\end{align}
where $\VV_{/ss}$ denotes the set of nodes without sink and source. If $i$ is not a sink or source, we have $1 \leq \deg^-(i)\leq D-1$. Notice that $(D-\deg^-(i) )\deg^-(i)$ is a quadratic function of $\deg^-(i)$, thus $S(G')$ is lower bounded by 
\begin{align}
    S(G')\geq  \frac{1}{M} \sum_{i\in \VV'} (D-1)=\frac{2}{D} (1-\frac{N_+ +N_-}{N})(D-1),
    \label{eq:right-hand-side-inequality}
\end{align}
and the equality of the inequality is attained if $\deg^-(i)=1$ or $D-1$ for all $i\in\VV_{/ss}$. Additionally, $S(G')$ is upper bounded by
\begin{align}
    S(G')\leq  \left\{ \begin{array}{ll}
 \frac{1}{M}\sum_{i\in \VV_{/ss}} \frac{D^2}{4}=\frac{D}{2}(1-\frac{N_+ +N_-}{N}) & \textrm{for even D;}\\
 \frac{1}{M}\sum_{i\in \VV_{/ss}} \frac{D^2-1}{4}=(\frac{D}{2}-\frac{1}{2D})(1-\frac{N_+ +N_-}{N}) & \textrm{for odd D,}
  \end{array} \right.
  \label{eq: SG-upperbound}
\end{align}
which splits into even $D$ and odd $D$ cases because $\deg^-(i)$ can only take integer numbers. The lemma is proved by taking these bounds of $S(G')$ to Eq.~\eqref{eq:averaged-in-degree-of-heads}. $\qed$

We have the following two corollaries derived directly from this lemma.
\begin{corollary}\label{corollary:averaged-head-in-degree=3=regular}
    Let $G$ be a directed $3$-regular graph with $N$ nodes. Then
    \begin{align}
        I_h = \frac{2}{3}+\frac{4(N_++N_-)}{3N}.
    \end{align}
\end{corollary}
This corollary holds because when $D=3$, both the lower bound and the upper bound of $I_h$ equal this value.

\begin{figure}
    \centering
    \includegraphics[width=0.4\textwidth]{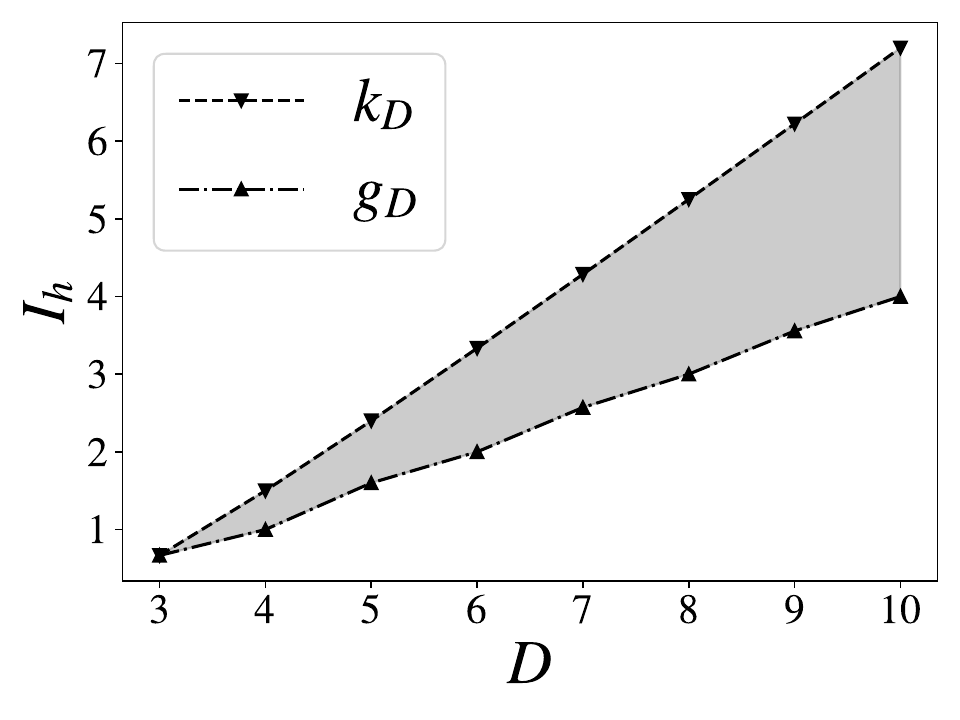}
    \caption{The lower bound and upper bound of the averaged heads in-degree $I_h$ in $D$-regular graph. The allowed value of $I_h$ is in the shadowed region.}
    \label{fig:Ih_upper_lower_bound}
\end{figure}

\begin{corollary}\label{corollary:I_h_infinit_D_regular_graph}
    Let $G$ be a directed $D$-regular graph with $N$ nodes. Assume that $N_+$ and $N_-$ are constants to $N$, as $N\to \infty$, we have
    \begin{align}
        I_h \leq k_D\equiv D-3+\frac{2}{D}
        \label{eq:I_h_upper_bound}
    \end{align}
    and 
    \begin{align}
        I_h \geq g_D\equiv \left\{ \begin{array}{ll}
 \frac{D}{2}-1 & \textrm{for even D.}\\
  \frac{D}{2}-1+\frac{1}{2D}& \textrm{for odd D.}
  \end{array} \right.
    \end{align}
    \label{eq:I_h_lower_bound}
\end{corollary}
The DAG of the bipolar-$ZY$ ansatz has $N_+=1$ and $N_-=1$ satisfying the requirement. This corollary will be used in providing the performance guarantees of the bipolar-$ZY$ ansatz in Supplementary Note~\ref{app:Proof of the lower bound on alpha 0}. In Fig.~\ref{fig:Ih_upper_lower_bound}, we plot the allowed value of $I_h$ given by $g_D$ and $k_D$. We see that there is a gap between $g_D$ and $k_D$ as $D\geq 4$.

Another quantity similar to the averaged heads in-degree is the averaged tails out-degree, defined by
\begin{align}
    O_t\equiv \frac{1}{M}\sum_{(i,j)\in \EE'}(\deg^{+}(i)-1).
    \label{eq:Ot-definition}
\end{align}
It will be utilized to analyze the performance guarantees of the bipolar-$ZY_p$ ansatz with $p=2$. Similar to $I_h$, $O_t$ can be rewritten into the summation of nodes
\begin{align}
    O_t=\frac{1}{M} \sum_{i\in\VV} \deg^{+}(i)(\deg^{+}(i)-1).
\end{align}
This leads to 
\begin{equation}
    \begin{aligned}
    O_t&=\frac{1}{M} \sum_{i\in\VV} \deg^{+}(i)(D-\deg^{-}(i)-1)=D-1-S(G')=I_h.
\end{aligned}
\end{equation}
Thus, $O_t$ has the same upper and lower bounds given by Lemma~\ref{lemma:averaged heads in-degree}.

\section{Biconnected decomposition for the MaxCut problem}\label{app:biconnected-decomposition}

\begin{figure}
    \centering
    \includegraphics[width=1\textwidth]{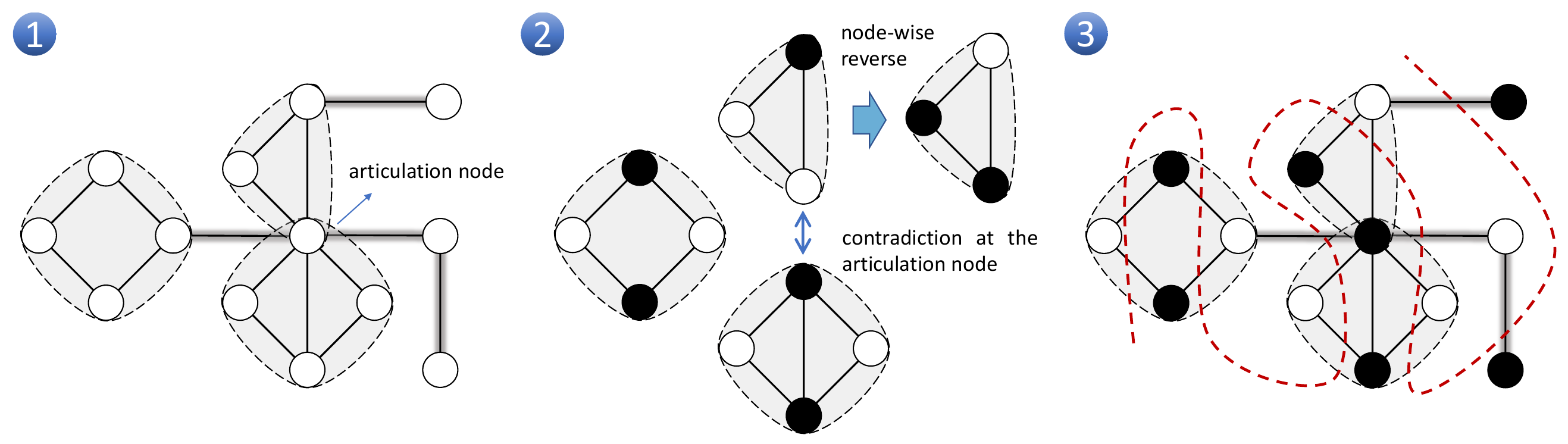}
    \caption{Illustration of the three steps to solve the MaxCut problem of a connected graph $G$. In the first step, we decompose $G$ into bridges and biconnected blocks, denoted by shaded edges and shaded regions, respectively. Secondly, we solve the MaxCut problem of each biconnected block using the bipolar-$ZY$ ansatz. Finally, combining the MaxCut solutions of each biconnected component gives the MaxCut solution of $G$. The red dashed line cuts the most number of edges of $G$.}
    \label{fig:biconnected_decomposition}
\end{figure}

In Sec.~II B of the main text, for an arbitrary connected graph $G$, we firstly use the biconnected decomposition to preprocess $G$. Then, we construct the bipolar-$ZY$ ansatz to solve the MaxCut of each biconnected component. Finally, combining solutions of the biconnected components gives the MaxCut solution of $G$. These three steps are illustrated in Fig.~\ref{fig:biconnected_decomposition}. The main text has discussed the second step of this procedure. In this note, we discuss the first and the third steps.

In the first step, the biconnected decomposition decomposes $G$ into bridges and biconnected blocks. 
\begin{definition}[Bridge]
    A bridge of $G$ is an edge whose removal disconnects $G$.
\end{definition}
\begin{definition}[Biconnected graph]
    A biconnected graph $G^{(2)}$ is a connected and ``nonseparable'' graph, meaning that if any one node were to be removed, $G^{(2)}$ will remain connected.
\end{definition}
\begin{definition}[Biconnected block]
    
    A biconnected block of $G$ is a biconnected subgraph of $G$ with the maximal possible number of nodes. 
\end{definition}
Any connected graph can be decomposed into bridges and biconnected blocks using depth-first search in $\OO(M+N)$ time~\cite{Hopcroft1973,cormen2001introduction}. Fig.~\ref{fig:biconnected_decomposition} shows the biconnected decomposition of an exemplary graph. The bridges are the shaded edges and biconnected blocks are the shaded regions.

In the third step, the MaxCut problem of $G$ can be solved by solving the MaxCut of all $G$'s biconnected blocks. Because the maximum cut number of $G$ and those of $G$'s biconnected components satisfy the following proposition.
\begin{proposition}\label{theorem:1}
    Assume that $B$ and $G_b$ are the set of bridges and biconnected blocks of $G$, respectively. The maximum cut number of $G$ can be calculated by
\begin{align}
    \max_{x\in G}~\mathrm{cut}(x) = |B|+\sum_{g\in G_b} \max_{x'\in g}~\mathrm{cut}(x'),
    \label{eq:cut-number-summation}
\end{align}
where $|B|$ is the total number of bridges in $G$, $\mathrm{cut}(x)$ denotes the cut number of the graph provided by a bi-partition $x$ of the graph nodes.
\end{proposition}
\noindent\textit{proof.}
This proposition holds because using biconnected decomposition, any connected graph is decomposed into a block-cut tree whose each node represents either a biconnected component (a bridge or a biconnected block) or an articulation node. The block-cut tree has no frustration such that the MaxCut solution of each biconnected component can be combined straightforwardly. If there are any contradictions of nodes bi-partition at the articulation node, as illustrated in the second step of Fig.~\ref{fig:biconnected_decomposition}, it can be solved by noting that $x$ and its node-wise reverse $\bar{x}$ have the same cut numbers. Thus, a MaxCut solution of $G$ can be retrieved by combining MaxCut solutions of all $G$'s biconnected components, and the summation in the Eq.~\eqref{eq:cut-number-summation} holds.  $\qed$

According to the proposition, all bridges can be cut exactly after combining the solutions. Thus, the approximation ratio of the whole graph $G$ is no smaller than the minimum approximation ratio among all biconnected blocks. Therefore, when we study the performance guarantees of our proposed quantum algorithm, we only need to focus on the worst-case approximation ratio of the biconnected blocks.

\section{Local analysis of bipolar-$ZY_p$ ansatz on $D$-regular graphs}\label{app:Proof of the lower bound on alpha 0}

\begin{figure}
    \centering
    \includegraphics[width=0.47\textwidth]{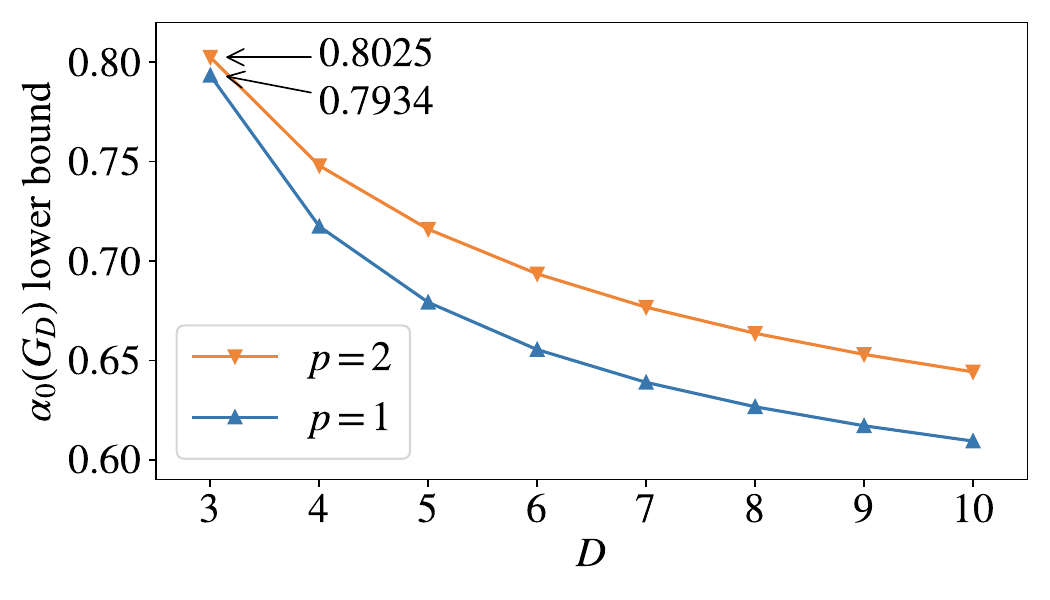}
    \caption{Lower bounds of the $0$-local approximation ratio of the bipolar-$ZY_p$ ansatz for $D$-regular graphs. Here we present the results $p=1$ and $p=2$ marked using upper triangles and lower triangles, respectively. }
    \label{fig:ratio_dependence}
\end{figure}

In this note, we provide the performance guarantees of the bipolar-$ZY_p$ ansatz using the $k$-local analysis. In Subsections~\ref{app:p1 for real in- and out-degree} and \ref{app:$0$-local analysis of the bipolar-$ZY_2$ ansatz}, we study the case $p=1$ and $p=2$ respectively using $0$-local analysis for $D$-regular graphs $G_D$. The derived lower bound of $\alpha_0(G_D)$ are plotted in Fig.~\ref{fig:ratio_dependence}. Particularly, the lower bounds $\alpha_0(G_3)\geq 0.7934$ with $p=1$ and $\alpha_0(G_3)\geq 0.8025$ with $p=2$ shown in the plot are presented in Table~I of the main text. Additionally, We see that the lower bounds of $\alpha_0$ decrease as $D$ increases for both $p=1,2$. The lower bounds are tight for $D=3$ but are not in cases of $D>3$, as will be explained in the following two subsections.

In Subsections~\ref{app:1-local-analysis-3-regular} and \ref{app:Hamiltonian-expectation}, we conduct $1$-local analysis for the bipolar-$ZY_1$ ansatz on $3$-regular graphs, where we derived $\alpha_1(G_3)\geq 0.7934$, 
which is the same as the $0$-local lower bound of $\alpha_0(G_3)$. This example presents the complexity of $k$-local analysis, where one has to consider all the $k$-local subgraphs and their possible orientations to search for the worst case. Consequently, the $k$-local analysis complexity exhibits rapid combinatorial growth with $k$.

In Subsection~\ref{app:Oriented tree conjecture}, we provide $1$-local analysis for the bipolar-$ZY_2$ and bipolar-$ZY_3$ with a conjecture that the worst case consisted of oriented trees. This conjecture is made based on the analysis results in Subsection \ref{app:1-local-analysis-3-regular}. With this conjecture, we provide the $1$-local analysis results for the bipolar-$ZY_2$ and bipolar-$ZY_3$ ansatz presented in Table~I of the main text.

\subsection{$0$-local analysis of the bipolar-$ZY_1$ ansatz}\label{app:p1 for real in- and out-degree}
For the bipolar-$ZY_1$ ansatz, we prove the following proposition.
\begin{proposition}\label{proposition:0-local-ZY-1-bound}
    Let $G_D$ be an infinite $D$-regular graph. The bipolar-$ZY_1$ ansatz on $G_D$ has the lower bound of $\alpha_0$
    \begin{align}
        \alpha_0(G_D) \geq \max_{\theta} \frac{1}{2}\left[1+(1-\frac{2}{D})(\cos\theta)^{D-3}\sin\theta+\frac{2}{D}(\cos\theta)^{D-2}\sin\theta\right].
    \end{align}
    \label{eq:app-0-local-ZY-1-bound}
\end{proposition}
It can be checked that the special case of $D=3$ can be derived directly from Theorem~1 by noting that $\lim_{N\to \infty} k_{N_+}=2/3$ for the bipolar-$ZY_1$ ansatz. 

\noindent \textit{proof.} Similar to the proof of Theorem~1, since the bipolar-$ZY_1$ ansatz is a specific lightcone-$ZY_1$ ansatz, $\alpha_0(G_D)$ is lower bounded by 
\begin{align}
    \alpha_0(G_D)\geq \max_{\theta} \frac{1}{2} \left[1+\frac{1}{M}\sum_{(i,j)\in\EE} f_{\theta}(k_{ij})\right],
    \label{eq:alpha0-lower-bound-app-D}
\end{align}
where we define $f_{\theta}(k)\equiv\cos^k\theta \sin\theta$ and $k_{ij}=\deg^-(j)-1$ for the bipolar-$ZY_1$ ansatz. For a directed $D$-regular graph, $k_{ij}$ are non-negative integers taking values $[0,D-1]$. When we maximize $\theta$, it can be checked that $\theta\in (0,\pi/2)$. In this region, $f_{\theta}(k)$ is convex and monotonically decreasing as $k$ increases. The convexity leads to
\begin{align}
    \frac{1}{M}\sum_{(i,j)\in\EE'}f_{\theta}(k_{ij})\geq f_{\theta}( \frac{1}{M}\sum_{(i,j)\in\EE'}k_{ij})=f_{\theta}(I_h),
    \label{eq:convex-f-k}
\end{align}
where $I_h$ is the averaged heads in-degree defined by Eq.~\eqref{eq:It-Ih-definition}. According to the monotonicity of $f_{\theta}(k)$ and the upper bound of $I_h$ given in Corollary~\ref{corollary:I_h_infinit_D_regular_graph}, we have 
\begin{align}
    \frac{1}{M}\sum_{(i,j)\in\EE'}f_{\theta}(k_{ij})\geq f_{\theta}(k_D).
    \label{eq:f_theta_inequality}
\end{align}
where $k_D\equiv D-3+2/D$ gives a lower bound of $\alpha_0(G_D)$. Taking this lower bound to Eq.~\eqref{eq:alpha0-lower-bound-app-D}, the maximization of $\theta$ gives
\begin{align}
     \alpha_0(G_D)\geq \frac{1}{2}\left[1+\sqrt{\frac{(k_D)^{k_D}}{(k_D+1)^{k_D+1}}} \right]. 
\end{align}

However, this lower bound is not tight. Because $k_D$ is not integer for $D>2$, the equality in Eq.~\eqref{eq:f_theta_inequality} cannot be attained for integer $k_{ij}$. Instead, since $k_D$ is located between two adjacent integers
\begin{align}
    D-3< k_D <D-2,
\end{align}
for $2<D<\infty$, $\frac{1}{M}\sum_{(i,j)\in\EE'}f_{\theta}(k_{ij})$ is lower bounded by the worst case where $k_{ij}$ takes either $D-2$ or $D-3$. Assume that the number of edges with $k_{ij}=D-3$ is $n_0$, and with $k_{ij}=D-2$ is $n_1$, and denote the ratios of each type of edges as $r_0\equiv n_0/M$, $r_1\equiv n_1/M$. We have
\begin{align}
    \frac{1}{M}\sum_{(i,j)\in\EE'}f_{\theta}(k_{ij})\geq r_{0} f_{\theta}(D-3)+r_{1} f_{\theta}(D-2),
    \label{eq:f-theta-integer-lower-bound}
\end{align}
$r_0,r_1$ satisfy the following equations and we solve that
\begin{align}
\left\{ \begin{array}{l}
r_{0}+r_{1}=1;\\
r_{0}(D-3)+r_{1}(D-2)\leq k_D.
  \end{array} \right.
\Rightarrow 
\left\{ \begin{array}{l}
r_{0}\geq 1-\frac{2}{D};\\
r_{1}\leq \frac{2}{D}.
  \end{array} \right.
\end{align}
Taking these values to Eq.~\eqref{eq:f-theta-integer-lower-bound} and combining with Eq.~\eqref{eq:alpha0-lower-bound-app-D}, we derive the lower bound of $\alpha_0(G_D)$ in the proposition. $\qed$

One can numerically maximize Eq.~\eqref{eq:app-0-local-ZY-1-bound} to give a lower bound of $\alpha_0(G_D)$. In case of $D=3$, we derive that
\begin{align}
    \alpha_0(G_3)\geq 0.7934,
\end{align}
which corresponds to the variational parameter $\theta=0.93\in (0,\pi/2)$. Numerical results for arbitrary $D$ are plotted in Fig.~\ref{fig:ratio_dependence} with $p=1$.

We have two remarks about this proposition. First, this lower bound is tight for $D=3$, but not tight for $D>3$. During the proof of the proposition, the lower bound is achieved if the heads in-degree are either $D-2$ or $D-3$, i.e., $\deg^-(i)=D-1$ or $D-2$ for $i\in \VV$. However, during the proof of Lemma~\ref{lemma:averaged heads in-degree}, the upper bound of $I_h$ is achieved if $\deg^-(i)=1$ or $D-1$ (See Eq.~\eqref{eq:right-hand-side-inequality}). These two requirements are consistent only if $D=3$. Thus, the lower bound in the proposition is not tight for $D>3$. 

Second, different from the case of $D=3$, the bipolar-$ZY_1$ ansatz on $D$-regular graphs with $D>3$ can be further optimized by fine-tuning the internal structure of the bipolar DAG. This is because the averaged heads in-degree $I_h$ are not uniquely determined for different bipolar orientations of the $D$-regular graph, as illustrated by Fig.~\ref{fig:Ih_upper_lower_bound}. Consequently, one way of further optimizing the bipolar-$ZY_1$ ansatz of the $D$-regular graph is choosing a bipolar orientation of the graph with the minimum $I_h$.

\subsection{$0$-local analysis of the bipolar-$ZY_2$ ansatz}\label{app:$0$-local analysis of the bipolar-$ZY_2$ ansatz}
\begin{figure}
    \centering
    \includegraphics[width=0.5\textwidth]{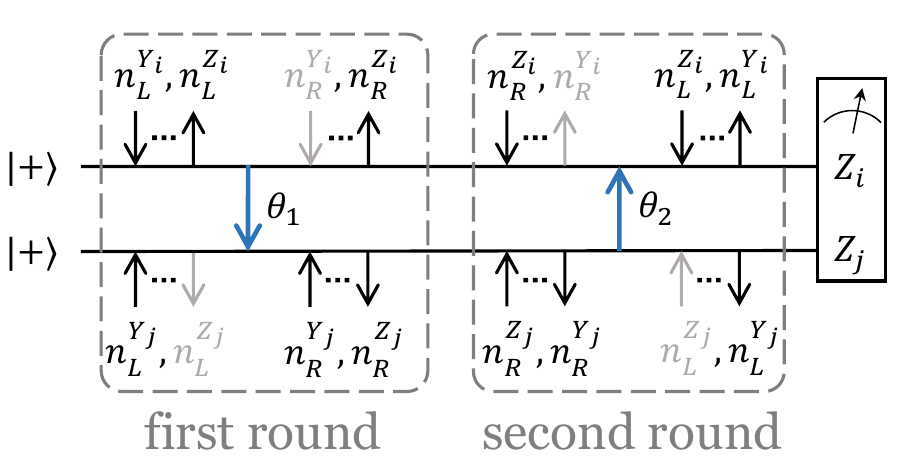}
    \caption{$0$-local analysis for bipolar-$ZY_2$ ansatz. The blue arrows denote the $ZY$ gates of the center edge $(i,j)$. The gray-colored numbers indicate that their values are zero by adopting the $Z$-right-$Y$-left structure.}
    \label{fig:p2_0_local}
\end{figure}

For the bipolar-$ZY_2$ ansatz, according to Algorithm~1, the second round takes the reversed topological order of the first round, such that the $Z$-right-$Y$-left structure defined in Supplementary Note~\ref{app:theorem-2-proof} is respected within each round. It results in the general structure of the $0$-local circuit as illustrated in Fig.~\ref{fig:p2_0_local}. In the first round, the numbers of gates adjacent to the qubit $i$ and $j$ are labeled identically to that in Fig.~\ref{fig:0-local-3-regular}(\textbf{a}). For the second round, the numbers of gates are denoted using the same numbers as in the first round. In this $0$-local circuit, $n_R^{Y_i}=n_L^{Z_j}=0$ according to the $Z$-right-$Y$-left structure. The variational parameters of the first and the second round are $\theta_1$ and $\theta_2$, respectively.

Using these gate number, the $Z_iZ_j$ expectation of the $0$-local circuit reads
\begin{align}
    \bra{\phi_2(\theta_1,\theta_2)} Z_i Z_j \ket{\phi_2(\theta_1,\theta_2)}_{0}=-\sin\theta_1 \cos^{k_{j}'} \theta_1 \cos^{k_{j}} \theta_2-\sin\theta_2 \cos^{k_{i}} \theta_2 \cos^{k_{i}'} \theta_1,
    \label{eq:ZZ-0-local-analysis-p2}
\end{align}
where the exponents are defined by
\begin{equation}
    \begin{aligned}
    k_j &\equiv n_R^{Z_i}+n_L^{Z_i}+n_R^{Z_j}+1=\deg^{+}(i)+\deg^{+}(j);\\
    k_j' &\equiv n_R^{Y_j} +n_L^{Y_j} = \deg^-(j)-1;\\
    k_i &\equiv n_L^{Z_i}+n_R^{Z_i} =\deg^{+}(i)-1;\\
    k_i' &\equiv n_R^{Z_i}+n_L^{Z_i}+n_L^{Y_i}+1 = \deg^+(i)+\deg^-(i)=D.
\end{aligned}
\end{equation}
Here, due to the $Z$-right-$Y$-left structure of each round, we can rewrite these exponents as the in- and out-degrees of the center node $i$ and $j$. Then, similar to the case of $p=1$. the $0$-local approximation ratio $\alpha_0(G_D)$ of the bipolar-$ZY_2$ ansatz is lower bounded by 
\begin{equation}
    \begin{aligned}
    \alpha_0(G_D)\geq &\max_{\theta_1,\theta_2} \frac{1}{2M} \sum_{(i,j)\in\EE}\left[1+\bra{\phi_2(\theta_1,\theta_2)} Z_i Z_j \ket{\phi_2(\theta_1,\theta_2)}_{0}\right]\\
    =&\max_{\theta_1,\theta_2} \frac{1}{2M} \sum_{(i,j)\in\EE}\left\{ 1+(\cos\theta_2)^{\deg^{+}(i)-1}\left[\sin\theta_1 (\cos\theta_2)^{D}\left(\cos\theta_1/\cos\theta_2\right)^{\deg^-(j)-1}+\sin\theta_2(\cos\theta_1)^D\right]\right\},
    \label{eq:alphaG-cut-fraction-p2}
\end{aligned}
\end{equation}
where we have used $\deg^+(j) = D-\deg^-(j)$ in the derivation of the second line. We see that the lower bound depends only on the head in-degree $(\deg^-(j)-1)$ and the tail out-degree $(\deg^+(i)-1)$ defined in Supplementary Note~\ref{app:Geometry constraints}.  Due to the convexity of the exponential function, the second line is lower bounded by
\begin{align}
    \alpha_0(G_D)\geq \max_{\theta_1,\theta_2} \frac{1}{2} \left\{ 1+(\cos\theta_2)^{O_t}\left[\sin\theta_1 \cos^D\theta_2\left(\cos\theta_1/\cos\theta_2\right)^{I_h}+\sin\theta_2\cos^D\theta_1\right]\right\},
    \label{eq:alphaG-cut-fraction}
\end{align}
where $O_t$ and $I_h$ are the averaged tails out-degree and averaged heads in-degree defined by Eq.~\eqref{eq:It-Ih-definition} and \eqref{eq:Ot-definition}, respectively. In Corollary~\ref{corollary:I_h_infinit_D_regular_graph}, for an infinite $D$-regular graph with a bipolar orientation, we provide the upper bound of $O_t$ and $I_h$ 
\begin{align}
    O_t=I_h\leq k_D,
\end{align}
Thus, the $0$-local approximation ratio is lower bounded by
\begin{align}
    \alpha_0(G_D)\geq \max_{\theta_1,\theta_2} \frac{1}{2} \left\{ 1+\left[\sin\theta_1 \cos^D\theta_2\left(\cos\theta_1\right)^{k_D}+\sin\theta_2\cos^D\theta_1(\cos\theta_2)^{k_D}\right]\right\}.
    \label{eq:0-local-p2-result}
\end{align}

\begin{figure}
    \centering
    \includegraphics[width=0.95\textwidth]{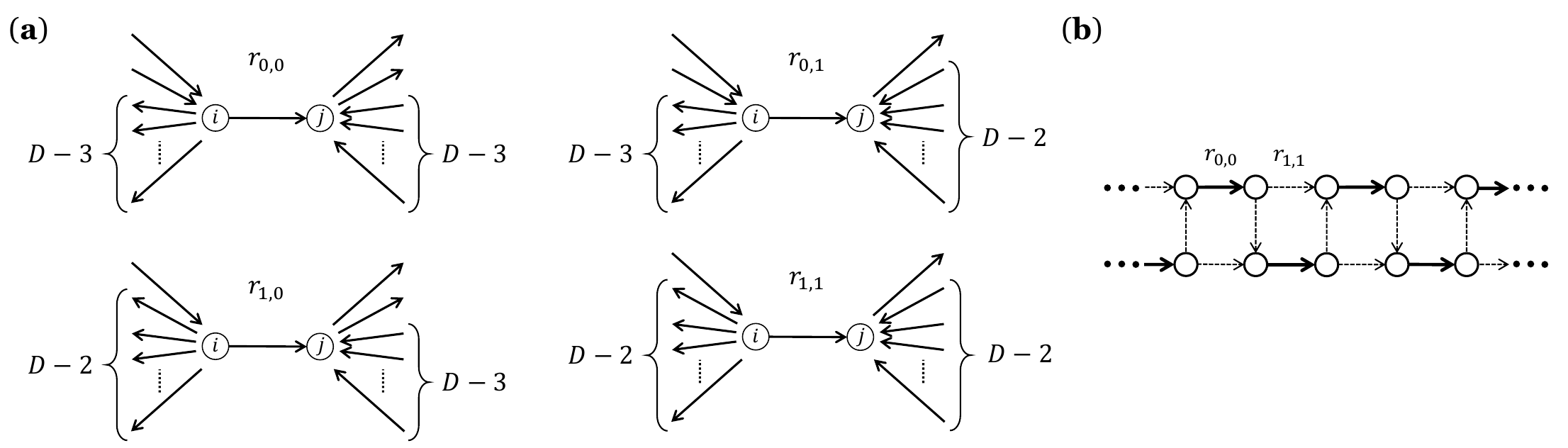}
    \caption{(\textbf{a}) Four kinds of edges that have $\deg^+(i)-1$ and $\deg^-(j)-1$ equal to $D-3$ or $D-2$. The worst-case graph of $\alpha_0(G_D)$ consists of these types of edges. (\textbf{b}) The worst-case $3$-regular graph of solving its MaxCut using the bipolar $ZY_2$ ansatz. The dashed edges and solid edges denote the types of $r_{1,1}$ and $r_{0,0}$ in (\textbf{a}), respectively.}
    \label{fig:D-regular-subgraphs}
\end{figure}

Similar to the case of $p=1$, this lower bound can be tightened by requiring integer $\deg^+(i)-1$ and $\deg^-(j)-1$. Here, because $D-3 < k_D < D-2$ for $2<D<\infty$, the worst case has both $\deg^+(i)-1$ and $\deg^-(j)-1$ taking values either $D-2$ or $D-3$. These four types of directed edges are illustrated in Fig.~\ref{fig:D-regular-subgraphs}(\textbf{a}). We denote the numbers of these four types of edges as $n_{0,0},n_{1,0},n_{0,1}$ and $n_{1,1}$, and denote their ratios as $r_{0,0}\equiv n_{0,0}/M,r_{1,0}\equiv n_{1,0}/M,r_{0,1}\equiv n_{0,1}/M$ and $r_{1,1}\equiv n_{1,1}/M$. Using these notations, the second line of Eq.~\eqref{eq:alphaG-cut-fraction-p2} is lower bounded by the worst-case result
\begin{equation}
    \begin{aligned}
    \alpha_0(G_D)\geq \frac{1}{2}+\min_{\substack{r_{0,0},r_{1,0}\\ r_{0,1},r_{1,1}}}\max_{\theta_1,\theta_2}&\frac{1}{2}\left[r_{0,0}(\sin\theta_1(\cos\theta_2)^{D} (\cos\theta_1)^{D-3}+\sin\theta_2(\cos\theta_1)^{D}(\cos\theta_2)^{D-3})\right.\\
    &+r_{0,1}(\sin\theta_1(\cos\theta_2)^{D-1} (\cos\theta_1)^{D-2}+\sin\theta_2(\cos\theta_1)^{D}(\cos\theta_2)^{D-3})\\
    &+r_{1,0}(\sin\theta_1(\cos\theta_2)^{D+1} (\cos\theta_1)^{D-3}+\sin\theta_2(\cos\theta_1)^{D}(\cos\theta_2)^{D-2})\\
    &\left. + r_{1,1}(\sin\theta_1(\cos\theta_2)^{D} (\cos\theta_1)^{D-2}+\sin\theta_2(\cos\theta_1)^{D}(\cos\theta_2)^{D-2})\right],
    \label{eq:3-regular-min-max-optimization}
\end{aligned}
\end{equation}
where the four ratios satisfy the following constraints
\begin{align}
    \left\{ \begin{array}{c}
    r_{1,0}+r_{0,0}+r_{0,1}+r_{1,1}=1;\\
         (r_{0,0}+r_{1,0})(D-3) +(r_{0,1}+r_{1,1})(D-2)\leq k_D;\\
         (r_{0,0}+r_{0,1})(D-3) +(r_{1,0}+r_{1,1})(D-2)\leq k_D.\\
    \end{array} \right.
\end{align}
Thus, the worst case can be obtained by a min-max optimization with the above constraints.

For the $3$-regular graph, for example, the result of the min-max optimization is 
\begin{align}
    \alpha_0(G_3)\geq 0.8025,
\end{align}
which occurs at $r_{0,1}=r_{1,0}=0, r_{0,0}=\frac{1}{2},r_{1,1}=1$. An example of this worst case is shown in Fig.~\ref{fig:D-regular-subgraphs}(\textbf{b}). It can be checked that $1/3$ directed edges in the directed graph are of the type $r_{0,0}$, and $2/3$ edges are of the type $r_{1,1}$. Therefore, this lower bound of $\alpha_0(G_D)$ is tight when $D=3$. However, the lower bound is not tight in the case of $D>3$, due to the same reason as we have argued in the previous subsection.

\subsection{$1$-local analysis of the bipolar-$ZY_1$ ansatz on $3$-regular graphs}\label{app:1-local-analysis-3-regular}

\begin{figure}
    \centering
    \includegraphics[width=0.8\textwidth]{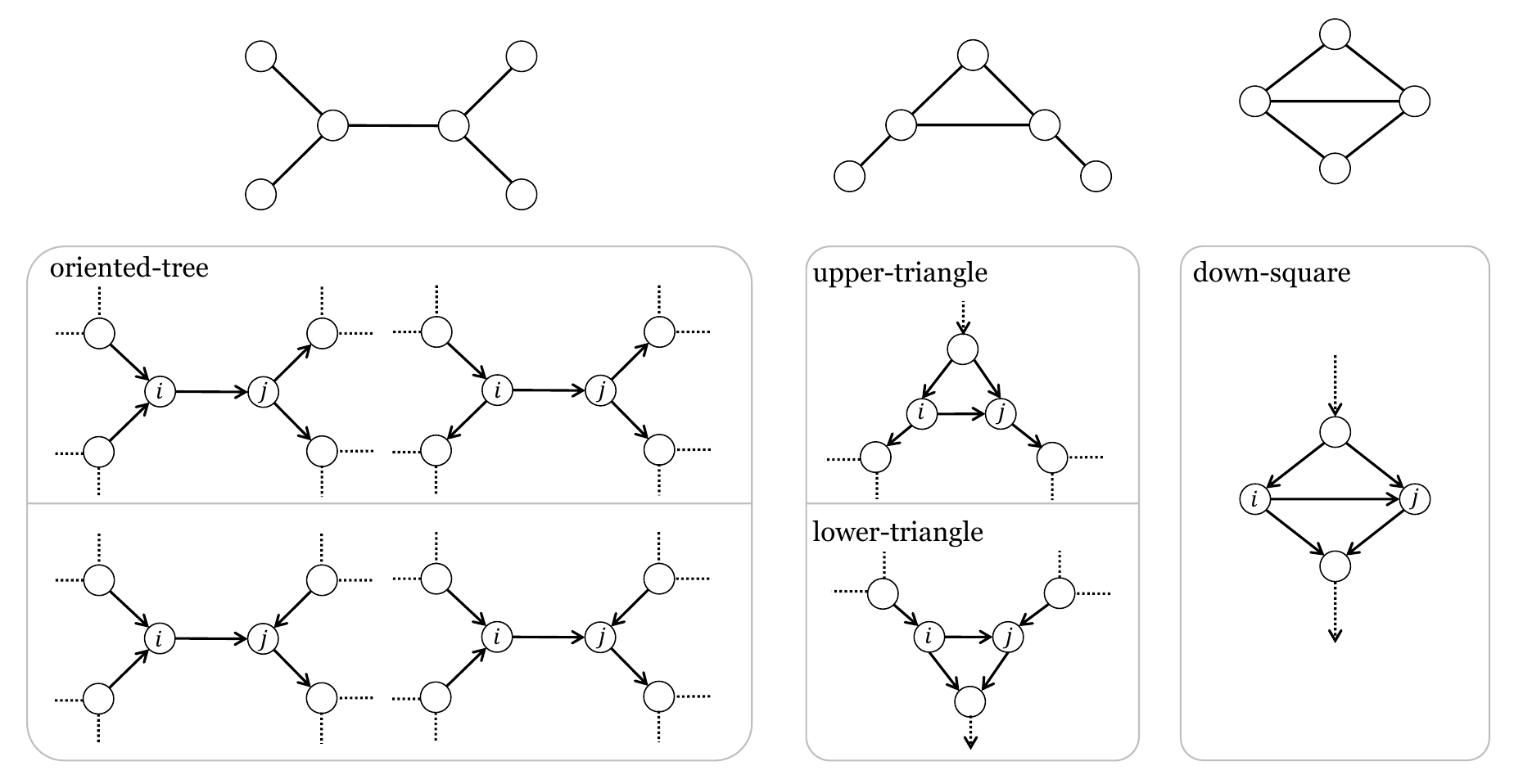}
    \caption{$1$-local directed subgraphs in directed acyclic $3$-regular graphs, which we named oriented-trees, upper-triangle, lower-triangle, and down-square. The dashed lines without orientations mean the orientations are arbitrary and are irrelevant to the 1-local expectation $\langle Z_iZ_j\rangle_1$. Since the $3$-regular graph is bipolar-oriented, subgraphs with source or sink nodes are not incorporated. Their contributions to the approximation ratio are negligible for $3$-regular graphs with an infinite number of nodes.}
    \label{fig:directed-acyclic-subgraphs}
\end{figure}

In this note, we provide the $1$-local performance guarantee of the bipolar-$ZY_1$ on $3$-regular graphs. Recall that the $1$-local performance guarantee is derived by finding the worst-case approximation ratio 
\begin{align}
    \alpha_1({G_3}) \equiv\frac{-\min_{\theta}\bra{\phi_1(\theta)} H_{\text{MC}}\ket{\phi_1(\theta)}_1}{C_{\max}}.
    \label{eq:approximation ratio-definition-app}
\end{align}
In $1$-local analysis, we find the worst case by considering all possible 1-local directed subgraphs that are allowed by the bipolar-oriented $3$-regular graphs. The possible undirected 1-local subgraphs are shown in the first row of Fig.~\ref{fig:directed-acyclic-subgraphs}. These subgraphs are used in analyzing the performance guarantees of $\text{QAOA}_1$~\cite{farhi2014quantum} on $3$-regular graphs, and we call them \textit{tree}, \textit{triangle} and \textit{square}, respectively. To analyze the bipolar $ZY_1$ ansatz, we additionally orient these 1-local subgraphs with the following two requirements.
\begin{itemize}
    \item[1.] The oriented subgraph has no directed cycles.
    \item[2.] The oriented subgraph has no source or sink nodes.
\end{itemize}
We have the second requirement because the bipolar DAG has only one source and one sink, their contribution to the analysis results can be neglected when we consider $3$-regular graphs with a large number of nodes. Directed 1-local subgraphs satisfying these requirements are shown in the second row of Fig.~\ref{fig:directed-acyclic-subgraphs}. In this figure, the solid directed edges correspond to center edges in the $1$-local analysis. Directed graphs of the tree are called \textit{oriented-trees}, and the other three directed subgraphs of the triangle and square are called \textit{upper-triangle}, \textit{lower-triangle} and \textit{down-square}, respectively. 

To calculate the Hamiltonian expectation in Eq.~\eqref{eq:approximation ratio-definition-app}, suppose a 3-regular graph with $N$ vertices contains $n_{\bigtriangleup}$ upper-triangles, $n_{\bigtriangledown}$ lower-triangles and $n_{\Diamond}$ down-squares, and denote $r_{\bigtriangleup}\equiv n_{\bigtriangleup}/N, r_{\bigtriangledown}\equiv n_{\bigtriangledown}/N,r_{\Diamond}\equiv n_{\Diamond}/N$. The number of oriented trees in directed $3$-regular graphs can be expressed in terms of $n_{\bigtriangleup},n_{\bigtriangledown}$ and $n_{\Diamond}$ according to the geometrical properties of directed $3$-regular graphs. Thus, the numerator of Eq.~\eqref{eq:approximation ratio-definition-app} can be expressed in terms of $n_{\bigtriangleup},n_{\bigtriangledown}$ and $n_{\Diamond}$. We denote 
\begin{align}
    \widetilde{\LL}_1(\theta,p=1)= -\bra{\phi_1(\theta)}H_{\text{MC}}\ket{\phi_1(\theta)}_1 \equiv NF(r_{\bigtriangleup}, r_{\bigtriangledown}, r_{\Diamond},\theta).
\end{align}
The explicit expression of $F(r_{\bigtriangleup}, r_{\bigtriangledown}, r_{\Diamond},\theta)$ will be derived in the following subsection.

For the maximum cut number $C_{\text{max}}$ of $3$-regular graphs, a graph with $n_{\bigtriangleup}+n_{\bigtriangledown}$ triangles and $n_{\Diamond}$ squares have at least $n_{\bigtriangleup}+n_{\bigtriangledown}+n_{\Diamond}$ edges that cannot be cut. Thus,  $C_{\text{max}}$ is upper bounded by 
\begin{align}
    C_{\text{max}} \leq (3N/2 - n_{\bigtriangleup} - n_{\bigtriangledown} - n_{\Diamond})
\end{align}
where $3N/2$ is the total number of edges of the $3$-regular graph. Thus, the worst-case approximation ratio of the bipolar-$ZY_1$ ansatz can be derived by the min-max optimization
\begin{align}
    \alpha_1(G_3) \geq \min_{r_{\bigtriangleup}, r_{\bigtriangledown}, r_{\Diamond}\geq 0}\max_{\theta} \frac{N_{\text{cut}}(\theta)}{C_{\text{max}}}=\min_{r_{\bigtriangleup}, r_{\bigtriangledown}, r_{\Diamond}\geq 0}\max_{\theta}\frac{F(r_{\bigtriangleup}, r_{\bigtriangledown}, r_{\Diamond},\theta)}{3/2-r_{\bigtriangleup} - r_{\bigtriangledown} - r_{\Diamond}}.
    \label{eq:1-local-min-max}
\end{align}
By numerically performing the min-max optimization, we find the worst case is achieved at $r_{\bigtriangleup}=r_{\bigtriangledown}=r_{\Diamond}=0$ and the optimized result is $\alpha_1(G_3)\geq 0.7934$. This result is listed in Table~I of the main text. 

\begin{figure}
    \centering
    \includegraphics[width=1\textwidth]{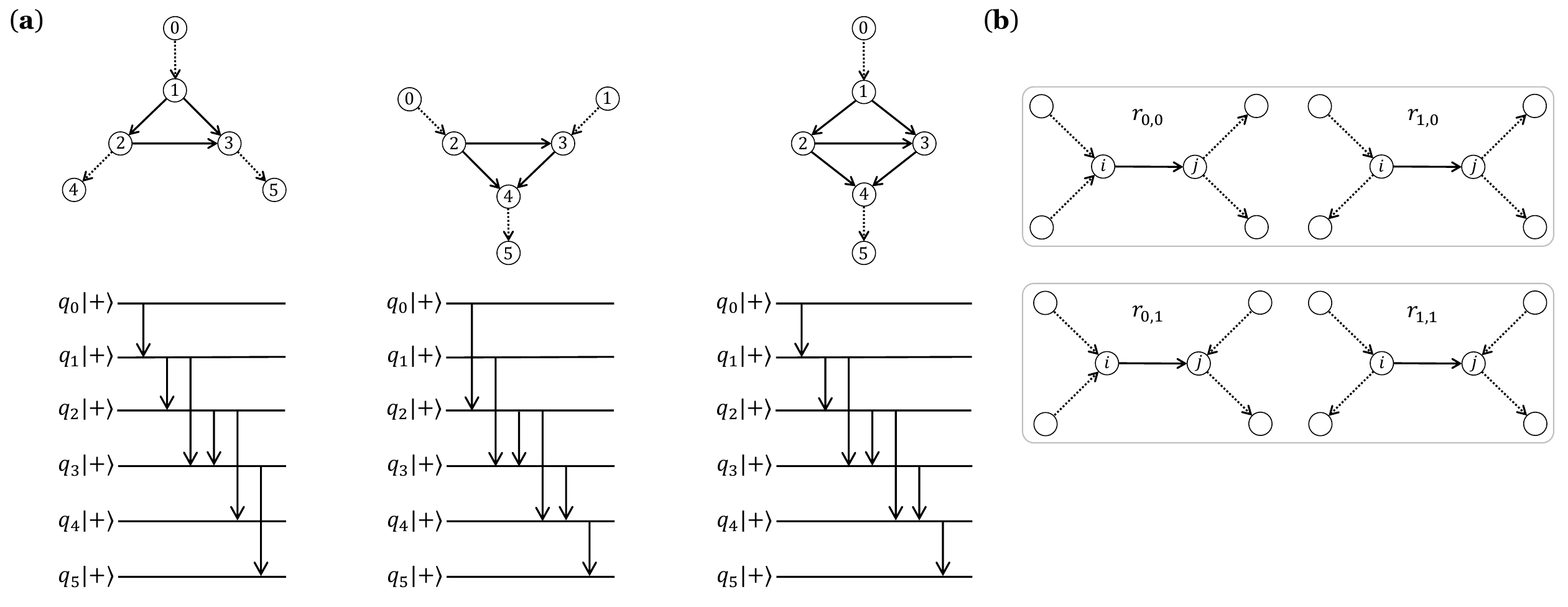}
    \caption{(\textbf{a}) 1-local subgraphs upper-triangle, lower-triangle, and down-square (upper panels) and their $ZY$ sub-circuits (lower panels). In the 1-local subgraphs, the solid directed edges are center edges in the $1$-local analysis, and the dashed edges denote other edges in the $1$-local subgraphs. (\textbf{b}) 1-local oriented trees in directed acyclic 3-regular graphs. The two boxes denote two kinds of center edges according to their head in-degrees, i.e. $\deg^{-}(j)-1=0$ for the upper box and $\deg^{-}(j)-1=1$ for the lower box.}
    \label{fig:subcircuits}
\end{figure}

\subsection{Hamiltonian expectation of $3$-regular graphs in $1$-local analysis}\label{app:Hamiltonian-expectation}

In this subsection, we explicitly calculate the MaxCut Hamiltonian expectation of $3$-regular graphs using $1$-local analysis. The 1-local circuits of the upper-triangle, lower-triangle, and down-square are illustrated in Fig.~\ref{fig:subcircuits}(\textbf{a}). The center edges of these subgraphs are 
\begin{equation}
    \begin{aligned}
    \EE_{\bigtriangleup}\equiv& \{(1,2),(1,3),(2,3)\};\\
    \EE_{\bigtriangledown}\equiv& \{(2,3),(2,4),(3,4)\};\\
    \EE_{\Diamond}\equiv& \{(1,2),(1,3),(2,3),(2,4),(3,4)\},
\end{aligned}
\end{equation}
as illustrated by the solid directed edges in the first row of Fig.~\ref{fig:subcircuits}(\textbf{a}). Then, the $Z_iZ_j$ expectation of the center edges can be calculated explicitly
\begin{equation}
    \begin{aligned}
\widetilde{\LL}_{\bigtriangleup}(\theta)&=\sum_{(i,j)\in \EE_{\bigtriangleup}}\widetilde{\LL}_{1,(i,j)}(\theta) = \frac{1}{2}(3+\sin\theta+2\cos\theta\sin\theta-2\cos\theta\sin^2\theta);\\
\widetilde{\LL}_{\bigtriangledown}(\theta)&=\sum_{(i,j)\in \EE_{\bigtriangledown}}\widetilde{\LL}_{1,(i,j)}(\theta) = \frac{1}{2}(3+3\cos\theta\sin\theta-2\cos^2\theta\sin^2\theta);\\
\widetilde{\LL}_{\Diamond}(\theta)&=\sum_{(i,j)\in \EE_{\Diamond}}\widetilde{\LL}_{1,(i,j)}(\theta) = \frac{1}{2}(5+\sin\theta+4\cos\theta\sin\theta-2\cos\theta\sin^2\theta-2\cos^2\theta\sin^2\theta),
\label{eq:triangle-exp}
\end{aligned}
\end{equation}
where $\widetilde{\LL}_{1,(i,j)}(\theta)$ denotes the expectation of $(1-Z_iZ_j)/2$ given the corresponding sub-circuit.


Center edges in the oriented trees of Fig.~\ref{fig:directed-acyclic-subgraphs} also contribute to the expected cut number. The four oriented trees are classified into two kinds, as illustrated in Fig.~\ref{fig:subcircuits}(\textbf{b}). The first kind in the upper panel of Fig.~\ref{fig:subcircuits}(\textbf{b}) has head in-degree $\deg^{-}(j)-1=0$. The second kind in the lower panel has $\deg^{-}(j)-1=1$. The $(1-Z_iZ_j)/2$ expectations of the center edge $(i\to j)$ for the first and the second kinds are 
\begin{align}
    \widetilde{\LL}^{(0)}(\theta)&=\frac{1}{2}(1+\sin\theta)
    \label{eq:tree-0-exp}
\end{align}
and 
\begin{align}
    \widetilde{\LL}^{(1)}(\theta)&=\frac{1}{2}(1+\sin\theta\cos\theta),
    \label{eq:tree-1-exp}
\end{align}
respectively.

Denote the numbers of the first and the second kinds of the oriented trees as $n_0$ and $n_1$, respectively. They can be expressed in terms of $n_{\bigtriangleup},n_{\bigtriangledown}$ and $n_{\Diamond}$ by considering the geometrical properties of directed $3$-regular graphs. According to Corollary~\ref{corollary:averaged-head-in-degree=3=regular}, the total heads in-degree of a $3$-regular bipolar DAG is
\begin{align}
    MI_h = N+2(N_++N_-) = N+4.
\end{align}
Counting the total heads in-degree of center edges in the oriented trees, upper-triangle, lower-triangle, and down-square leads to the equation
\begin{align}
    n_1+2n_{\bigtriangleup}+3n_{\bigtriangledown}+4n_{\Diamond}=N+4.
\end{align}
Additionally, the total number of center edges are 
\begin{align}
    n_0+n_1+3n_{\bigtriangleup}+3n_{\bigtriangledown}+5n_{\Diamond}=3N/2.
\end{align}
Solving these two equations gives that 
\begin{equation}
    \begin{aligned}
n_0 &= \frac{N}{2}-4-n_{\bigtriangleup} -n_{\Diamond};\\
    n_1 &= N+4-2n_{\bigtriangleup}-3n_{\bigtriangledown}-4n_{\Diamond}.
\end{aligned}
\end{equation}

Gathering the expectations of all kinds of center edges, the expected cut number of the bipolar-$ZY_1$ ansatz reads
\begin{align}
    \widetilde{\LL}_1(\theta,1) = n_{\bigtriangleup}\widetilde{\LL}_{\bigtriangleup}(\theta)+n_{\bigtriangledown}\widetilde{\LL}_{\bigtriangledown}(\theta)+n_{\Diamond}\widetilde{\LL}_{\Diamond}(\theta)+n_0 \widetilde{\LL}^{(0)}(\theta)+n_1 \widetilde{\LL}^{(1)}(\theta).
\end{align}
Denote $r_{\bigtriangleup}\equiv n_{\bigtriangleup}/N, r_{\bigtriangledown}\equiv n_{\bigtriangledown}/N,r_{\Diamond}\equiv n_{\Diamond}/N$. In the case of an infinite number of nodes, we have
\begin{equation}
    \begin{aligned}
    F(r_{\bigtriangleup},r_{\bigtriangledown},r_{\Diamond},\theta) =\lim_{N\rightarrow\infty} \frac{\widetilde{\LL}_1(\theta,1)}{N} =& r_{\bigtriangleup}\widetilde{\LL}_{\bigtriangleup}(\theta)+r_{\bigtriangledown}\widetilde{\LL}_{\bigtriangledown}(\theta)+r_{\Diamond}\widetilde{\LL}_{\Diamond}(\theta)\\
    &+(\frac{1}{2}-r_{\bigtriangleup} -r_{\Diamond}) \widetilde{\LL}^{(0)}(\theta)+(1-2r_{\bigtriangleup}-3r_{\bigtriangledown}-4r_{\Diamond}) \widetilde{\LL}^{(1)}(\theta)
\end{aligned}
\end{equation}
where $\widetilde{\LL}_{\bigtriangleup}, \widetilde{\LL}_{\bigtriangledown}, \widetilde{\LL}_{\Diamond},\widetilde{\LL}^{(0)}$ and $\widetilde{\LL}^{(1)}$ are functions of $\theta$ given by Eq.~\eqref{eq:triangle-exp},\eqref{eq:tree-0-exp} and \eqref{eq:tree-1-exp}. This formula is utilized to perform the min-max optimization in Eq.~\eqref{eq:1-local-min-max}.

\subsection{Oriented tree conjecture}\label{app:Oriented tree conjecture}

According to the $1$-local analysis results of the bipolar-$ZY_1$ on $3$-regular graphs, the worst case is achieved at $r_{\bigtriangleup}=r_{\bigtriangledown}=r_{\Diamond}=0$ after the min-max optimization in Eq.~\eqref{eq:1-local-min-max}. This result means that the worst case consisted only of the oriented tree subgraphs in Fig.~\ref{fig:directed-acyclic-subgraphs}. For this reason, in the $1$-local analysis of the bipolar-$ZY_2$ and bipolar-$ZY_3$ on $3$-regular graphs, we conjecture that the worst case also consisted of oriented tree subgraphs shown in Fig.~\ref{fig:directed-acyclic-subgraphs}. 

In the $1$-local analysis of the bipolar-$ZY_2$ and bipolar-$ZY_3$, we calculate the analytic expression of the expected cut number using Mathematica, and then search for the ratio of oriented trees with the minimum $\alpha_1(G_3)$. The procedure is similar to the $0$-local analysis of the bipolar-$ZY_2$ ansatz in Subsection~\ref{app:$0$-local analysis of the bipolar-$ZY_2$ ansatz}. The Mathematica code for calculating the expected cut number is available in~\cite{Wang2025_lightcone}.

\section{Proof of Theorem~2}\label{app:exact-lower-bound-ZY1}

In this note, we provide the exact performance guarantee of bipolar-$ZY_1$ ansatz given by Theorem~\ref{theorem:exact-ZY1-bound-app}, which we recall for convenience
\begin{customthm}{2}\label{theorem:exact-ZY1-bound-app}
For infinite $3$-regular graphs $G_3$, the bipolar-$ZY_1$ ansatz has the performance guarantee lower bounded by 
    \begin{align}
        \alpha \geq 0.7926.
        \label{eq:exact-lower-bounds-ZY1-app}
    \end{align}
\end{customthm}

\begin{figure}
    \centering
    \includegraphics[width=0.8\textwidth]{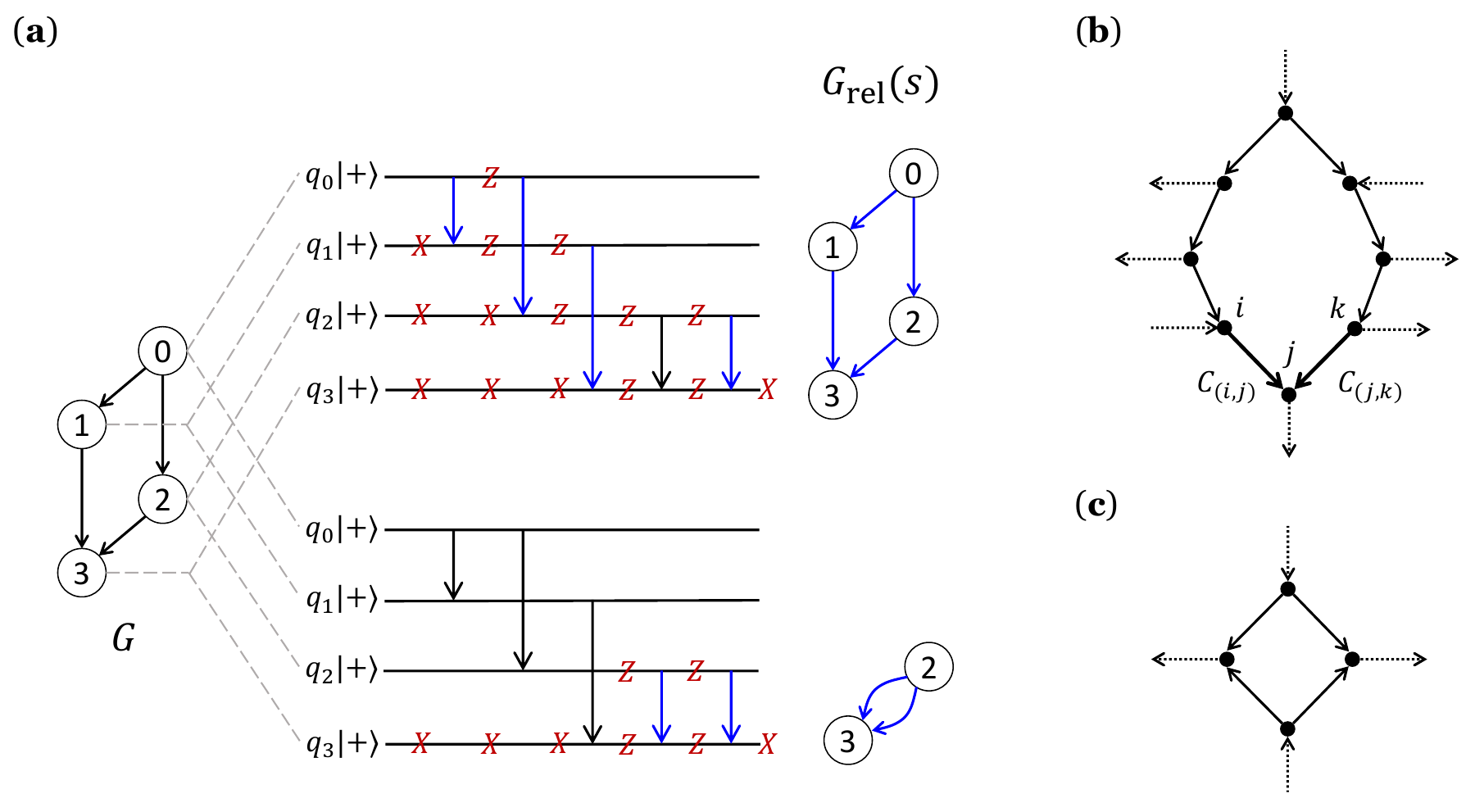}
    \caption{(\textbf{a}) Extended Pauli paths that have non-zero contributions to the $Z_iZ_j$ expectation. The upper panel illustrates an example that the relevant gate graph $G_{\text{rel}}(s)$ contains a cycle of the problem graph. The lower panel illustrates an example that the $G_{\text{rel}}(s)$ is a parallel edge cycle. (\textbf{b}) An oriented cycle with one cycle source and one cycle sink. The expectation of $C_{(i,j)}$ and $C_{(j,k)}$ give the cycle contribution shown in Eq.~\eqref{eq:cycle-contribution}. In this example, the cycle length $|\CC|=8$ and the cycle in-degree $I_{\CC}=3$.  (\textbf{c}) An oriented cycle with two cycle sources and two cycle sinks has no cycle contribution to the expected cut number.}
    \label{fig:oriented-cycle}
\end{figure}

$\alpha$ is the minimum approximation ratio $\alpha(G_3)$ among $3$-regular graphs $G_3$. Evaluating the exact approximation ratio $\alpha(G_3) = \max_{\theta}(-\bra{\phi_1(\theta)}H_{\text{MC}}\ket{\phi_1(\theta)})/C_{\max}$ requires to calculate the exact cut number $\sum_{(i,j)\in\EE}\bra{\phi_1(\theta)}C_{(i,j)}\ket{\phi_1(\theta)}$, where $C_{(i,j)}\equiv(1-Z_iZ_j)/2$. In Supplementary Note~\ref{app:truncation-error-of-k-local}, we mentioned that the expectation of $Z_iZ_j$ consists of Pauli paths. An extended Pauli path $s$ has a non-zero contribution to the expectation if its relevant gate graph $G_{\text{rel}}(s)$ is two-edge-connected. For the bipolar-$ZY_1$ ansatz, since each edge in the problem graph corresponds to only one $ZY$ gate, the relevant gate graph either contains at least one cycle of the problem graph $G$ or is a parallel edge cycle. The former and the latter cases are illustrated in the upper panel and lower panel of Fig.~\ref{fig:oriented-cycle}(\textbf{a}), respectively. The parallel edge cycles are considered in the $0$-local analysis that has been studied in Supplementary Note~\ref{app:theorem-2-proof}. In this note, we provide the exact performance guarantee by explicitly calculating the cycle contributions to the expected cut number.

\vspace{0.2cm}

\noindent\textbf{Cycle contributions ---} Consider a cycle $\CC$ as a subgraph of $G$. After bipolar orientation, the cycle is oriented. We define the cycle source (sink) of $\CC$ to be the node in $\CC$ with two cycle edges of $\CC$ leaving (entering) the node. Since the oriented cycle is acyclic, the cycle has at least one cycle source and one cycle sink. Figure~\ref{fig:oriented-cycle}(\textbf{b}) and (\textbf{c}) show examples of oriented cycles in $3$-regular graphs with one cycle source and one cycle sink, and with two cycle sources and two cycle sinks. In these two figures, a dashed-directed edge connects a node in $\CC$ with the rest of $G$ or with a node in $\CC$ itself (such that the edge is a chord of $\CC$). Here, we assume that nodes in cycles are not a source or sink of $G$, since $G$ has only one source and one sink after its bipolar orientation. The contribution of such cases can be neglected as the size of $G$ is very large. Thus all nodes have out-degree and in-degree $1$ or $2$ in Fig.~\ref{fig:oriented-cycle}(\textbf{b}) and (\textbf{c}).

The contributions of Pauli paths due to the cycle $\CC$ can be derived explicitly. We denote the sub-circuit of $G's$ bipolar-$ZY_1$ ansatz on $\CC$ as 
\begin{align}
    U_{ZY}^{\CC}(\theta)=\prod_{(i,j)\in\CC\cup \text{dashed edges}}e^{-i\theta Z_i Y_j/2},
\end{align}
whose gate sequence and $ZY$ orientation follow the topological order of the bipolar-oriented $\CC$. It can be checked that in Fig.~\ref{fig:oriented-cycle}(\textbf{b}), among all directed edges in $\CC$, only $C_{(i,j)}$ and $C_{(j,k)}$, where $j$ is the unique cycle sink, have non-zero cycle contribution. In Fig.~\ref{fig:oriented-cycle}(\textbf{c}), \textit{no} directed edges in $\CC$ has non-zero cycle contribution. In general, one can derive the cycle contributions
\begin{equation}
    \begin{aligned}
    &\sum_{(i,j)\in\CC} \bra{+}^{\otimes N} U_{ZY}^{\CC\dagger} C_{(i,j)}U_{ZY}^{\CC} \ket{+}^{\otimes N}-\sum_{(i,j)\in\CC} \bra{+}^{\otimes N} U_{ZY}^{\CC\dagger} C_{(i,j)}U_{ZY}^{\CC} \ket{+}^{\otimes N}_0\\=& \left\{ \begin{array}{ll}
 -(-\sin\theta)^{|\CC|-1}\cos^{I_{\CC}}\theta, & \textrm{$\CC$ has one cycle source and one cycle sink}\\
 0, & \textrm{$\CC$ has more than one cycle source or more than one cycle sink.}
  \end{array} \right.
  \label{eq:cycle-contribution}
\end{aligned}
\end{equation}
where $|\CC|$ is the cycle length, $I_{\CC}$ is defined to be the total in-degree of $\CC$, i.e., the total number of dashed edges entering nodes in $\CC$. The subtracted terms $\langle\cdot \rangle_0$ come from the $0$-local analysis, i.e., the parallel edge cycles of each edge in $\CC$. Figure~\ref{fig:oriented-cycle}(\textbf{b}) illustrates the first case in the above equation with $|\CC|=8$ and $I_{\CC}=3$, and Figure~\ref{fig:oriented-cycle}(\textbf{c}) illustrates the second. In the following content, we construct the worst case of solving the MaxCut problem using the bipolar-$ZY_1$ ansatz according to the cycle contributions to the expected cut number.

\vspace{0.2cm}

\noindent\textbf{The worst-case candidates ---} The exact $\sum_{(i,j)\in\EE} C_{(i,j)}=-H_{\text{MC}}$ expectation corresponds to the cut number of $G$. Thus the worst-case candidates require the expectation to be as small as possible. According to the $0$-local result given by Proposition~\ref{proposition:0-local-ZY-1-bound}, for $3$-regular graph, $\theta=0.93$ (rad) after the maximization of the lower bound. Due to the dominance of the $0$-local expectation, the exact performance guarantee of $3$-regular graph should have $\theta$ not far from $0.93$. In the neighborhood of this value, $\cos\theta$ and $\sin\theta$ are positive. Thus, the sign of $-(-\sin\theta)^{|\CC|-1}\cos^{I_{\CC}}\theta$ is determined by the parity of the cycle length, i.e., odd-length cycles negatively contribute to the expected cut number, and even-length cycles positively contribute to the cut number. Therefore, the expected cut number is smaller if the graph $G$ contains more odd-length cycles, and the bipolar-$ZY_1$ ansatz has worse performance. 

However, the existence of odd-length cycles in a graph $G$ also makes the exact cut number $C_{\text{max}}$ smaller, i.e., the denominator of the approximation ratio smaller. Because an odd-length cycle has at least one edge uncut. In contrast, even-length cycles can have all edges cut. In other words, one uncut edge belongs to at least one odd-length cycle. We focus on one uncut edge of the whole graph, such that the exact maximum cut number $C_{\max}\leq M-1$, and consider all odd-length cycles sharing this uncut edge, whose other edges are cut. In the subgraph consisting of these odd-length cycles, two or more uncut edges make $C_{\max}$ smaller and thus is not the worst case. Then, to minimize the expected cut number for the worst case, we require the total negative contribution of these odd-length cycles to be as large as possible. On the one hand, it means that the uncut edge should be adjacent to as many odd-length cycles as possible. However, on the other hand, since $\sin\theta\leq 1$, the length of odd-length cycles should be as small as possible for a larger absolute value of $-(-\sin\theta)^{|\CC|-1}\cos^{I_{\CC}}\theta$.

\begin{figure}
    \centering
    \includegraphics[width=0.98\textwidth]{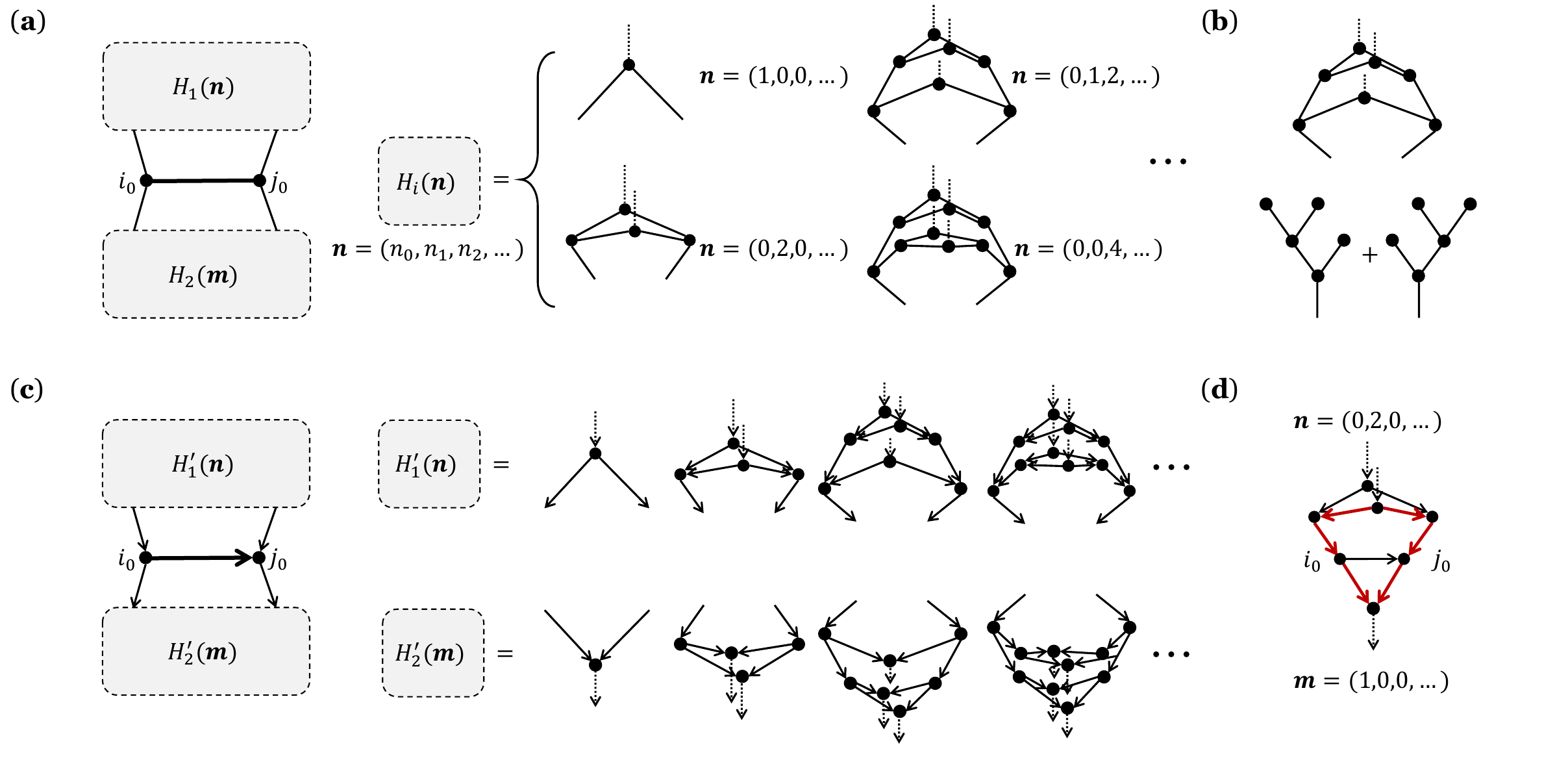}
    \caption{(\textbf{a}) The general pattern of the maximum number of odd-length cycles sharing one uncut edge $(i_0,j_0)$ with cycle lengths characterized by $\bos{n}=(n_0,n_1,n_2,\ldots)$. (\textbf{b}) The subgraph $H_i(\bos{n})$ consists of two binary trees with shared leaf nodes, such that the number of odd-length cycles is maximized. (\textbf{c}) Orientation of odd-length cycles in (\textbf{a}). This orientation has the most negative cycle contributions to the expected cut number. (\textbf{d}) An even-length cycle (red-highlighted) consisted of one odd-length cycle in $H_1(\bos{n})$ and one odd-length cycle in $H_2(\bos{m})$ that contributes to $f_{\CC}(\bos{n}',\bos{m}',\theta)$ in Eq.~\eqref{eq:fcnmtheta}.}
    \label{fig:odd-length-cycles}
\end{figure}

For $3$-regular graphs, the length of the odd-length cycles sharing one uncut edge limits their numbers. For example, an odd-length cycle with the minimum length is the triangle, whereas only 2 triangles can share the same uncut edge (This case is studied in the 1-local analysis of the bipolar-$ZY_1$ ansatz. See the down-square in Fig.~\ref{fig:directed-acyclic-subgraphs}). To determine which cycle length gives the worst case, we construct candidates of the worst case with some fixed cycle lengths and the corresponding maximal number of odd-length cycles sharing one uncut edge. These candidates are illustrated in Fig.~\ref{fig:odd-length-cycles}(\textbf{a}). In Fig.~\ref{fig:odd-length-cycles}(\textbf{a}), the uncut edge $(i_0,j_0)$ is connected with two 3-regular subgraphs $H_1(\bos{n})$ and $H_2(\bos{m})$, where $\bos{n}=(n_0,n_1,\ldots n_k\ldots)$, $\bos{m}=(m_0,m_1,\ldots m_k\ldots)$ are non-negative integers which characterize the number of odd-length cycles adjacent to $(i_0,j_0)$. For example, $\bos{n}=(1,0,0,\ldots)$, $\bos{m}=(0,2,0,0,\ldots)$ means there is one cycle with length $|\CC|=3$ in $H_1$ and two cycles with length $|\CC|=5$ in $H_2$. $k\in\mathbf{N}$ in $n_k$ is related to the cycle length by 
\begin{align}
    k = \frac{|\CC|-3}{2}.
\end{align}
Fig.~\ref{fig:odd-length-cycles}(\textbf{a}) has the maximal number of odd-length cycles since $H_i(\bos{n})$ consists of two binary trees with shared leaf nodes (See Fig.~\ref{fig:odd-length-cycles}(\textbf{b})). The tree height is proportional to the cycle length. For a given cycle length and tree height, the binary tree has the maximal number of branches, and thus those subgraphs correspond to the maximal number of odd-length cycles sharing one uncut edge.

\vspace{0.2cm}

\noindent\textbf{Proof of Theorem~\ref{theorem:exact-ZY1-bound-app} ---} Consider the worst-case candidates consisted of $H_1(\bos{n}),H_2(\bos{m})$ in Fig.~\ref{fig:odd-length-cycles}(\textbf{a}). Since the subgraphs $H_1(\bos{n}),H_2(\bos{m})$ are $3$-regular, $\bos{n},\bos{m}$ satisfy the following relation
\begin{align}
    \sum_{k=0}^{\infty} \frac{n_k}{2^k}=1;\quad \sum_{k=0}^{\infty} \frac{m_k}{2^k}=1,
\end{align}
which can be checked by examples shown in Fig.~\ref{fig:odd-length-cycles}(\textbf{a}). For convenience, we denote $n_k'\equiv n_k/2^k$,$m_k'\equiv m_k/2^k$, such that the corresponding vectors $\bos{n}',\bos{m}'$ are normalized as 
\begin{align}
    \sum_{k=0}^{\infty} n_k'=1,\quad \sum_{k=0}^{\infty} m_k'=1;\quad n_k',m_k'\geq0.
    \label{eq:normalizatin-condition}
\end{align}
Additionally, the number of edges in the whole graph of Fig.~\ref{fig:odd-length-cycles}(\textbf{a}) is a function of these numbers of cycles
\begin{align}
    M_0(\bos{n}',\bos{m'})=4\sum_{k=0}^{\infty} (n_k+m_k)-3=\sum_{k=0}^{\infty} 2^{k+2}(n_k'+m_k')-3.
    \label{eq:number-of-edges}
\end{align}


The number of odd-length cycles in Fig.~\ref{fig:odd-length-cycles}(\textbf{a}) reaches the maximum while remaining only one uncut edge $(i_0,j_0)$. For this reason, the bipolar-$ZY_1$ ansatz should have the worst behavior if the target 3-regular graph $G$ contains the subgraph in Fig.~\ref{fig:odd-length-cycles}(\textbf{a}). Additionally, if we orient the subgraph in Fig.~\ref{fig:odd-length-cycles}(\textbf{a}), the worst-case acyclic orientation should orient more odd-length cycles with one cycle source and one cycle sink, and more even-length cycles with more than one cycle source or more than one cycle sink. Such that the cycles the most negatively contribute to the expected cut number. Figure~\ref{fig:odd-length-cycles}(\textbf{c}) illustrates the worst-case orientations of the subgraphs in Fig.~\ref{fig:odd-length-cycles}(\textbf{a}). Constructing the bipolar-$ZY_1$ ansatz following this orientation, the total cycles contribution to the expected cut number reads
\begin{equation}
\begin{aligned}
    f_{\CC}(\bos{n}',\bos{m}',\theta)\equiv &-\sum_{k_1=0}^{\infty} 2^{k_1}(\sin\theta)^{2k_1+2}(\cos\theta)^{2k_1+1} n_{k_1}'\\
    &-\sum_{k_2=0}^{\infty} 2^{k_2}(\sin\theta)^{2k_2+2}(\cos\theta)^2 m_{k_2}'\\
    &+\sum_{k_1,k_2=0}^{\infty}2^{k_1+k_2} (\sin\theta)^{2(k_1+k_2)+3}(\cos\theta)^{2k_1+2} n_{k_1}' m_{k_2}',
\end{aligned}
\label{eq:fcnmtheta}
\end{equation}
where the negative contribution in the first line comes from the odd-length cycles in $H_1(\bos{n})$ with $|\CC|=2k_1+3$ and $I_{\CC}=2k_1+1$. The second line comes from the odd-length cycles in $H_2(\bos{m})$ with $|\CC|=2k_2+3$ and $I_{\CC}=2$. The third line comes from the even-length cycles. Each even-length cycle is a combination of one odd-length cycle in $H_1(\bos{n})$ and one odd-length cycle in $H_2(\bos{m})$. Figure~\ref{fig:odd-length-cycles}(\textbf{d}) gives an example of even-length cycles with $\bos{n}=(0,2,0,\ldots)$ and $\bos{m}=(1,0,0,\ldots)$.

Next, we combine these cycle contributions with the $0$-local results given by Proposition~\ref{proposition:0-local-ZY-1-bound}. According to Proposition~\ref{proposition:0-local-ZY-1-bound}, we define
\begin{align}
    g(\theta)\equiv\frac{1}{2}(1+\frac{1}{3}\sin\theta+\frac{2}{3}\sin\theta\cos\theta)
\end{align}
to be the expected cut fraction given by the $0$-local analysis. Assuming that the target graph $G$ has $M$ edges and includes the worst-case subgraph in Fig.~\ref{fig:odd-length-cycles}(\textbf{a}) characterized by $\bos{n}',\bos{m}'$. The uncut edge leads to the maximum cut number $C_{\max}\leq M-1$ such that the approximation ratio of the bipolar-$ZY_1$ ansatz on the target graph $G$ is lower bounded by 
\begin{align}
     \alpha(G_3)\geq \alpha(G)\geq \min_{M,\bos{n}',\bos{m}'} \max_{\theta}\{\frac{Mg(\theta)+f_{\CC}(\bos{n}',\bos{m}',\theta)}{M-1}\}
\end{align}
Additionally, since $G$ contains the subgraph $H_1(\bos{n}), H_2(\bos{m})$, $M$ is larger than the number of edges in the subgraph, i.e., 
\begin{align}
    \infty>M\geq M_0(\bos{n}',\bos{m'}),
\end{align}
where $M_0(\bos{n}',\bos{m'})$ is given by Eq.~\eqref{eq:number-of-edges}. Thus $\alpha(G_3)$ is lower bounded by

\begin{align}
    \alpha(G_3)\geq \min[\max_{\theta}\{g(\theta) \},~\min_{\bos{n}',\bos{m}'}\max_{\theta}\{\frac{M_0g(\theta)+f_{\CC}(\bos{n}',\bos{m}',\theta)}{M_0-1} \}].
    \label{eq:alpha_c_lower_bound}
\end{align}
The first case $\max_{\theta}\{g(\theta) \}$ indicates that $G$ contains no worst-case candidates because the worst-case candidates are not worse than tree graphs without cycles. The second case indicates that $G$ consists only of the worst-case candidates. (Although the worst-case candidates are not strictly 3-regular graphs, one can add edges to the candidates to make them 3-regular. But this only makes the expected cut number larger and the inequality in Eq.~\eqref{eq:alpha_c_lower_bound} still holds.) The first case can be derived straightforwardly as $\max_{\theta}\{g(\theta) \}=0.7934$. Thus, it left to find the lower bound of the second case.

The numerator and the denominator in the second line of Eq.~\eqref{eq:alpha_c_lower_bound} are linear functions of $\bos{n}',\bos{m}'$, and $\bos{n}',\bos{m}'$ can be interpreted as probability distribution functions satisfying the normalization condition in Eq.~\eqref{eq:normalizatin-condition}. Thus we have 
\begin{align}
    \min_{\bos{n}',\bos{m}'}\max_{\theta}\frac{M_0g(\theta)+f_{\CC}(\bos{n}',\bos{m}',\theta)}{M_0-1}=\min_{\bos{n}',\bos{m}'}\max_{\theta}\frac{\sum_{k_1 k_2=0}^{\infty} F(k_1,k_2,\theta)n_{k_1}' m_{k_2}'}{\sum_{k_1 k_2=0}^{\infty} G(k_1,k_2)n_{k_1}' m_{k_2}'},
    \label{eq:ratio-to-probability-distribution}
\end{align}
where 
\begin{equation}
    \begin{aligned}
    F(k_1,k_2,\theta)\equiv &g(\theta)(2^{k_1+2}+2^{k_2+2}-3)\\
    &- (\sin\theta)^{2k_1+2} (\cos\theta)^{2k_1+1}2^{k_1}- (\sin\theta)^{2k_2+2} (\cos\theta)^{2}2^{k_2}\\
    &+ 2^{k_1+k_2} (\sin\theta)^{2(k_1+k_2)+3}(\cos\theta)^{2k_1+2};\\
    G(k_1,k_2)\equiv &2^{k_1+2}+2^{k_2+2}-4.
\end{aligned}
\end{equation}
Then, the right-hand-side of Eq.~\eqref{eq:ratio-to-probability-distribution} is lower bounded by 
\begin{align}
    \min_{\bos{n}',\bos{m}'}\max_{\theta}\frac{\sum_{k_1 k_2=0}^{\infty} F(k_1,k_2,\theta)n_{k_1}' m_{k_2}'}{\sum_{k_1 k_2=0}^{\infty} G(k_1,k_2)n_{k_1}' m_{k_2}'}\geq \min_{\bos{n}',\bos{m}'}\max_{\theta}\min_{k_1,k_2} \frac{F(k_1,k_2,\theta)}{G(k_1,k_2)}.
    \label{eq:laldjfoasdjf}
\end{align}

Let us have a more close look at $F(k_1,k_2,\theta)/G(k_1,k_2)$. As a function of $\theta$, $F(k_1,k_2,\theta)/G(k_1,k_2)$ is a linear combination of triangular functions of $\theta$, which always has a maximal value at a point $\theta_m$. On the one hand, in the neighborhood of $\theta_m$, $F(k_1,k_2,\theta)/G(k_1,k_2)$ is a concave function of $\theta$ for every fixed $k_1,k_2$.  On the other hand, as a function of $k_1$ and $k_2$, the leading and next-leading order terms of $\sin\theta$ in $F(k_1,k_2,\theta)/G(k_1,k_2)$ reads
\begin{align}
    \frac{F(k_1,k_2,\theta)}{G(k_1,k_2)}=\frac{1}{2}(1+\frac{1}{3}\sin\theta+\frac{2}{3}\sin\theta\cos\theta)(1+\frac{1}{2^{k_1+2}+2^{k_2+2}-4})+\OO(\sin^2\theta),
\end{align}
which is a convex function of $k_1,k_2$ in the domain $k_1,k_2\geq 0$ for every fixed $\theta\in(0,\pi/2)$. According to Von Neumann's minimax theorem~\cite{min_max_theorem}, the $\max_{\theta}$ and $\min_{k_1,k_2}$ in Eq.~\eqref{eq:laldjfoasdjf} can be exchanged, such that 
\begin{align}
    \min_{\bos{n}',\bos{m}'} \max_{\theta}\min_{k_1,k_2} \frac{F(k_1,k_2,\theta)}{G(k_1,k_2)} = \min_{\bos{n}',\bos{m}'} \min_{k_1,k_2}\max_{\theta} \frac{F(k_1,k_2,\theta)}{G(k_1,k_2)}=\min_{k_1,k_2}\max_{\theta} \frac{F(k_1,k_2,\theta)}{G(k_1,k_2)}.
    \label{eq:min-max-subfinal}
\end{align}
Thus, the worst case only has the degree of freedom of $k_1$ and $k_2$. In other words, we should focus on the cases where both $H_1(\bos{n})$ and $H_2(\bos{m})$ have homogeneous cycle lengths.

\begin{figure}
    \centering
    \includegraphics[width=0.45\textwidth]{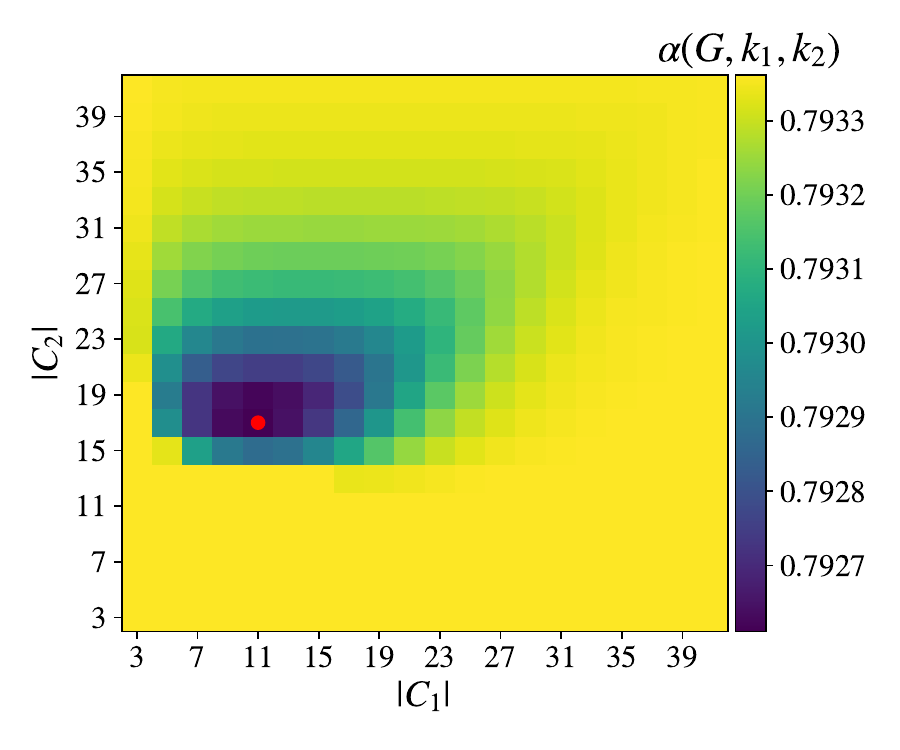}
    \caption{$\alpha(G,k_1,k_2)$ versus the cycle length $|\CC_1|=2k_1+3$ and $|\CC_2|=2k_2+3$. The global minimum $\alpha(G,k_1,k_2)=0.7926$ is at $|\CC_1|=11$ and $|\CC_2|=17$ (red dot in the plot). For large $|\CC_1|$ and $|\CC_2|$, $\alpha(G,k_1,k_2)$ asymptotically approaches the $0$-local value $0.7934$ (yellow region for large $|\CC_1|$ and $|\CC_2|$). Because the graph $G$ approaches a tree with large $|\CC_1|$ and $|\CC_2|$, and the $0$-local expectation dominates the expected cut number.}
    \label{fig:find-worst-case}
\end{figure}

Combining Eq.~(\ref{eq:alpha_c_lower_bound}-\ref{eq:min-max-subfinal}), we derive the lower bound of $\alpha(G_3)$
\begin{align}
    \alpha(G_3)\geq \min_{k_1,k_2} \alpha(G_3,k_1,k_2), 
    \label{eq:F-G-minimize}
\end{align}
where we define
\begin{align}
    \alpha(G_3,k_1,k_2) \equiv \min\left\{0.7934,~ \max_{\theta}\frac{F(k_1,k_2,\theta)}{G(k_1,k_2)}\right\}.
\end{align}
By numerically maximize $\theta$, $\alpha(G_3,k_1,k_2)$ as a function of $|\CC_1| = 2k_1+3$ and $|\CC_2| = 2k_2+3$ is plotted in Fig.~\ref{fig:find-worst-case}. The global minimum of $\alpha(G_3,k_1,k_2)$ corresponds to the worst case where $|\CC_1|=11$ and $|\CC_2|=17$ and
\begin{align}
     \alpha(G_3)\geq 0.7926.
\end{align}
We take this value as the exact performance guarantee of the bipolar-$ZY_1$ ansatz on 3-regular graphs. $\qed$

\section{Properties of the multi-angle bipolar-$ZY_1$ ansatz}
In this section, we discuss the properties of the multi-angle and the angle-relaxed bipolar-$ZY_1$ ansatz. We show that the multi-angle bipolar-$ZY_1$ ansatz can achieve the MaxCut solution of all graphs, and the angle-relaxed bipolar-$ZY_1$ ansatz derives the MaxCut solution of 2-regular graphs. In these discussions, we demonstrate the importance of the multi-angle relaxation to improve the solution accuracy of the light-cone VQA.

\subsection{Multi-angle bipolar-$ZY_1$ achieves MaxCut of all graphs}

\begin{figure}
    \centering
    \includegraphics[width=0.5\textwidth]{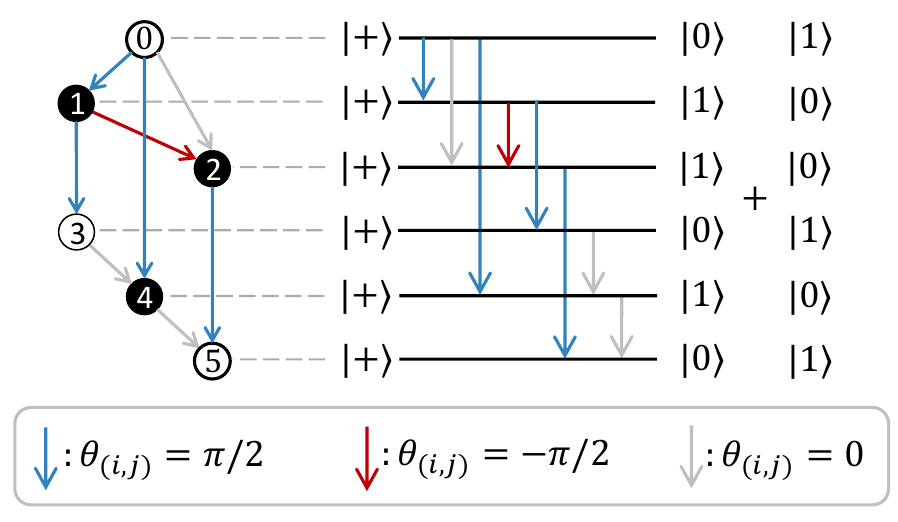}
    \caption{Parameter setting of the multi-angle bipolar-$ZY_1$ ansatz to achieve the MaxCut solution of an arbitrary graph $G$. The blue, red, and gray directed edges denote $ZY$ gates with free parameters $\pi/2$, $-\pi/2$, and $0$, respectively. The subgraph with blue and red-colored edges is a maximal tree of $G$.}
    \label{fig:multi-angle-bipolar-ZY}
\end{figure}

In this note, we show that the multi-angle bipolar-$ZY_1$ ansatz can achieve the exact MaxCut solution of any graph. For a biconnected graph $G$, the variational ansatz state of the multi-angle bipolar-$ZY_1$ ansatz reads
\begin{align}
    \ket{\phi_{\text{multi}}(\bos{\theta})} = \prod _{(i,j)\in G}e^{-i\theta_{(i,j)}Z_iY_j/2}\ket{+}^{\otimes N}
\end{align}
where the ansatz has $ZY$ orientation $(i\to j)$ and the gate sequence follows the topological order of $G$'s DAG. We assume that $G$ has one of its MaxCut solutions represented by a classical bit-string $\bos{x}=(x_0x_1\ldots x_{N-1}),x_j\in\{0,1\}$, and we define the desired solution state 
\begin{align}
    \ket{\bos{x}}\equiv \frac{1}{\sqrt{2}}(\ket{x_0x_1\ldots x_{N-1}}+\ket{\bar{x}_0\bar{x}_1\ldots \bar{x}_{N-1}}),
\end{align}
where $\bar{x}_j$ is the $0\leftrightarrow 1$ reverse of $x_j$. This solution state can be generated exactly by setting $\theta_{(i,j)}=0,\pm \pi/2$ in the multi-angle bipolar-$ZY_1$ ansatz by noting that the $ZY$ gate can be rewritten as 
\begin{equation}
    \begin{aligned}
    e^{-i\theta_{(i,j)}Z_iY_j/2}&=\cos\frac{\theta_{(i,j)}}{2}-i\sin\frac{\theta_{(i,j)}}{2} Z_iY_j\\
    &=\cos\frac{\theta_{(i,j)}}{2}(\ket{0}\bra{0}_i+\ket{1}\bra{1}_i)-i\sin\frac{\theta_{(i,j)}}{2} (\ket{0}\bra{0}_i-\ket{1}\bra{1}_i)\otimes Y_j \\
    &= \ket{0}\bra{0}_i e^{-i\theta_{(i,j)}Y_j/2}+\ket{1}\bra{1}_i e^{i\theta_{(i,j)}Y_j/2}\\
    &=\left\{ \begin{array}{ll}
 \ket{0}\bra{0}_i \otimes (\ket{1}\bra{+}_j+\ket{0}\bra{-}_j)+\ket{1}\bra{1}_i \otimes (\ket{0}\bra{+}_j-\ket{1}\bra{-}_j), & \textrm{$\theta_{(i,j)}=\pi/2$;}\\
  \ket{0}\bra{0}_i \otimes (\ket{0}\bra{+}_j-\ket{1}\bra{-}_j)+\ket{1}\bra{1}_i \otimes (\ket{1}\bra{+}_j+\ket{0}\bra{-}_j), & \textrm{$\theta_{(i,j)}=-\pi/2$.}
  \end{array} \right.
\end{aligned}
\end{equation}
The last equation means that: if the qubit $j$ is initialized in the $\ket{+}$ state, the $ZY$ gate rotates qubit $j$ aligning with the qubit $i$ to states $\ket{0_i0_j}$ or $\ket{1_i1_j}$ by taking $\theta_{(i,j)}=-\pi/2$, or anti-aligning with the qubit $i$ to states $\ket{0_i1_j}$ or $\ket{1_i0_j}$ by taking $\theta_{(i,j)}=\pi/2$. Since in the solution $\bos{x}$, two adjacent nodes $x_i x_j$ are either aligned or anti-aligned, the solution state $\ket{\bos{x}}$ can be generated exactly by applying one $ZY$ gate with $\theta_{(i,j)}=\pm\pi/2$ for each qubit.

These non-trivial $ZY$ gates can be chosen whose corresponding edges belong to the maximal tree of the bipolar-$ZY$'s DAG, i.e., a directed tree subgraph $T\subseteq G$ of the DAG connecting all the graph nodes. Then, it can be shown that $\ket{\bos{x}}=\ket{\phi_{\text{multi}}(\bos{\theta})}$ by taking free parameters
\begin{align}
    \theta_{(i,j)} = \left\{\begin{array}{ll}
        \frac{\pi}{2}, & \text{if $x_i\neq x_j$ and $(i,j)\in T$;} \\
         -\frac{\pi}{2}, & \text{if $x_i= x_j$ and $(i,j)\in T$;} \\
         0, & \text{if $(i,j)\notin T$.}
    \end{array}\right.
\end{align}
This parameter setting is illustrated in Fig.~\ref{fig:multi-angle-bipolar-ZY}, where the blue, red and gray directed edges denote $ZY$ gates with $\theta_{(i,j)}=\pi/2$, $-\pi/2$ and $0$, respectively. Although these parameters can be easily conceived given the MaxCut solution $\bos{x}$, finding these parameters using a classical optimizer without $\bos{x}$ would still require an exponential number of iterations with respect to the system size. This exponential scaling would result from the numerous local minima in the cost landscape.

\subsection{Angle-relaxed bipolar-$ZY_1$ achieves MaxCut of 2-regular graphs}\label{app:Angle-relaxed bipolar-$ZY_1$ achieves MaxCut of 2-regular graphs}

\begin{figure}
    \centering
    \includegraphics[width=0.25\textwidth]{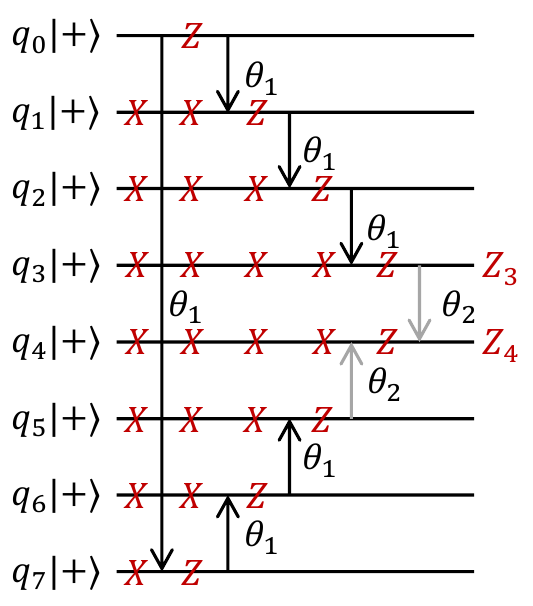}
    \caption{Angle-relaxed bipolar-$ZY_1$ ansatz on 2-regular graphs. The non-local Pauli path denoted by red $X$ and $Z$ letters gives the cycle contribution in the expected cut number $N_{\text{exp}}(\theta_1,\theta_2)$, which extends the whole 2-regular graph. This cycle contribution is essential to the exact solution for the 2-regular graph and the non-classicality of the ansatz.}
    \label{fig:bipolar-ZY-for-2-regular}
\end{figure}

The angle-relaxed bipolar-$ZY_1$ ansatz can solve the MaxCut problem of all 2-regular graphs, i.e., cycles of any length. By investigating 2-regular graphs, one can obtain intuitions on the importance of angle-relaxation, the classical simulability of the ansatz, and non-local correlations of the ansatz state. We will see that the angle-relaxed bipolar-$ZY_1$ ansatz can automatically distinguish the 2-regular graphs with even or odd lengths, and exactly solve their MaxCut with a uniform circuit structure and variational parameters. Additionally, QAOA requires $\OO(L)$ ansatz rounds to exactly solve 2-regular graphs~\cite{mbeng2019quantum}, whereas the bipolar-$ZY$ requires only one round.

Given a 2-regular graph with $L$ nodes and $L$ edges, its MaxCut solution reads
\begin{align}
    N_{\text{sol}} = \left\{ \begin{array}{ll}
 L, & \textrm{even $L$;}\\
 L-1, & \textrm{odd $L$.}
  \end{array} \right.
\end{align}
The corresponding angle-relaxed bipolar-$ZY_1$ ansatz is illustrated in Fig.~\ref{fig:bipolar-ZY-for-2-regular}, with two free variational parameters $\theta_1$ and $\theta_2$ assigned to each $ZY$ gate according to the directed edges' in-degree $\deg^-(j)=1,2$ (Different from the 3 parameters $\theta_0,\theta_1,\theta_2$ assigned in Fig.~5(\textbf{b}) of the main text, we assign 2 parameters here because we find the role of $\theta_0$ and $\theta_1$ are the same in the single-round ansatz). The expected cut number of the 2-regular graph using the ansatz reads
\begin{equation}
    \begin{aligned}
    N_{\text{exp}}(\theta_1,\theta_2) \equiv& \frac{1}{2}\sum_{i=0}^{L-1}\bra{\phi_1(\theta_1,\theta_2)}(1-Z_iZ_{i+1})\ket{\phi_1(\theta_1,\theta_2)}\\
    =& \frac{1}{2}(L+\underbrace{(L-2)\sin\theta_1 +2\sin\theta_2\cos\theta_2}_{\text{local contribution}}+\underbrace{2(-\sin\theta_1)^{L-2}\sin\theta_2\cos\theta_2}_{\text{cycle contribution}}).
    \label{eq:exact-expected-cut-number-2-regular}
\end{aligned}
\end{equation}
In this formula, terms of $\sin\theta_1$ and $\sin\theta_2\cos\theta_2$ come from the $0$-local contribution of each edge, and the last term is the cycle contribution as discussed in Note~\ref{app:exact-lower-bound-ZY1}. Particularly, the cycle contribution comes from a highly non-local Pauli path expanding the whole cycle as illustrated in Fig.~\ref{fig:bipolar-ZY-for-2-regular}. The factor of $2$ appears because two $ZZ$ observables have cycle contributions ($Z_3Z_4$ and $Z_4Z_5$ in the example of Fig.~\ref{fig:bipolar-ZY-for-2-regular}). The maximized cut number reads
\begin{align}
    \max_{\theta_1,\theta_2}  N_{\text{exp}}(\theta_1,\theta_2)=L+\frac{(-1)^{L-2}-1}{2}
    =\left\{ \begin{array}{ll}
 L, & \textrm{even $L$};\\
 L-1, & \textrm{odd $L$}
  \end{array} \right.=N_{\text{sol}},
  \label{eq:max-N-exp}
\end{align}
with the optimized parameters 
\begin{align}
    \theta_1=\frac{\pi}{2}, ~ \theta_2=\frac{\pi}{4}.
    \label{eq:variational-parameters-2-regular}
\end{align}
We see that the angle-relaxed bipolar-$ZY_1$ ansatz exactly solves the MaxCut of 2-regular graphs with even or odd lengths. Specifically, the optimized parameters $\theta_1$ and $\theta_2$ have different values. If they are set to be equal, as in the uniform-angle ansatz, the exact solution $N_{\text{sol}}$ will not be obtained. This highlights the importance of the multi-angle relaxation.

\section{Light-cone VQA for the $\mathbb{Z}_2$ gauge model}

\begin{figure}
    \centering
    \includegraphics[width=0.25\textwidth]{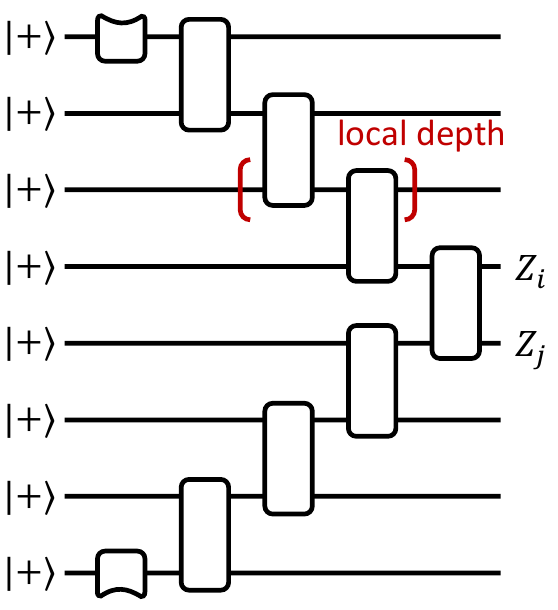}
    \cprotect\caption{Local depth of the light-cone ansatz. The lightcone-$ZY_p$ ansatz for $D$-regular graphs has a local depth of $pD$ in units of the $ZY$ gate depth, which is finite if both $p$ and $D$ are constant with respect to the system size.}
    \label{fig:finit-local-depth}
\end{figure}

The light-cone ansatz proposed in this work is applicable to general models beyond the MaxCut. Here we show how the light-cone ansatz can be extended to general models from four aspects: the BP-free property, the basic gate, the bipolar orientation, and the performance analysis.
\\\\
\textbf{The absence of BP ---} Corollary~1 in the main text shows that the light-cone ansatz for the MaxCut problem is BP-free with a constant rounds $p$ and a constant graph degree $D$. This conclusion can be extended to general physical models by considering the \textit{local depth} of the light-cone ansatz, i.e., the number of gates applied on a single qubit. The local depth of the lightcone-$ZY_p$ ansatz is $pD$, as illustrated in Fig.~\ref{fig:finit-local-depth}. Ref.~\cite{zhang_absence_2024} proves that variational circuits with a finite local depth are free from BPs. For the light-cone ansatz of general models, if the ansatz rounds $p$ and the model spatial dimension are constant to the system size, then the light-cone ansatz has a finite local depth, and thus is free from the BP problem. The model spatial dimension is proportional to its lattice degree, and therefore proportional to the circuit local depth. For example, a $d$-dimensional model has the square lattice degree $2d$, and the circuit local depth is $2dp$. Quantum many-body systems in condensed matter physics and particle physics usually have a finite spatial dimension. Thus, for these finite-dimensional quantum many-body systems, the light-cone ansatz with a constant rounds is absent of BP. 
\\\\
\textbf{Basic gate and bipolar orientation ---} For the MaxCut problem, the light-cone ansatz is constructed using the $ZY$ gate $e^{-i\theta ZY/2}$ as the basic building block. This basic gate is chosen because the $ZY$ gate performs the imaginary-time evolution of the MaxCut Hamiltonian's $ZZ$ interaction term
    \begin{align}
        e^{-\tau ZZ}\ket{++}\propto e^{-i\theta(\tau)ZY}\ket{++},
    \end{align}
    where $\tau$ is the imaginary evolution time.

\begin{figure}
    \centering
    \includegraphics[width=0.98\textwidth]{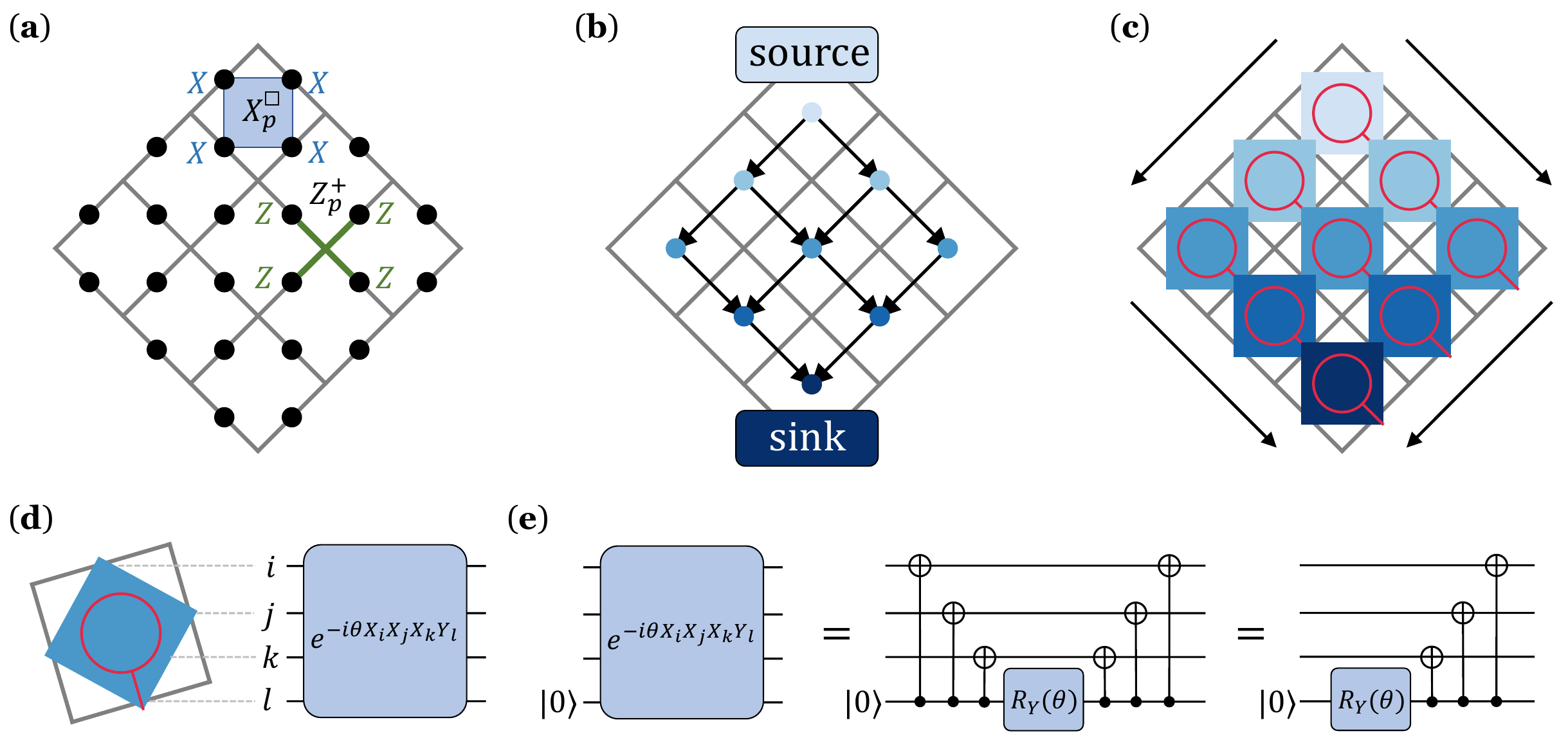}
    \cprotect\caption{(\textbf{a}) 2-$d$ square lattice of the $\mathbb{Z}_2$ gauge model. Each black dot denotes a qubit initialized to state $\ket{0}$. (\textbf{b}) Bipolar oriented graph of the $\mathbb{Z}_2$ gauge model. Each node of the graph denotes a plaquette of the original lattice. Two nodes are connected if the two plaquettes share common qubits. (\textbf{c}) The light-cone VQA for the $\mathbb{Z}_2$ gauge model can be obtained following the topological order of the bipolar oriented graph in (\textbf{b}). The color of the plaquettes---from lighter to darker---represents the sequence in which the corresponding quantum gates are applied. Each red oriented cycle denotes a plaquette rotation $e^{-i\theta XXXY}$ as depicted in (\textbf{d}). (\textbf{e}) Since the initial state is $\ket{0}^{\otimes N}$, the plaquette rotation can be equivalently realized by three CNOT gates and one $R_Y$ rotation. The three CNOT gates and the $R_Y$ rotation are used to construct the weight-adjustable loop ansatz (WALA) in Ref.~\cite{PhysRevB.107.L041109,Cochran_2025}.}
    \label{fig:Z-2-gauge-theory}
\end{figure}

    The imaginary-time evolution can be used to design a heuristic ansatz for more general models. As a concrete example, we consider the ground state preparation of the $\mathbb{Z}_2$ gauge model. The $\mathbb{Z}_2$ gauge model on a 2-$d$ square lattice has one qubit at each link of the lattice, as illustrated in Fig.~\ref{fig:Z-2-gauge-theory}(\textbf{a}). The Hamiltonian of these qubits reads
    \begin{align}
    H_{\mathbb{Z}_2} = -\sum_{\text{plaquette}} X_{p}^{\square}-\sum_{\text{vertex}} Z_{v}^{+},
    \label{eq:HZ2-Hamiltonian}
\end{align}
where the plaquette operator $X_p^{\square} = \prod_{i\in p}X_i$ are products of Pauli-$X$ operator on four links of the plaquette $p$, and the vertex operator $Z_{v}^{+}=\prod_{i\in v}$ are products of Pauli-$Z$ operator emanating from the vertex $v$. To prepare the ground state of this Hamiltonian from the initial state $\ket{0}^{\otimes N}$, using the method described in Ref.~\cite{PhysRevA.111.032612}, we find that the imaginary-time evolution of the plaquette operator $X_p^{\square} = XXXX$ can be realized by the unitary gate 
\begin{equation}
    \begin{aligned}
    e^{\tau XXXX}\ket{0000} &= [\cosh(\tau)+\sinh(\tau) XXXX]\ket{0000}=\cosh(\tau)\ket{0000}+\sinh(\tau) \ket{1111}\\
    &\propto [\sin(\theta)-i\cos(\theta) XXXY]\ket{0000}=e^{-i\theta XXXY}\ket{0000}.
\end{aligned}
\end{equation}
Thus, we use the plaquette rotation $e^{-i\theta XXXY}$ as the basic gate for the light-cone VQA of $\mathbb{Z}_2$ gauge model. 

The bipolar orientation we used for the MaxCut problem can be applied to the $\mathbb{Z}_2$ gauge model. Denote a plaquette in the 2-$d$ lattice as one node in a graph, and two nodes are connected if the two plaquettes share common qubits, as illustrated in Fig.~\ref{fig:Z-2-gauge-theory}(\textbf{b}). The resulting graph is a 2-$d$ lattice that has a natural bipolar orientation with a single source and a single sink in Fig.~\ref{fig:Z-2-gauge-theory}(\textbf{b}). Following the topological order of the bipolar oriented graph, we apply plaquette rotations $e^{-i\theta XXXY}$ sequentially to the initial state $\ket{0}^{\otimes N}$, where the qubit of the $Y$ letter is given explicitly by the oriented red cycle in Fig.~\ref{fig:Z-2-gauge-theory}(\textbf{c}, \textbf{d}). The resulting variational ansatz is the light-cone ansatz for the $\mathbb{Z}_2$ gauge model with one parameter $\theta$. This light-cone ansatz is equivalent to the weight-adjustable loop ansatz (WALA) used in Ref.~\cite{PhysRevB.107.L041109,Cochran_2025}, as depicted in Fig.~\ref{fig:Z-2-gauge-theory}(\textbf{e}), and the WALA has been proved to be able to exactly prepare the ground state of $H_{\mathbb{Z}_2}$ at $\theta=\pi/4$. Thus, the light-cone ansatz can also exactly prepare the ground state of the $\mathbb{Z}_2$ gauge model at $\theta=\pi/4$. This example shows that the light-cone VQA can be applied to more general models, and the light-cone ansatz is able to accurately prepare the models' ground states.
\\\\
\textbf{Performance analysis ---} The performance analysis proposed in the main text can also be extended to general models. Here we take the light-cone ansatz of the $\mathbb{Z}_2$ gauge model as an example. The performance analysis calculates the Hamiltonian expectation of the observable in the Heisenberg picture. For the plaquette operator $X_{p}^{\square}$ and vertex operator $Z_{v}^{+}$ in Eq.~\eqref{eq:HZ2-Hamiltonian}, their expectation of the light-cone ansatz reads
\begin{align}
    \bra{0}_l e^{i\theta X_iX_jX_kY_l} X_iX_jX_kX_l e^{-i\theta X_iX_jX_kY_l}\ket{0}_l = \bra{0}_l \cos(2\theta) X_iX_jX_kX_l+\sin(2\theta) Z_l\ket{0}_l = \sin(2\theta),
\end{align}
where $i,j,k,l$ are qubits in the plaquette $p$. Since all plaquette rotations commute with all vertex operators $Z_{v}^{+}$, we have
\begin{align}
    \bra{0}_l e^{i\theta X_iX_jX_kY_l} Z_iZ_j Z_mZ_n e^{-i\theta X_iX_jX_kY_l}\ket{0}_l = 1.
\end{align}
where $i,j,m,n$ are qubits emanating from the vertex $v$. Thus, $H_{\mathbb{Z}_2}$ has the expectation
\begin{align}
    \langle H_{\mathbb{Z}_2}\rangle_{\theta} = -\sum_{\text{plaquette}} \sin(2\theta)-\sum_{\text{vertex}} 1,
    \label{eq:HZ2-expectation}
\end{align}
with the minimum value achieved at $\theta=\pi/4$. This is the minimum possible expectation of $H_{\mathbb{Z}_2}$, since the spectral norm of $H_{\mathbb{Z}_2}$ is upper bounded by 
\begin{align}
    \| H_{\mathbb{Z}_2}\|\leq \sum_{\text{plaquette}} \|X_{p}^{\square} \|+\sum_{\text{vertex}}\| Z_{v}^{+} \| = \sum_{\text{plaquette}} 1+\sum_{\text{vertex}}1,
    \label{eq:spectral-norm-bound}
\end{align}
and the expectation $\langle H_{\mathbb{Z}_2}\rangle_{\theta}$ at $\theta=\pi/4$ exactly achieves this upper bound. In other words, similar to the MaxCut problem, we define the approximation ratio of the $\mathbb{Z}_2$ gauge model to be the estimated ground state energy over the true ground state energy $E_0$
\begin{align}
    \alpha_{\mathbb{Z}_2}\equiv \left|\frac{\min_{\theta}\langle H_{\mathbb{Z}_2}\rangle_{\theta}}{E_0} \right|\leq 1,
\end{align}
and this approximation ratio is lower bounded by the performance analysis results
\begin{align}
    \alpha_{\mathbb{Z}_2}\geq \left|\frac{\min_{\theta}\langle H_{\mathbb{Z}_2}\rangle_{\theta}}{\|H_{\mathbb{Z}_2}\|}\right|\geq 1,
\end{align}
where the first inequality uses property of the spectral norm $|E_0|\leq \|H_{\mathbb{Z}_2}\|$ and the second one takes Eq.~\eqref{eq:HZ2-expectation} and \eqref{eq:spectral-norm-bound}. Thus, the performance analysis shows that $\alpha_{\mathbb{Z}_2}=1$, confirming the ground state of $H_{\mathbb{Z}_2}$ exactly prepared by the light-cone VQA.

\section{Classical simulability of the light-cone VQA}

In this section, we consider the classical time complexity of simulating the bipolar-$ZY_1$ ansatz using two classical methods: matrix product state (MPS) and Pauli propagation. 
\\\\
\noindent\textbf{Matrix product state ---} The time complexity of simulating quantum circuits using MPS is characterized by the entanglement entropy of the quantum circuit. The entanglement entropy of the multi-angle bipolar-$ZY_1$ ansatz is defined by $S(\bos{\theta})\equiv -\tr[\rho_A(\bos{\theta})\log \rho_A(\bos{\theta})]$, where $\rho_A(\bos{\theta})=\tr_B[\ket{\phi_{\text{multi}}(\bos{\theta})}\bra{\phi_{\text{multi}}(\bos{\theta})}]$ is the reduced density matrix of the half-chain partition $A\cup B=[N]$ of the $N$-qubit system. Figure~\ref{fig:entanglement_entropy}(\textbf{a}) plots $S(\bos{\theta}_i)$ as a function of SLSQP iteration steps $i$ for one random 3-regular graph with the number of nodes $N=6,8,\ldots,18$. The time complexity of MPS is characterized by the entropy peak during the SLSQP iteration steps, as labeled by the cross in Fig.~\ref{fig:entanglement_entropy}(\textbf{a}) for a random 3-regular graph with $N$ nodes. For each number of nodes $N$, we generate $50$ random 3-regular graphs with their entropy peaks plotted in Fig.~\ref{fig:entanglement_entropy}(\textbf{b}). We see that the median of these entropy peaks among 50 graphs increases linearly with the number of qubits. Since the time complexity of simulating quantum circuits using MPS scales as $T_{\text{MPS}}\sim e^{S(\bos{\theta})}$~\cite{ORUS2014117}, the linearly increasing behavior indicates that the multi-angle bipolar-$ZY_1$ ansatz is hard to simulate by MPS for a large number of qubits.
\\\\
\noindent\textbf{Pauli propagation without truncation ---} 
Pauli propagation method~\cite{Angrisani2025,lerch2024efficientquantumenhancedclassicalsimulation,rudolph2025paulipropagationcomputationalframework} evaluates the observable expectation of the variational ansatz by summing over contributions from Pauli paths, as introduced in Supplementary Note~\ref{app:truncation-error-of-k-local}. The efficiency of the Pauli propagation method is guaranteed for the locally scrambling circuits with appropriate truncation schemes~\cite{Angrisani2025}. Here, we consider the time complexity of the Pauli propagation to evaluate the MaxCut Hamiltonian expectation with three schemes: Pauli propagation without truncation, with Pauli path weight truncation~\cite{Angrisani2025}, and with coefficient magnitude truncation~\cite{lerch2024efficientquantumenhancedclassicalsimulation}.

\begin{figure}
    \centering
    \includegraphics[width=0.98\textwidth]{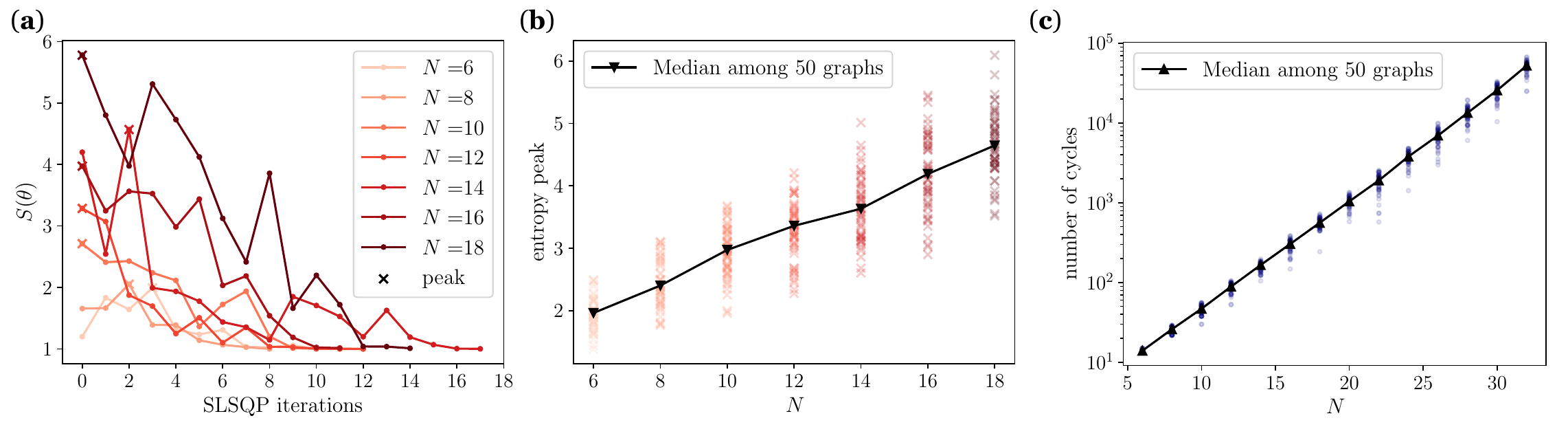}
    \cprotect\caption{(\textbf{a}) Entanglement entropy of the multi-angle bipolar-$ZY_1$ ansatz during classical optimizations using \verb|SLSQP| optimizer. The cross denotes the entropy peak in each optimization process. (\textbf{b}) Entropy peaks of $50$ random 3-regular graphs as a function of the graph nodes $N$ up to $N=18$, with their median values denoted by lower-triangles. (\textbf{c}) The number of cycles of $50$ random 3-regular graphs versus the graph nodes $N$ up to $N=32$, with the median values at each $N$ denoted by upper-triangles. The plot is in the log-$y$ scale.}
    \label{fig:entanglement_entropy}
\end{figure}   

For the Pauli propagation method without truncation, according to the evaluation formula Eq.~\eqref{eq:ZZ-summary}, the time complexity of the method is at least proportional to the number of non-zero Pauli paths contributing to the MaxCut Hamiltonian expectation. In Supplementary Note~\ref{app:exact-lower-bound-ZY1}, we have shown that a non-zero Pauli path of the bipolar-$ZY_1$ ansatz corresponds to a cycle of the graph. Thus, the number of non-zero Pauli paths is proportional to the number of cycles of the graph. For 3-regular graphs, it has been shown that the number of cycles can grow exponentially with the number of nodes~\cite{arman2017maximumnumbercyclesgraph}. In Fig.~\ref{fig:entanglement_entropy}(\textbf{c}), we numerically estimate the number of cycles in random 3-regular graphs. Up to $N=32$ nodes, the median number of cycles over 50 random graphs grows exponentially with $N$, indicating that the Pauli propagation method without truncation is hard to evaluate the Hamiltonian expectation for very large graphs.
\\\\
\noindent\textbf{Pauli propagation with Pauli weight truncation ---} Pauli propagation with Pauli path weight truncation discards high-weight Pauli paths in Eq.~\eqref{eq:ZZ-summary}. Applying this method to the bipolar-$ZY_1$ ansatz, non-zero contributions from cycles with large lengths are truncated, for example, with lengths growing linearly with the system size. In this way, the MaxCut Hamiltonian expectation can be efficiently but approximately evaluated. However, for the bipolar-$ZY_1$ ansatz, truncating contributions by cycles with linearly growing length could introduce large truncation errors. We demonstrate this using the example of $2$-regular graphs that have been studied in Supplementary Note~\ref{app:Angle-relaxed bipolar-$ZY_1$ achieves MaxCut of 2-regular graphs}.

Applying the Pauli propagation method to the bipolar-$ZY_1$ ansatz on 2-regular graphs, Pauli paths with weights larger than an arbitrary constant are truncated. Thus, the cycle contribution in Eq.~\eqref{eq:exact-expected-cut-number-2-regular} is truncated in the Pauli propagation method due to its weight $L$ Pauli path as shown in Fig.~\ref{fig:bipolar-ZY-for-2-regular}. The resulting approximate cut number reads
\begin{equation}
    \begin{aligned}
    N_{\text{exp}}^{\text{tr}}(\theta_1,\theta_2) =\frac{1}{2}(L+(L-2)\sin\theta_1 +2\sin\theta_2\cos\theta_2).
\end{aligned}
\end{equation}
Maximizing this cut number gives
\begin{align}
    \max_{\theta_1,\theta_2}  N_{\text{exp}}^{\text{tr}}(\theta_1,\theta_2)=L-\frac{1}{2}.
\end{align}
Compared with the exact cut number provided by the bipolar-$ZY_1$ ansatz, the $L$-parity sensitive term $(-1)^{L-2}$ in Eq.~\eqref{eq:max-N-exp} is truncated by the Pauli propagation method, leading to a constant error
\begin{align}
    |\max_{\theta_1,\theta_2}  N_{\text{exp}}^{\text{tr}}(\theta_1,\theta_2)-\max_{\theta_1,\theta_2}  N_{\text{exp}}(\theta_1,\theta_2)|=\frac{1}{2},
\end{align}
for both even-$L$ and odd-$L$ cycles. This constant error exemplifies the limitation of the classical simulation by low-weight Pauli paths~---~the $L$-parity sensitive term is truncated in the Pauli propagation method, but this term is crucial to give the exact solution and distinguish even- and odd-length cycles. 
\\\\
\noindent\textbf{Pauli propagation with coefficient magnitude truncation ---} Pauli propagation with coefficient magnitude truncation discards Pauli paths with small enough coefficient $c_s(\theta)$ defined in Eq.~\eqref{eq:Pauli-path-coefficient}. This truncation scheme works for small rotation angles in the bipolar-$ZY_1$ ansatz. Ref.~\cite{lerch2024efficientquantumenhancedclassicalsimulation} shows that if all variational angles $\theta_i$ are independent random variables uniformly distributed in the small interval $[-r,r]$, where $r=\OO(1/\sqrt{N})$, then the truncation error can be well controlled with high probability. However, for the bipolar-$ZY_1$ ansatz, the optimized rotation angles $\theta_i$  have a constant magnitude with respect to the system size $N$. For example, in Theorem~1, the uniform-angle bipolar-$ZY_1$ ansatz has the optimized angle $\theta=0.93\sim \Theta(1)$ to maximize the lower bound in Eq.~\eqref{eq:alpha_0_lower-bound-app}. In that case, the Pauli propagation with coefficient magnitude truncation might make it hard to evaluate the expected cut number of the bipolar-$ZY_1$ ansatz with an adequately controlled truncation error.

\section{Details on the hardware implementation}\label{app:Circuit compilation and post-processing of the hardware demonstration}
In this note, we present the details of the hardware implementation given in Sec.~III B of the main text, including the optimized bipolar-$ZY_1$ circuit depth by BFS-type bipolar orientation, circuit compilation, classical optimizer, post-processing, and hardware calibration details. 

\begin{figure}

    \centering

    \includegraphics[width=0.95\textwidth]{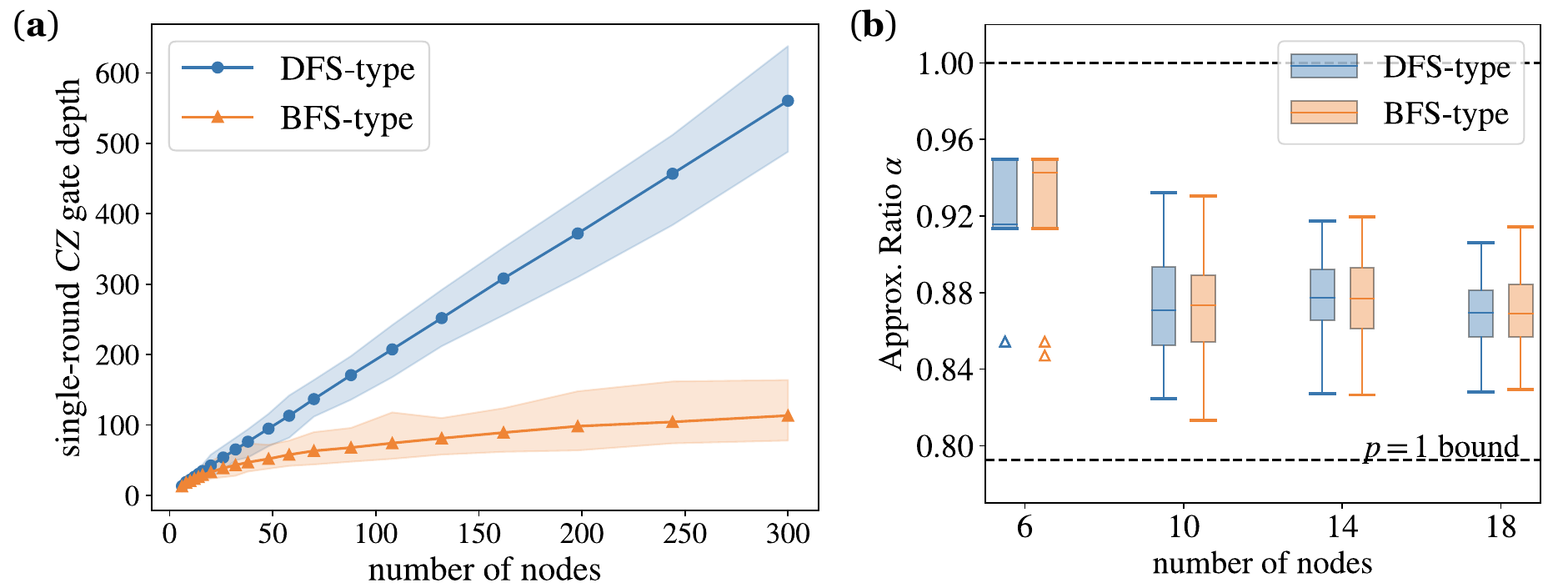}

    \caption{(\textbf{a}) $CZ$ gate depth and (\textbf{b}) approximation ratios of the bipolar-$ZY_1$ ansatz versus the number of graph nodes using the breadth-first-search (BFS)-type bipolar orientation algorithm and the depth-first-search (DFS)-type algorithm. Each data point in (\textbf{a}) is the average depth, and the colored band denotes the maximum and minimum depth among 200 randomly generated $3$-regular graphs. The bar plot in (\textbf{b}) is obtained by performing classical optimization using the bipolar-$ZY_1$ ansatz of 50 randomly generated $3$-regular graphs.}

    \label{fig:bipolar-orientation-depth}

\end{figure}

\vspace{0.2cm}

\noindent\textbf{BFS-type bipolar orientation ---} To reduce the circuit depth of the bipolar-$ZY_1$ ansatz, in the hardware implementation, we adopt a bipolar orientation algorithm based on breadth-first search (BFS)~\cite{PAPAMANTHOU2008224}. For a given problem graph $G$, the circuit depth of its bipolar-$ZY_1$ ansatz is proportional to the longest path length between source $s$ and sink $t$ of the bipolar-oriented graph. Different from the bipolar orientation based on DFS proposed by R. E. Tarjan~\cite{Tarjan1982}, the BFS-type bipolar orientation should have a shorter longest path between $s$ and $t$, and thus smaller circuit depth. This intuition is numerically verified by evaluating the $CZ$ gate depth of the bipolar-$ZY_1$ ansatz. Figure~\ref{fig:bipolar-orientation-depth} plots the $CZ$ gate depth and empirical approximation ratios of the bipolar-$ZY_1$ ansatz versus the number of nodes of random $3$-regular graphs, where the BFS-type and the DFS-type bipolar orientation algorithm are used. We see in Fig.~\ref{fig:bipolar-orientation-depth}(\textbf{a}) that the DFS-type bipolar orientation has $CZ$ gate depth that increases linearly with the number of nodes, whereas the BFS-type has a depth that increases almost logarithmically with the number of nodes. Additionally, Fig.~\ref{fig:bipolar-orientation-depth}(\textbf{b}) shows that the BFS-type and DFS-type bipolar orientation have almost the same empirical performance on the solution accuracy. For this reason, we construct the bipolar-$ZY_1$ ansatz using the BFS-type bipolar orientation in our hardware implementation.

\vspace{0.2cm}

\noindent\textbf{Circuit compilation ---} The above constructed bipolar-$ZY$ ansatz is a logical circuit. It remains to be compiled into instruction set architecture (ISA) circuits for execution on IBM's quantum chip~\cite{Qiskit}. To minimize both the circuit depth and gate numbers in the ISA circuits, we adopt the following steps for the bipolar-$ZY$ and QAOA ansatz:

\begin{itemize}

    \item For the bipolar-$ZY_1$ ansatz with a given biconnected graph $G=(\VV,\EE)$, first, we randomly choose 20 pairs of source and sink nodes of $G$, construct the logical circuit using BFS-type bipolar orientation and the procedure in Algorithm~1 of the main text. Second, each logical circuit is repeatedly compiled into 50 ISA circuits using IBM's preset random transpiler with \verb|optimization_level=3|. Finally, we select the ISA circuit with the minimum $CZ$ gate depth from the resulting $20\times 50$ ISA circuits.

    \item Each round of QAOA ansatz has driving terms $e^{-i\gamma Z_iZ_j}$ for each edge $(i,j)\in\EE$. To minimize the circuit depth, the greedy edge-coloring algorithm~\cite{Bravyi_2020} is applied, such that the driving terms can be performed in parallel as much as possible. Additionally, for each $\text{QAOA}_p$ ansatz with $p=1,2,3$, we repeatedly compile the logical circuit 50 times and select the ISA circuit with the minimum $CZ$ gate depth, as in the case of the bipolar-$ZY_1$ ansatz.

\end{itemize}

\vspace{0.2cm}

\noindent\textbf{Circuit execution ---} During the execution of all quantum circuits, dynamic decoupling with super-Hahn sequence~\cite{Ezzel2023} is used to suppress the decoherence error during the idle periods of the qubits. In the $148$-qubit demonstration, we additionally use randomized compiling~\cite{PhysRevX.11.041039} to mitigate the coherent noise during the execution of the variational ans\"atze.

 \vspace{0.2cm}

\noindent\textbf{Classical optimizer ---} We use the classical optimizer \verb|COBYLA| for the optimization of the bipolar-$ZY_1$ and $\text{QAOA}_p$ ansatz. Their variational parameters are initialized using the small constant initialization~\cite{riveradean2021avoiding}. Specifically, the initial values are chosen randomly within the range $[0,0.01]$ for all the ans\"atze in the $72$-qubit and $148$-qubit demonstrations, with two exceptions: the $\text{QAOA}_2$ and $\text{QAOA}_3$ in the $148$-qubit demonstration. These two ans\"atze have much more quantum gates than the others, and have stronger quantum noise during the execution. For them, we use larger random initial values within the range $[0.1,0.15]$. During the optimization, we use Conditional Value at Risk (CVaR) with a confidence level of $0.05$ as the objective function, which helps to accelerate optimization for combinatorial optimization problems~\cite{Barkoutsos2020improving}.



\vspace{0.2cm}

\noindent\textbf{Classical post-processing ---} The post-processing of the hardware results is implemented using a greedy algorithm as proposed in Ref.~\cite{sachdeva2024quantumoptimizationusing127qubit}. After sampling $01$ bit strings from the optimized variational ans\"atze, the greedy algorithm iteratively flips each $01$ bit string on an individual qubit if doing so improves the cost. This method can be regarded as performing local moves on the results until a local minimum is reached. The implemented post-processing algorithm is outlined in Algorithm~\ref{algo:Greedy Post-Processing}.

\begin{algorithm}[H]
\caption{Greedy post-processing}\label{algo:Greedy Post-Processing}
\begin{algorithmic}
 \State \textbf{Input:} An $n$-bit string $\boldsymbol{x}$, cost function $C$.
 \State \textbf{Output:} Bit string $\boldsymbol{x}_0$ at local minima.
 \State Initial $\boldsymbol{x}_0 \leftarrow \varepsilon$
\While{$\boldsymbol{x}_0 \neq \boldsymbol{x}$}
\State $\boldsymbol{x}_0 \leftarrow \boldsymbol{x}$
    \For{$j$ in Shuffle$(n)$}
        \State $\boldsymbol{x}^\prime \leftarrow$ Flip $\boldsymbol{x}$ in bit $j$.
        \If{$C(\boldsymbol{x}^\prime)<C(\boldsymbol{x})$}
        \State $\boldsymbol{x} \leftarrow \boldsymbol{x}^\prime$
        \EndIf
    \EndFor
\EndWhile
\end{algorithmic}
\end{algorithm}

\begin{figure*}
    \centering
    \includegraphics[width=0.95\textwidth]{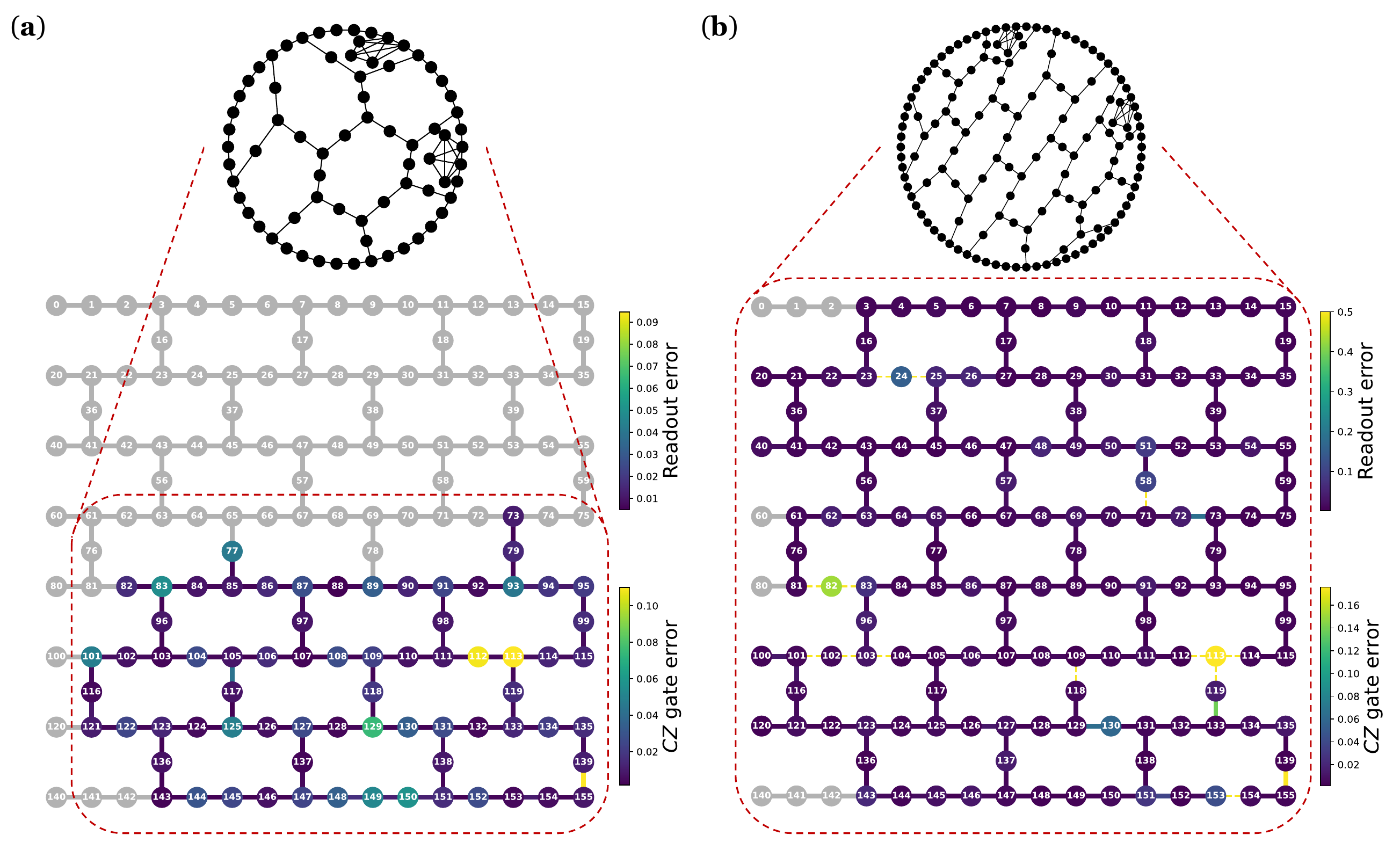}
    \cprotect\caption{(\textbf{a}) The $72$-node graph (upper panel) and its layout on \verb|ibm_fez|'s coupling map (lower panel). The coupling map is colored to represent the readout error for each qubit and the two-qubit $CZ$ gate error for each qubit connection. (\textbf{b}) The $148$-node graph (upper panel) and its layout on \verb|ibm_marrakesh|'s coupling map (lower panel). The couplings with yellow dashed lines denote that the data is unavailable from the IBM quantum platform.}
    \label{fig:graph-circuit}
\end{figure*}
\vspace{0.2cm}
\noindent\textbf{Hardware calibration details ---} The hardware implementation results in Fig.~7 of the main text are run on \verb|ibm_fez| and \verb|ibm_marrakesh| with coupling maps shown in Fig.~\ref{fig:graph-circuit}(\textbf{a}) and (\textbf{b}). The $72$-node and $148$-node graphs are mapped to the colored qubits in Fig.~\ref{fig:graph-circuit}(\textbf{a}) and (\textbf{b}), respectively. The coupling maps are colored to represent the readout error rate for each qubit and the two-qubit $CZ$ gate error rate for each qubit connection. Other single-qubit properties of \verb|ibm_fez| and \verb|ibm_marrakesh| are summarized in Table~\ref{table:qubit-properties}. All hardware data are obtained from the IBM cloud quantum platform~\cite{Qiskit} and more details are available in~\cite{Wang2025_lightcone}.

\begin{table}
    \centering
    
        \begin{tabular}{c|cccc}
\hline
 \verb|ibm_fez|& median & mean & min & max \\
\hline\hline
$\sqrt{X}$ error (\%)& 0.024 & 0.032$\pm$0.0198 & 0.013 & 0.098 \\
Readout error (\%)& 1.52 & 2.10$\pm$1.77 & 0.50 & 9.46 \\
$T_1 (\mu s)$ & 123.49 & 124.98$\pm$40.01 & 43.78 & 201.40 \\
$T_2 (\mu s)$ & 88.44 & 87.45$\pm$51.90 & 5.60 & 225.27 \\
\hline
\end{tabular}
\quad
        \begin{tabular}{c|cccc}
\hline
\verb|ibm_marrakesh|& median & mean & min & max \\
\hline\hline
$\sqrt{X}$ error (\%)& 0.026 & 0.070$\pm$0.357 & 0.010 & 4.195 \\
Readout error (\%)& 0.93 & 2.44$\pm$5.76 & 0.17 & 50.00 \\
$T_1 (\mu s)$ & 193.23 & 203.60$\pm$73.87 & 6.65 & 439.29 \\
$T_2 (\mu s)$ & 101.84 & 124.44$\pm$93.59 & 0.18 & 517.57 \\
\hline
\end{tabular}
     \cprotect\caption{Summary of single-qubit properties in \verb|ibm_fez| (left table) and \verb|ibm_marrakesh| (right table) used in the hardware implementation.}
    \label{table:qubit-properties}
\end{table}

\end{document}